\documentclass[useAMS,usenatbib,usegraphicx]{mn2e}
\usepackage{verbatim}
\usepackage{xspace}
\usepackage{rotating}
\usepackage{array}

\def\kms{km~s$^{-1}$\xspace}

\def\msun{$\rm M_{\sun}$\xspace}
\def\Msun{$\rm M_{\sun}$\xspace}

\newcommand{\oiii}{[O$\:${\small III}]\xspace}
\newcommand{\hbeta}{H$\beta$\xspace}
\newcommand{\otohb}{[O$\:${\small III}]/H$\beta$\xspace}
\newcommand{\mm}{$\mu$m\xspace}
\newcommand{\mum}{$\mu$m\xspace}
\newcommand{\bve}{$(B-V)_{\mathrm e}$\xspace}
\newcommand{\bv}{$B-V$\xspace}
\newcommand{\hi}{H$\:${\small I}\xspace}

\title[Molecular gas and star formation in early-type galaxies]{Molecular gas and star formation in early-type galaxies}
\author[A. F. Crocker et al.]{Alison F. Crocker$^{1,2}$, Martin
  Bureau$^{2}$, Lisa M. Young$^{3,4}$, Francoise Combes$^{5}$\\
$^{1}$Department of Astrophysics, University of Massachusetts, 710
  North Pleasant Street, Amherst, MA, 01003\\
$^{2}$Sub-Department of Astrophysics, University of Oxford, Denys
  Wilkinson Building, Keble Road, Oxford OX1 3RH \\
$^{3}$Physics Department, New Mexico Institute of Mining and
Technology, Socorro, NM 87801, U.S.A. \\
$^{4}$Adjunct Astronomer at the National Radio Astronomy Observatory ???????\\
$^{5}$Observatoire de Paris, LERMA, 61 Av. de l'Observatoire, 75014,
Paris, France}

\begin{document}

\date{}

\pagerange{\pageref{firstpage}--\pageref{lastpage}} \pubyear{}

\maketitle

\label{firstpage}

\begin{abstract}   

We present new mm interferometric and optical integral-field unit
(IFU) observations and construct a sample of 12 elliptical (E) and
lenticular (S0) galaxies with molecular
gas which have both CO and optical maps. The
galaxies contain $2 \times 10^7$ to $5 \times 10^9$ \Msun of molecular
gas distributed primarily in central discs or rings (radii 0.5 to
4~kpc). The molecular gas distributions are always coincident with 
distributions of optically-obscuring dust that reveal tightly-wound
spiral structures in many cases. The ionised gas
always approximately corotates with the molecular gas, evidencing a
link between these
two gas components, yet star formation is
not always the dominant ionisation source. The galaxies
with less molecular gas tend to have 
\otohb emission-line ratios at high values not expected for star formation. Most
E/S0s with molecular gas have young or intermediate age stellar
populations based on optical colours, ultraviolet colours and absorption
linestrengths. The few that appear purely old lie
close to the limit where such populations would be
undetectable based on the mass fractions of expected young to observed old
stars. The 8\mum polycyclic aromatic hydrocarbon (PAH) and 24\mum emission yield similar star formation rate estimates of E/S0s, but the total infrared
overpredicts the rate due to a contribution to dust heating
from older stars. The radio-far infrared
relation also has much more scatter than for other star-forming
galaxies. However, despite these biases and additional scatter, the
derived star formation rates locate the E/S0 galaxies within the
large range of the Schmidt-Kennicutt and constant efficiency star
formation laws. Thus the star formation
process in E/S0s is not overwhelmingly different than in other
star-forming galaxies, although one of the more reliable tracers (24\mum) points to a possible  lower star-formation
efficiency at a given gas surface density.

\end{abstract}


\section{Introduction}

Traditionally, much attention has been given to the homogeneity of
elliptical and lenticular (hereafter early-type or E/S0)
galaxies. Most famously, these 
galaxies lie on a surface within the volume space of effective radius,
mean surface brightness and central velocity dispersion \citep[the
  Fundamental   Plane;
  e.~g.][]{faber87,dressler87,djorgovski87}. Early-type galaxies also
follow a tight colour-magnitude relation, brighter galaxies being
redder \citep[e.~g.][]{baum59, visvanathan77}. The small 
scatter in this relation requires the stellar content of the galaxies
to be fairly uniform and
predominantly old \citep[e.~g.][]{bower92}. The interpretation of absorption
linestrength indices shows that age, metallicity and alpha-element enhancement
all depend smoothly on galaxy mass, more massive galaxies being
older, more metal-rich and more alpha-element enhanced
\citep{thomas05}.  Initial observations showed that early-type
galaxies as a group lack a significant cool or cold interstellar
medium (ISM), with both
atomic hydrogen (H$\:${\small I}) and carbon-monoxide (CO) non-detections
\citep{faber76, johnson79}. 

However, observations over the past few decades have also gradually
revealed the individuality of early-type galaxies. In terms of radial
structure, some E/S0s have cuspy cores or missing light at small
radii, while others appear coreless 
\citep[e. g.][]{kormendy09}. Some have very little ordered rotation
while others rotate quickly \citep[quantifiable with the specific
  angular-momentum proxy $\lambda_{\mathrm
  R}$;][]{emsellem07}. Stellar populations and all measurements of the
gas phase also show a wide spread. Despite their generally uniform
optical colours, a range in stellar population ages (particularly
towards young ages) is found for E/S0 galaxies through the analysis of
absorption linestrengths \citep[e.~g.][]{trager00} and ultraviolet-optical
colours \citep[e.~g.][]{yi05}. Ionised gas exhibits a remarkable
complexity in early-type galaxies, with different distributions,
luminosities, line ratios and equivalent widths not obviously
correlated to any particular galaxy property \citep{sarzi06,sarzi09}. 
Molecular gas, \hi and warm dust have all now been detected in a
significant fraction of early-type galaxies, with no strong
dependence on galaxy luminosity or environment, except for the lack of
atomic hydrogen in clusters \citep[e.~g.][Oosterloo et al. in preparation, Young et al. in preparation]{knapp89, morganti06, combes07}.  

Of course, several of these varying properties are likely to be related. In
particular, galaxies with molecular gas reservoirs are
expected to be star-forming. This star formation should naturally lead
to ionised gas and young stellar populations.  However, little is 
currently known about the origin and evolution of the cold gas and the young
stars in E/S0s. Single-dish studies of the molecular gas show that the
gas mass does not correlate as well with the blue luminosity of
early-types as it does for spirals \citep{knapp96, combes07}. However,
the molecular gas mass is tightly correlated with the far-infrared
(FIR) luminosity, matching the correlation for spiral galaxies
\citep{wiklind89,wiklind95,combes07}. This match suggests that the
star formation efficiency is the same in early-type galaxies as it is
in spirals, implying that a given amount of molecular gas transforms
into the same amount of stars per unit time. However, caution is
necessary as theory predicts that the general interstellar radiation
field (ISRF) of early-type galaxies should be of sufficient strength
to heat the dust observed  without requiring any young stars
\citep{jura82}. In this case, the close relation between molecular gas
mass and FIR luminosity simply arises through a constant molecular
gas to dust ratio.  

Despite these single-dish based efforts, many questions regarding cold
gas and star formation in early-type 
galaxies remain. First, where does the gas come from? Predictions of
internal stellar mass loss would provide more gas than is
observed if this gas were capable of cooling
\citep[e.~g.][]{faber76}. Most galaxies have far too little gas to
match these naive predictions, but several galaxies with 
interferometric CO maps (revealing the direction of rotation) show
that the gas origin is at least consistent with such internal stellar mass loss
\citep*{young02, young08}. Of course, some clear cases in which the
misalignment of the  molecular gas rotation with respect to the stellar rotation
indicates external accretion are also documented \citep{young08,
  crocker08}. A large statistical sample with interferometric CO data is
thus needed to infer the relative importance of internal versus external
accretion processes. 

A second major question is: how does star formation proceed in E/S0
galaxies? As mentioned above, the good FIR-CO correlation need not
arise from star formation if other dust-heating sources are
present. Moreover, some  early-type galaxies do not seem to be forming
stars as efficiently as predicted. \citet{okuda05} argue that the
molecular gas observed in the centre of NGC~383 is stable to
gravitational collapse and thus not forming stars. Infrared
\citep*{young09} and radio \citep{lucero07} observations of NGC~2320
also do not obviously indicate star formation from its massive
molecular disc. Theoretical models from \citet{kawata07} suggest
that circumnuclear discs in massive galaxies are more stable than
those in less massive galaxies. Moreover, a recent cosmological
simulation shows that the potential of a massive spheroid can
significantly reduce the star formation efficiency of a gas disc
\citep[`morphological quenching';][]{martig09}. Further investigations into the connection between
molecular gas and star formation in early-type galaxies are thus needed.  

If star formation from molecular gas were the dominant property of
early-type galaxies, they probably would not be classified as
early-type. As such, the molecular gas and star formation are always
subtle features and require deeper and more specific observations to
investigate in detail. An important step in this direction is to
obtain resolved maps of the cold gas, the star formation and
the stellar populations. The goal of this study is thus to use
interferometric CO maps and optical integral-field unit (IFU) maps to
compare the molecular  gas to the ionised gas, stellar populations and
stellar kinematics in a sample of early-type galaxies. In doing so, we
thoroughly investigate the connection between the molecular gas and
star formation. Our moderate sample size allows us to focus on
individual galaxies when warranted, as well as to try to create a global picture of the
evolution of early-types with molecular gas.   

This paper presents some new mm interferometric data on five
galaxies, and investigates the
general molecular gas, ionised gas and stellar properties within the
sample of twelve early-type galaxies that have both CO maps and IFU
data (including literature data). Section 2 describes the sample and
Section 3 explains both the interferometric and IFU observations and
data reduction. We present the new CO maps of three early-type galaxies in
Section 4, as well as characterising two CO non-detections 
and two mm continuum detections. In Section 5, we switch from the analysis of
individual galaxies to a consideration of the sample as a whole,
investigating the  properties of the molecular gas alone and with
respect to the IFU data. Section~6 presents an analysis
of the origin of the cold gas and then delves into the star
formation properties of these galaxies, investigating the validity of various
star formation tracers and comparing our sample galaxies to the
predictions of the Schmidt-Kennicutt and constant efficiency star
formation laws. We summarise our findings in Section 7.  

\begin{figure}
\begin{center}
\includegraphics[width=7cm]{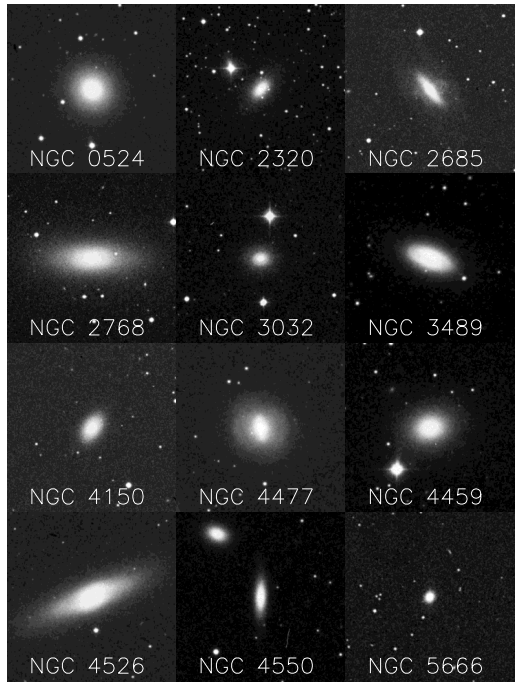}
\caption{DSS $R$-band images of the {\bf 12} sample E/S0 galaxies with
  molecular gas. Each image is 7\arcmin $\times$ 7\arcmin.}
\label{fig:dss}
\end{center}
\end{figure}

\section[] {Sample}

Our goal is to obtain as large a sample as possible of early-type galaxies with
both interferometric CO data and optical IFU maps. We
start from the IFU observations of the the {\tt SAURON}
instrument, which consist of the main {\tt SAURON} sample of
48 E and S0 galaxies and a handful of other E and S0 galaxies observed as
`specials' for particular projects (most of these are listed in Table~3 of
\citealp{cappellari07}). The selection of the main sample was designed
to be representative (as opposed to complete), with equal numbers of Es and S0s and cluster and
non-cluster members, additionally trying to uniformly populate
ellipticity-magnitude space \citep{dezeeuw02}.

Combining a detection campaign with the Institut de Radioastronomie
Millim\'etrique (IRAM) 30m telescope \citep{combes07} and data from the
literature, 13/48 galaxies of the main sample are detected in CO
from single-dish measurements. Follow-ups to obtain interferometric
molecular gas data have mapped 4 galaxies with the
Berkeley-Illinois-Maryland Array \citep[BIMA;][]{young08}
and 7 galaxies with the IRAM Plateau de Bure Interferometer
\citep[PdBI;][5 presented in this paper]{crocker08, crocker09}.  Two
galaxies observed with PdBI turn out to be non-detections (described
below). One galaxy, NGC~2685,  already has interferometric data in the literature
\citep{schinnerer02}. Thus out of the original 13 CO detections from
the main {\tt SAURON} early-type sample, 10 now have interferometric
maps, 2 are false single-dish detections and for only one, NGC~3156, we
have been unable to obtain interferometric data. 

No systematic survey of molecular gas in the {\tt SAURON} `specials' sample has been
undertaken, but a literature search reveals CO detections for three:
NGC~2320, NGC~5666 and NGC~5866 \citep{wiklind95, welch03}. NGC~2320
and NGC~5666 both have interferometric CO maps \citep{young02,
  young05}, so we also include these two galaxies in our sample. As they
have not gone through the same selection process, we briefly describe
both of them here. 

 NGC~2320 lies at 83~Mpc, more distant than the {\tt SAURON} sample
limit of 40~Mpc. It is classified as an elliptical in both the RC3
\citep{devaucouleurs91} and UGC \citep{nilson73} catalogues and is a
member of the Abell 569 cluster.  

NGC~5666 (at 35~Mpc) is a difficult galaxy to
classify. \citet{donzelli03} conclude that NGC~5666 has too much gas 
and spiral structure to be a true early-type, but also has too high a
bulge-to-total ratio and too old stellar populations for a normal Scd
galaxy. An $r^{1/4}$ law 
fits the Spitzer Multiband Imaging Photometer (MIPS) 24~\mum surface
brightness profile reasonably well (based on data from
\citealp{young09}), hinting at the dominance of a classic de
Vaucouleurs profile. We thus include NGC~5666 in our
sample, but remain aware of its uncertain morphology. In particular, we
note that it  did not meet the visual early-type selection criteria
for the complete Atlas3D sample, which follows upon the {\tt SAURON}
survey (Cappellari et al. in preparation).  

We therefore have a total of 12 galaxies with both IFU and CO maps (10 from the
main sample and 2 `specials'). Unfortunately
the CO distribution of NGC~2685 lies in a polar ring outside of the
region mapped with the {\tt SAURON} IFU,  so sometimes we will not be
able to discuss this galaxy with the rest of the sample. Additionally, the CO-detected but unmapped galaxy from the main {\tt SAURON} sample, NGC~3156, is included when information on the CO distribution is not required. The inclusion of the two `special' galaxies complicates the
sample, but we are primarily interested in
considering the range of molecular gas and star formation properties in E/S0s
and so want the largest sample size possible. When the statistics of
the properties are important, we will limit the sample to the
galaxies from the main {\tt SAURON} sample. However, more detailed work in this vein 
requires a much larger (and complete) sample. Digital Sky
Survey (DSS) $R$-band images for the 12 sample galaxies are shown in 
Fig.~\ref{fig:dss}.

\section[]{Observations and Data Reduction}

\subsection[]{CO observations and data analysis}

\begin{table}
 \caption{Observing parameters}
 \begin{tabular}{@{}lccccc}
  \hline
  Galaxy & Date & Config. &  Ant. & Receiver & {\it T}$_{\mathrm {sys}}$\\
  (NGC) & & & & & (K) \\
  \hline
  \phantom{0}524 & 24/08/07 & D & 5 & new: 3mm & $250$\\
   & 21/12/07 & C & 6 & new: 3mm & $200$\\
  3489 & 23/04/07 & C & 6 & new: 3mm & $225$ \\
   & 23/10/08 & D & 6 & new: 3mm & $250$ \\
   & 24/10/08 & D & 6 &  new: 3mm & $350$ \\ 
  4278 & 24/12/05 & C & 6 & old & 400, 300\\
   & 01/11/07 & D & 6 & new: 1mm & 200\\
  4477 & 09/01/06 & C & 5/6 & old & 250, 300\\
   & 30/04/06 & D & 6 & old & 225, 350\\
  7457 & 29/06/07 & D & 4/5 & new: 3mm & 200\\
   & 13/07/07 & D & 5 & new: 3mm & 250\\ 
  \hline
 \end{tabular}
\label{tab:obspar}
\end{table}

We observed five galaxies (NGC~524, NGC~3489, NGC~4278, NGC~4477
and NGC~7457) in the C and D configurations at the PdBI. Observing
parameters for individual galaxies are listed in
Table~\ref{tab:obspar}. The receivers were 
upgraded in December 2006. After this point, the CO(2-1) and CO(1-0)
lines could no longer be simultaneously observed. The correlator
configuration before December 2006 had a total bandwidth of
580 MHz and 1.25 MHz resolution. Since December 2006, the total bandwidth is
950~MHz with 2.5 MHz resolution. During the observations, the
correlator was regularly calibrated by a  noise source inserted in the
IF system.   
   
The data were reduced with the Grenoble Image and Line Data Analysis
System ({\sc GILDAS}) software packages {\sc CLIC} and {\sc MAPPING} \citep{GL}.
Using CLIC, we first calibrated the data using
the standard pipeline for bandpass calibration, phase calibration,
absolute flux calibration and amplitude calibration. Data obtained
during periods of bad weather were flagged and then ignored. After
calibration, we used {\sc MAPPING} to create data cubes with velocity planes
separated by 20 km s$^{-1}$ and natural weighting. The pixel size  
for each map is based upon the size of the beam, both of which are reported in Table~\ref{tab:mappar}.  
For the three galaxies with detected CO emission, NGC~524, NGC~3489 and NGC~4477, we cleaned using the H\"ogbom method \citep{hogbom74}. We stopped cleaning
in each velocity plane after the brightest residual pixel had a value
lower than the rms noise of the uncleaned datacube. We used the {\sc MIRIAD}
task {\tt contsub} to subtract the continuum present in NGC~524. 

\begin{table}
 \caption{Mapping parameters} 
 \begin{tabular}{@{}lcccr}
  \hline
  Galaxy & CO Line & Clean Beam & Pixel Size & {\it V}$_{\mathrm{sys}}$~~~~\\
  (NGC) & & & & (km s$^{-1}$) \\
  \hline
  \phantom{0}524 & (1-0) & $2 \farcs 8 \times 2\farcs 6$ & $0\farcs 68
  \times 0\farcs 68$ & 2379 \\
  3489 & (1-0) & $3\farcs 1 \times 2\farcs 9$ & $0\farcs 75
  \times 0\farcs 75$ & 713\\
  4278 & (2-1) & $1\farcs 7 \times 1\farcs 5$ & $0\farcs 40 \times
  0\farcs 40$ & 649\\
  4477 & (1-0) & $3\farcs 3 \times 2\farcs 6$ & $0\farcs 75
  \times 0\farcs 75$ & 1355\\
   & (2-1) & $1\farcs 6 \times 1\farcs 2$ & $0\farcs 375 \times
  0\farcs 375$ & 1355\\
  7457 & (1-0) & $4\farcs 2 \times 3\farcs 5$ & $1\farcs 00
  \times 1\farcs 00$ & 812\\ 
  \hline
 \end{tabular}
Note: Systemic velocities from NASA/IPAC
  Extragalactic Database (NED),
  except for NGC~3489 where it is taken from the {\tt SAURON} stellar
  kinematics \citep{emsellem04}.
\label{tab:mappar}
\end{table}

\begin{figure*}
\begin{center}
\begin{tabular}[c]{cccc}
NGC 524 & NGC 3489 & NGC 4477 & NGC 4477\\
CO(1-0) & CO(1-0) & CO(1-0) & CO(2-1) \\
\rotatebox{0}{\includegraphics[width=4.2cm]
  {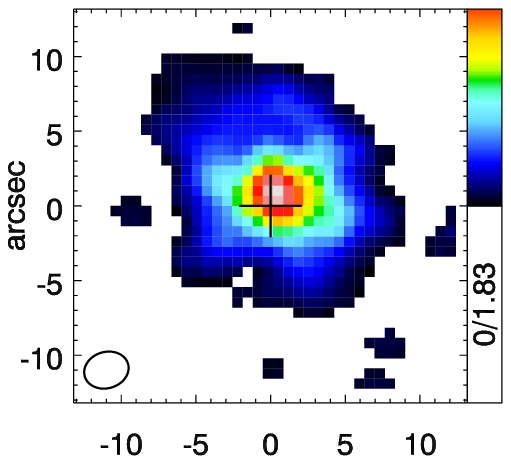}} &
\rotatebox{0}{\includegraphics[width=4.2cm]
  {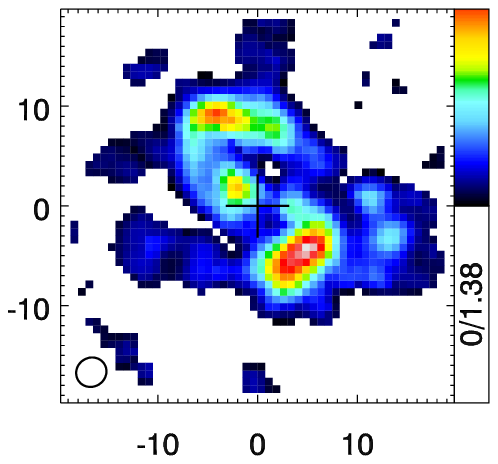}} &
\rotatebox{0}{\includegraphics[width=4.2cm]
  {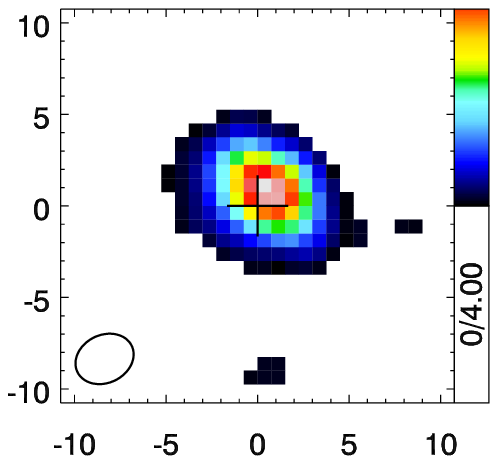}} &
\rotatebox{0}{\includegraphics[width=4.2cm]
  {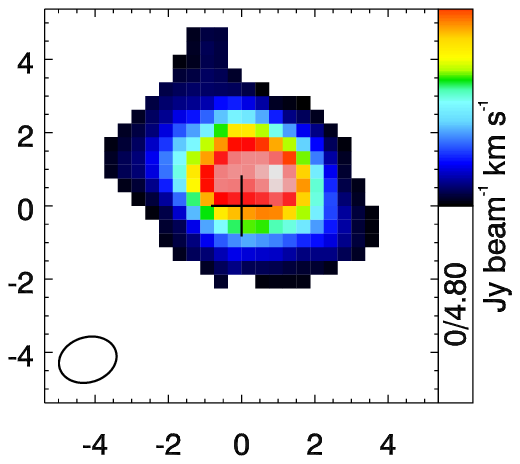}}
\\
\rotatebox{0}{\includegraphics[width=42mm]
  {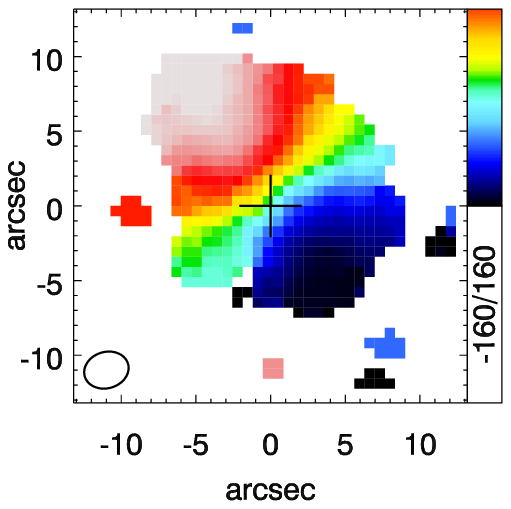}} &
\rotatebox{0}{\includegraphics[width=4.2cm]
  {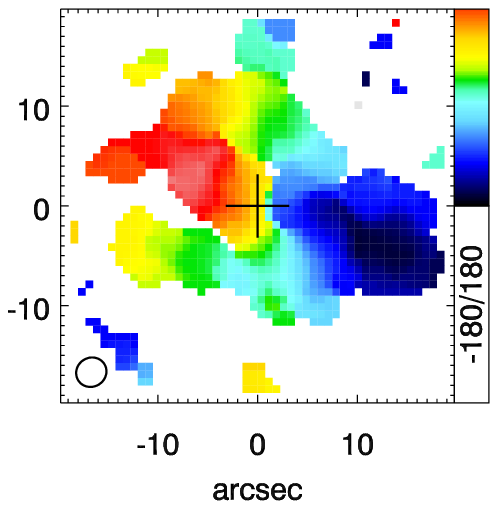}} &
\rotatebox{0}{\includegraphics[width=4.2cm]
  {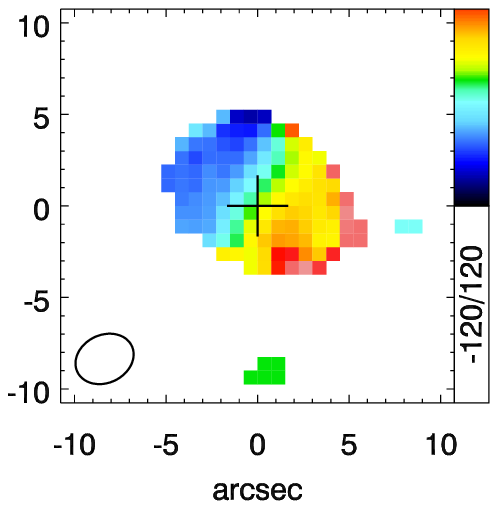}} &
\rotatebox{0}{\includegraphics[width=4.2cm]
  {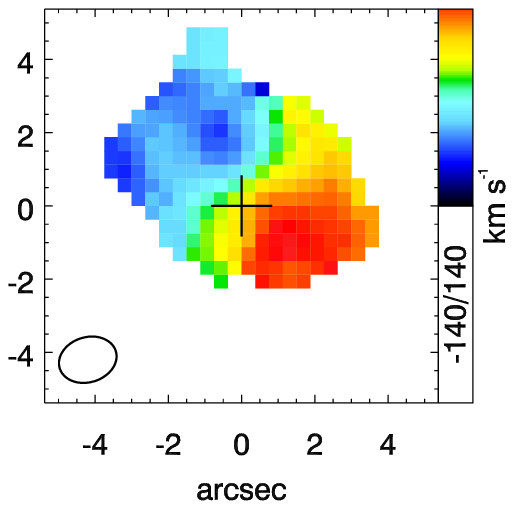}}\\
\\
\end{tabular}
\end{center}
\caption{{\em Top:} Integrated intensity maps (Jy beam$^{-1}$ km
  s$^{-1}$). {\em Bottom:} Mean velocity   maps (km s$^{-1}$),
  relative to the systemic velocity of each galaxy (see values 
  listed in Table~\ref{tab:mappar}). White pixels have been masked
  according to the method described in Section~3.1. The beam size is
  indicated in the lower left-hand corner of each panel and the cross marks the
  centre of the galaxy from the Two Micron All Sky Survey
  (2MASS). Note that the CO(2-1) map for NGC~4477 is on a smaller
  spatial scale than the CO(1-0) map. }  
\label{fig:COmaps}
\end{figure*}

For the galaxies with detected CO emission, we made integrated
intensity and mean velocity maps using the 
smoothed-mask method (see Fig.~\ref{fig:COmaps}). First a smoothed cube is created by Hanning
smoothing in velocity and smoothing with a 4-pixel circular
Gaussian spatially. Then, all values in this cube below three times the rms
noise are masked. This mask is then used on the original cube to
compute the moments. With the extent of the molecular gas defined by
the 0th moment map, we then integrate over this region for all
velocities to obtain a total spectrum for each galaxy. These spectra are shown in
Fig.~\ref{fig:spec}, where the shaded regions indicate the velocity
widths integrated over to obtain the total fluxes. 

We compute the molecular hydrogen masses from the total CO(1-0) fluxes
using the formula $M(\mathrm{H}_2) = (1.22 \times 10^4
$~M$_{{\tiny \sun}})D^2 \cdot S_{\mathrm{CO}}$, where {\it D }is the distance
measured in Mpc and  $S_{\mathrm{CO}}$ is the total CO(1-0) flux in Jy
km s$^{-1}$. This
formula comes from using the standard CO to H$_{2}$ conversion
 ratio $N(\mathrm{H}_2)/I(\mathrm{CO}) = 3 \times 10^{20}$ cm$^{-2}$
 (K  km s$^{-1}$)$^{-1}$, where $N(\mathrm{H}_{2})$ is
 the column density of H$_{2}$ and $I(\mathrm{CO})$ is the CO(1-0) intensity in
 K km s$^{-1}$ based on the main-beam temperature, $T_{\mathrm{mb}}$.

For the non-detected galaxies, we integrated the datacube over a
central 5\arcsec~radius aperture, obtaining the spectra shown in
Fig.~\ref{fig:spec}. We then measured the noise in each spectrum and
obtained a flux upper limit by the formula 3~$\sigma_{\mathrm
  {rms}}\sqrt{n}$, where $n$ is the number of channels in which emission is expected. The velocity
width used for NGC~4278 is -240 to 240 km s$^{-1}$ (25 channels) and
for NGC~7457 it is -140 to 140 km s$^{-1}$ (15 channels) about the
systemic velocity. These are based
on the maximum velocity seen in the ionised gas within the {\tt SAURON}
field-of-view. We convert these CO flux upper limits to molecular gas
masses using the same formula as for the detected galaxies. These
mass upper limits are listed in Table~\ref{tab:COFlux}.

To constrain the continuum emission, we created continuum {\it uv}
tables for each galaxy using all frequencies without any detected line
emission. For the galaxies with line emission, we selected channels at
least 40 km s$^{-1}$ away from the highest and lowest velocity
channels with any line emission. The very edges of the bandwidth were
also avoided. No continuum emission was detected in NGC~3489, NGC~4477 or
NGC~7457. In NGC~524 and NGC~4278, we fit the continuum with a point
source model in the {\it uv} plane. The fit results give the flux and
location of the point source. In both cases, a point source located at
the galaxy centre was a good fit. For NGC~4278, we have observations at
1~mm separated by a period of nearly two years. We created separate
{\it uv} tables for the two observations and find that the fluxes are
different by a factor of about 2, even after special care is given to
the flux calibrations.

\subsection[]{{\tt SAURON} IFU observations and data analysis}

The {\tt SAURON} IFU data have been presented in a series of papers: the
stellar kinematics in \citet{emsellem04}, the ionised gas in
\citet{sarzi06,sarzi09}, the absorption linestrengths in \citet{kuntschner06}
and the stellar populations in \citet{kuntschner09}. However, since
these papers have been published \citep[except][]{kuntschner09}, a new
stellar template library consisting of single stellar population (SSP)
models from Vazdekis et al. \citep[in preparation, based on MILES
stellar spectra from][]{sanchezblazquez06} and emission-free {\tt
SAURON} spectra has been used for the data analysis
\citep[see][]{sarzi09, kuntschner09}. The data we discuss and the
maps we present here are all based upon the new stellar template
library. 

NGC~2320 and NGC~5666 (outside of the main {\tt SAURON} sample) were also
run through the {\tt SAURON} reduction and 
analysis pipeline, with only minor modifications to accommodate the
higher redshift of NGC~2320 (in particular, the disappearance of the
$\lambda$5170 \AA~Mg triplet). As their {\tt SAURON} data have not
previously been presented, we present the {\tt SAURON} maps for these
galaxies in Appendix~1, along with some short notes on both galaxies.

\section[] {Millimetre Results}

We detect CO emission in NGC~524, NGC~3489 and NGC~4477, but not in
NGC~4278 or NGC~7457. Continuum emission is detected in both NGC~524
and NGC~4278. In NGC~4278, we infer an intrinsic variation of the
continuum emission associated with its AGN (Section~4.4). Here we
present the millimetre data for these five individual galaxies.  

\subsection[]{Comparison to single-dish observations}

\begin{figure*}
\begin{center}
\rotatebox{270}{\includegraphics[clip=true, width=4cm]
{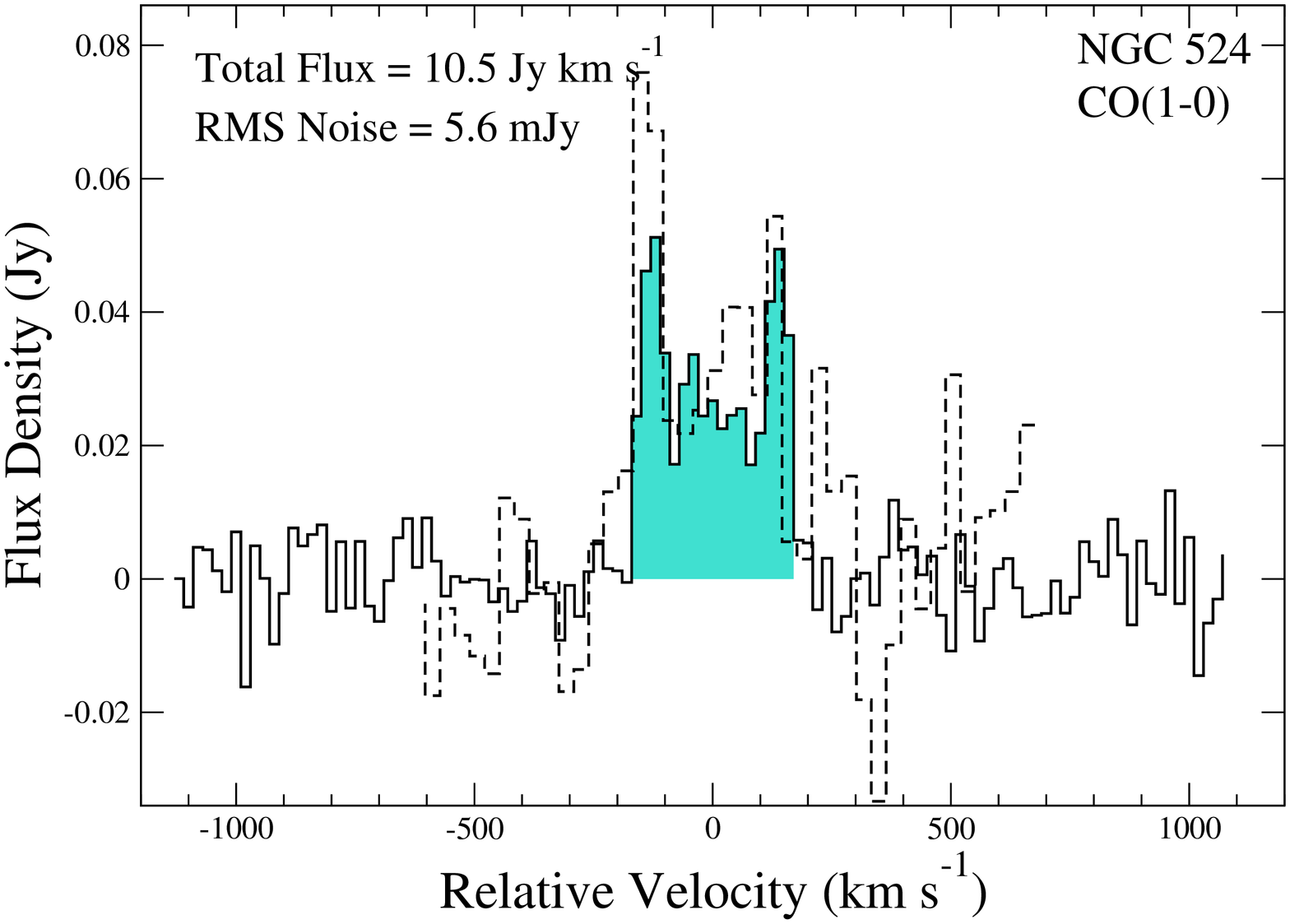}} 
 \rotatebox{270}{\includegraphics[clip=true,width=4cm]
{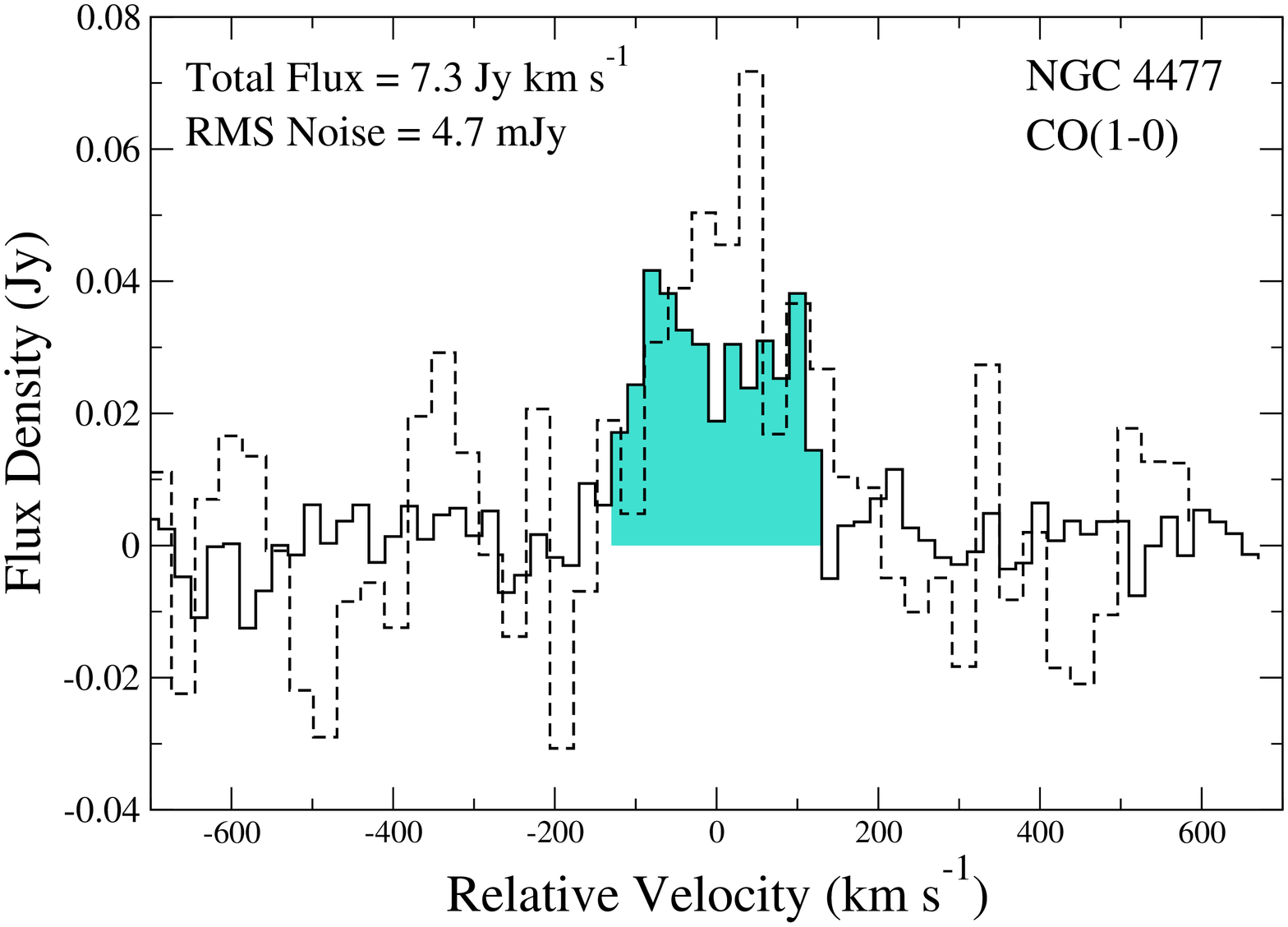}} 
\rotatebox{270}{\includegraphics[clip=true,width=4cm]
{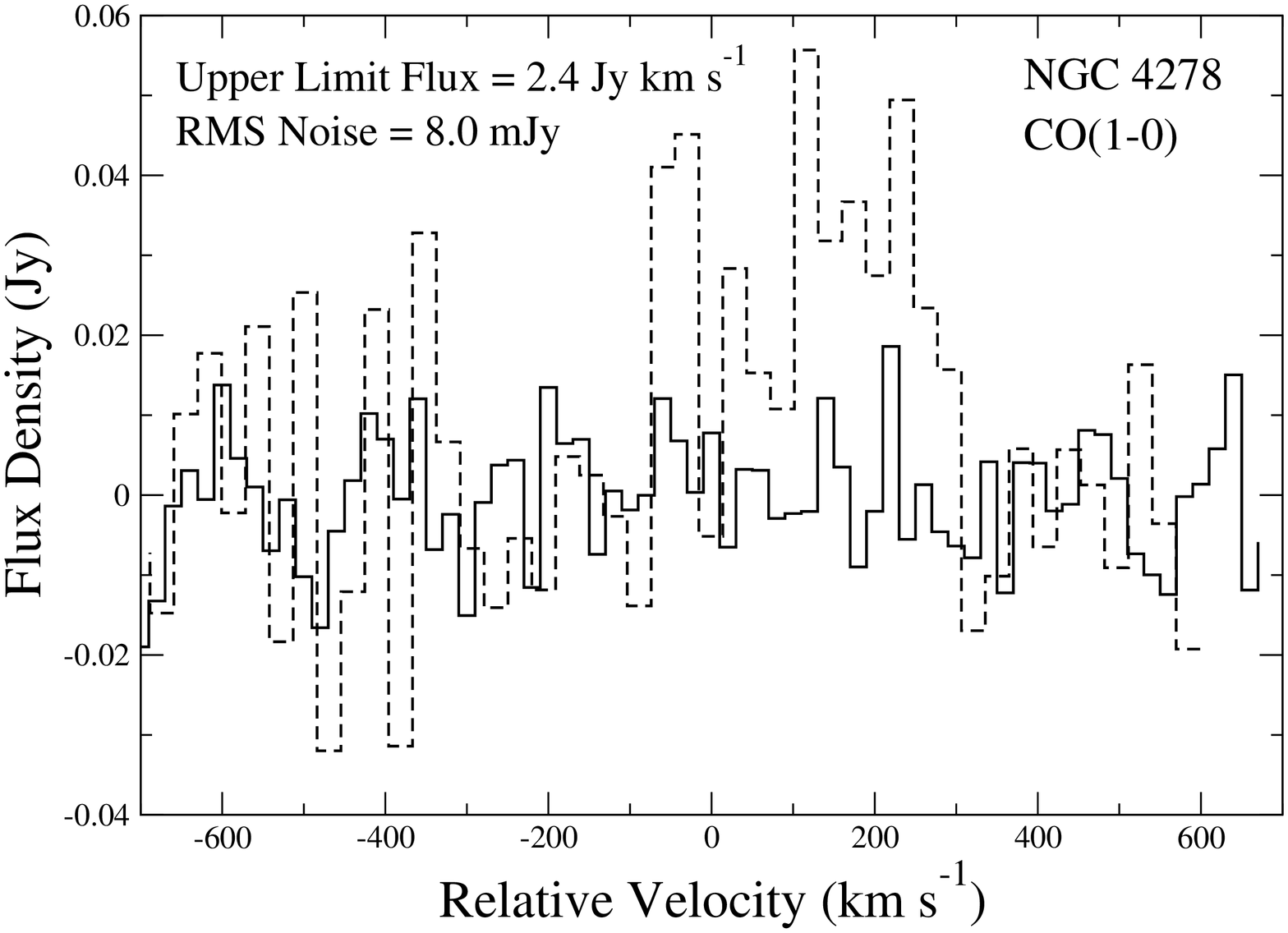}} 
\rotatebox{270}{\includegraphics[clip=true, width=4cm]
{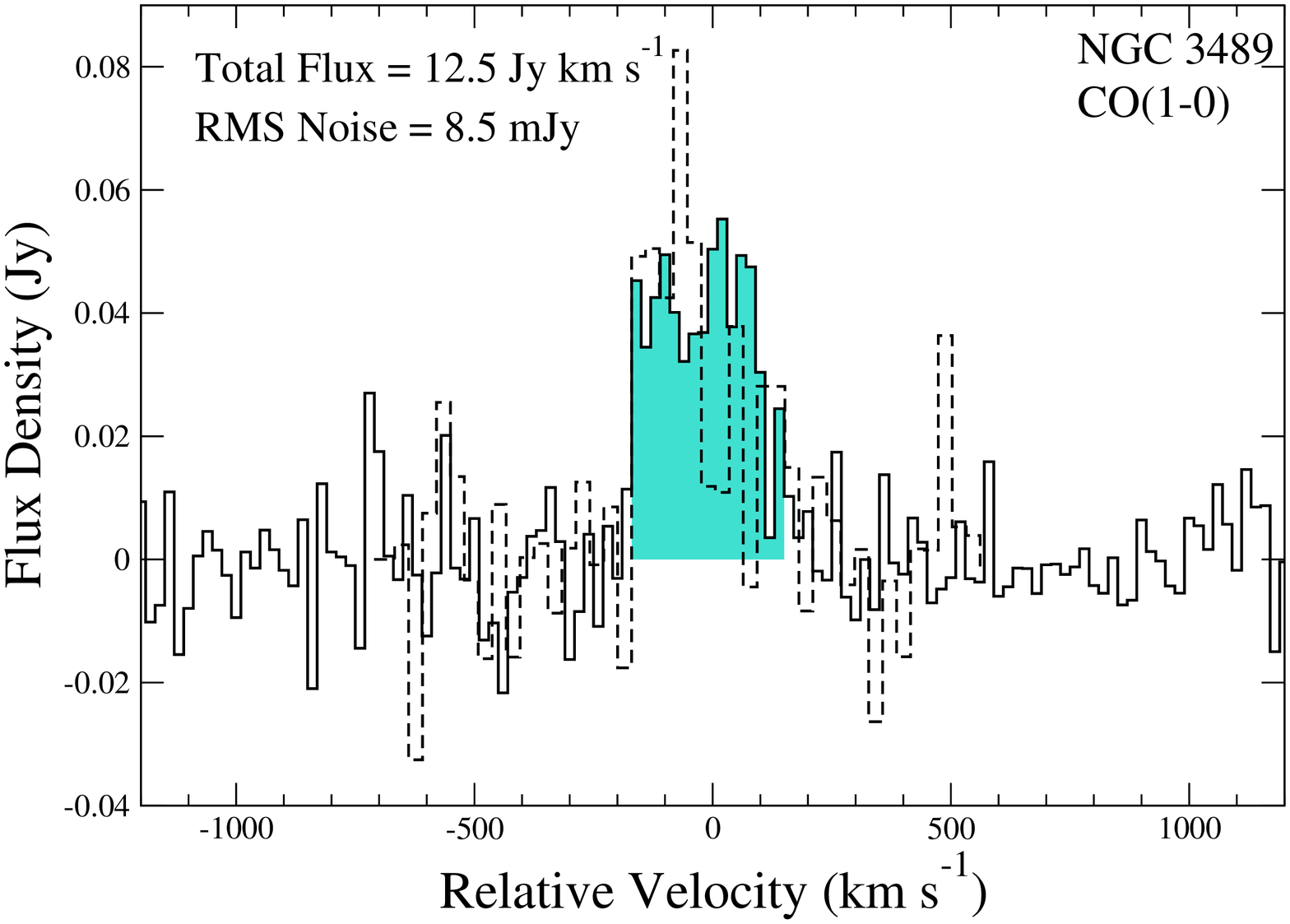}} 
\rotatebox{270}{\includegraphics[clip=true,width=4cm]
{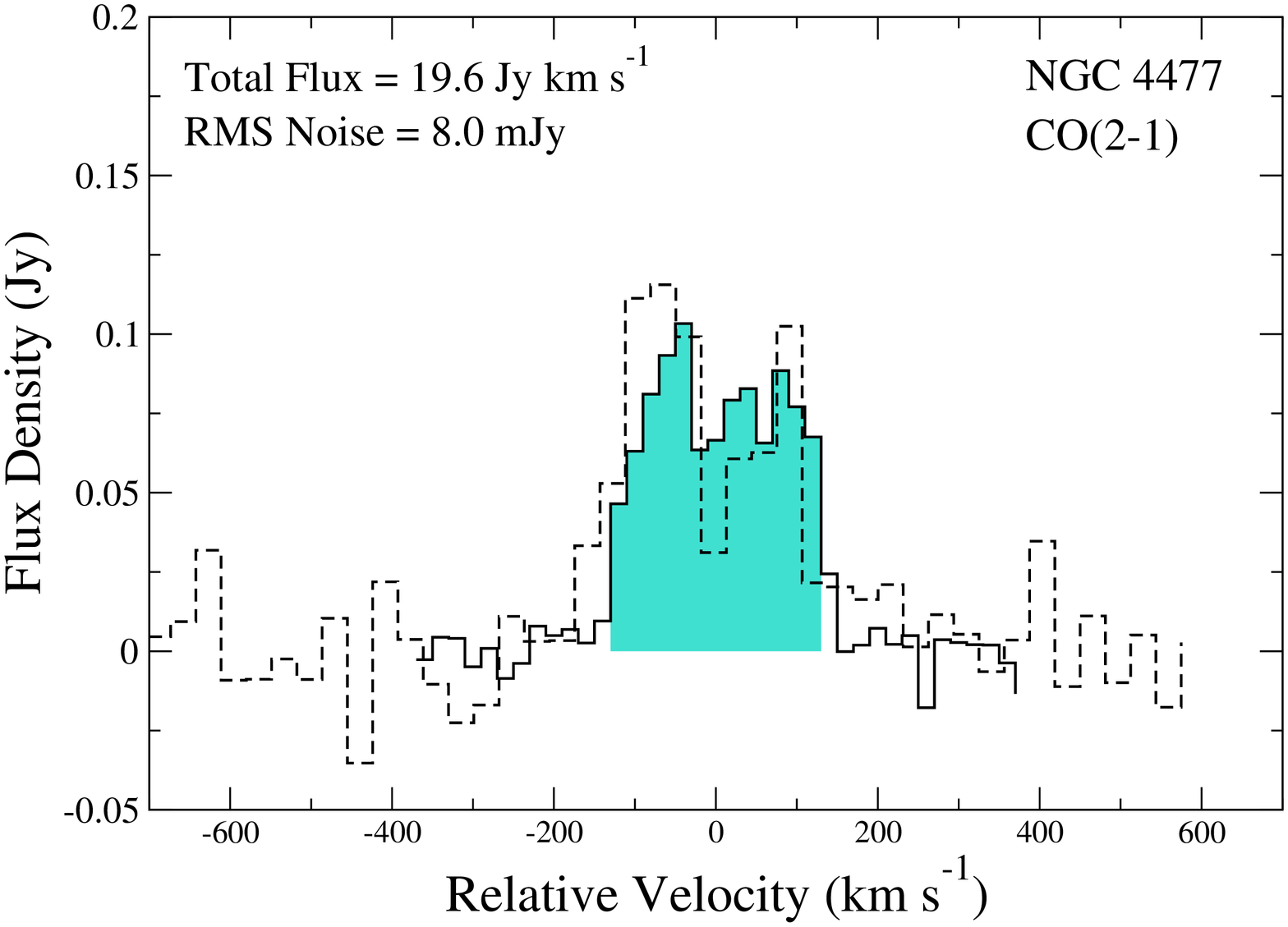}}
\rotatebox{270}{\includegraphics[clip=true,width=4cm]
{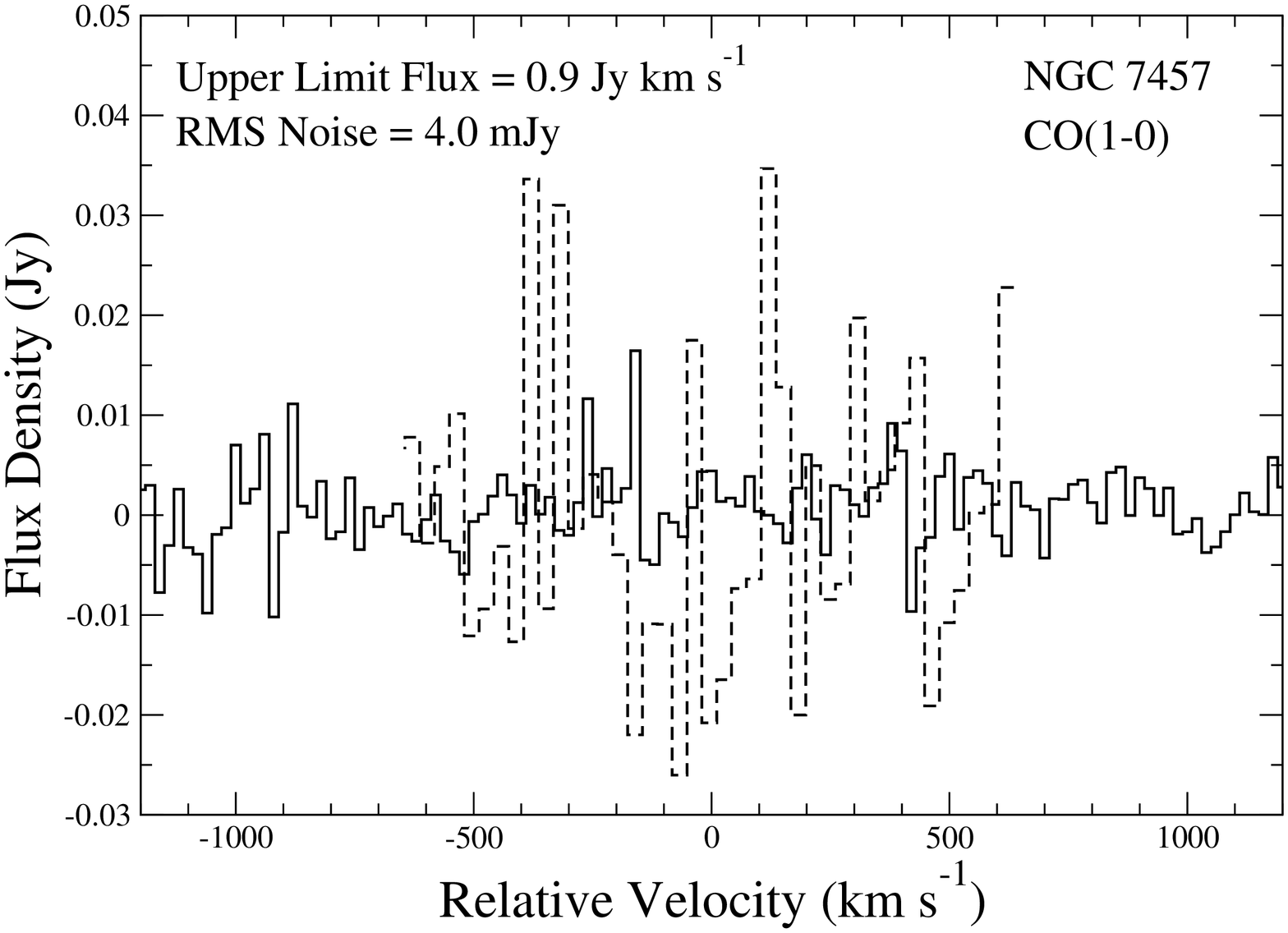}}
\caption{PdBI spectra integrated over the regions with detected emission
  (detections) or
  a central 5\arcsec~radius aperture (non-detections). For the
  detected galaxies, the shaded area represents the velocity range
  integrated over to obtain the total flux. The dashed lines show the single-dish data from the IRAM 30m telescope of \citet{combes07} and Young et al. (in preparation).}
\label{fig:spec}

\end{center}
\end{figure*}

Before discussing the results for each galaxy, we compare the interferometric spectra and fluxes to those obtained from the IRAM 30m single-dish data. The single-dish data of NGC~3489, NGC~4477 and NGC~4278 were published in \citet{combes07} while those of NGC~524 and NGC~7457 have been obtained more recently, to be published in Young et al. (in preparation). The single-dish spectra are plotted as dashed lines over the interferometric spectra in Fig.~\ref{fig:spec}. 

In light of the noise in both the single-dish and interferometric spectra and the 10-20\% calibration uncertainty in each, the agreement between the spectra is good. However, we note large ($\approx$50\%) disagreements between the molecular masses derived for the two galaxies NGC~524 and NGC~3489. For NGC~524, the interferometer recovers less flux than the single dish. The interferometer may have resolved out some flux, possible for the relatively extended gas distribution in this galaxy. Additionally, the velocity window we integrate over is narrower than that of Young et al. (in preparation), who include some channels at higher velocities with positive intensities. For NGC~3489, conversely, the interferometer recovers more flux than the single-dish observations. This flux discrepancy is due to the single-dish detection only being fit to the peak seen at negative velocities, while we see definite emission up to 140 \kms. Noise  is the most likely culprit, making the emission at negative velocities more prominent in the single-dish spectrum. A mild pointing error is also possible, especially given the extended CO distribution revealed in the interferometric map. 

Our interferometric spectra show no sign of CO(1-0) emission in either NGC~4278 or NGC~7457.  
The observations of NGC~4278 give an upper limit of $6.9
\times 10^{6}$ M$_{{\tiny \sun}}$ of molecular gas, three times less
than the $2.3 \times 10^7$ M$_{{\tiny \sun}}$ reported from the
single-dish observations. For this galaxy, the
single-dish detection is suspect because neither of the two CO lines `detected'
is at the galaxy's systemic velocity; the CO(1-0) is reported to be centred at 185 \kms and the CO(2-1) at -112 \kms. We are thus confident in our non-detection. NGC~7457 was reported to contain $3.8 \times 10^{7}$ M$_{{\tiny \sun}}$ of
molecular gas within the 15\arcsec~beam of the
Nobeyama 45m telescope by \citet{taniguchi94}. \citet{welch03} reported $4.7 \times 10^{6}$
M$_{{\tiny \sun}}$ within the 55\arcsec~beam of the NRAO 12m telescope
(both values corrected for our X$_{\mathrm{CO}}$ factor and distance).
However, both of these detections have extremely low signal-to-noise ratios and our observations at the PdBI give an upper limit of $1.8 \times 10^{6}$
M$_{{\tiny \sun}}$. We also note that this non-detection is corroborated by the new spectrum from the IRAM 30m of Young et al. (in preparation), shown as a dashed line in Fig.~\ref{fig:spec}. 

\subsection[]{Individual galaxies}
\subsubsection[] {NGC~524}

\begin{figure*}
\begin{center}
\begin{tabular}{cccc}
& NGC~524 & NGC~2768 & NGC~3032 \\
\begin{sideways}
\phantom{000000000000}arcsec
\end{sideways} &
\includegraphics[scale=0.68]{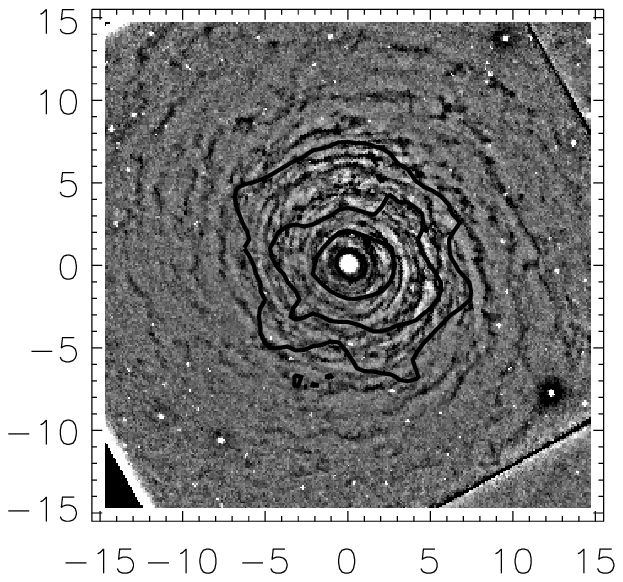} &
\includegraphics[scale=0.68]{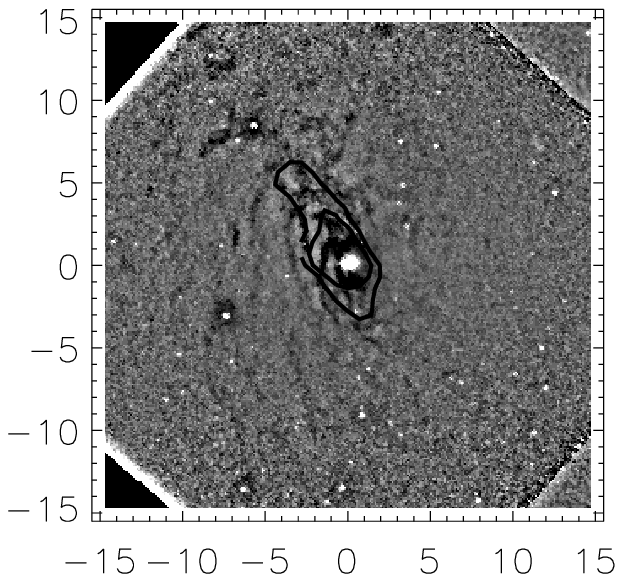} &
\includegraphics[scale=0.68]{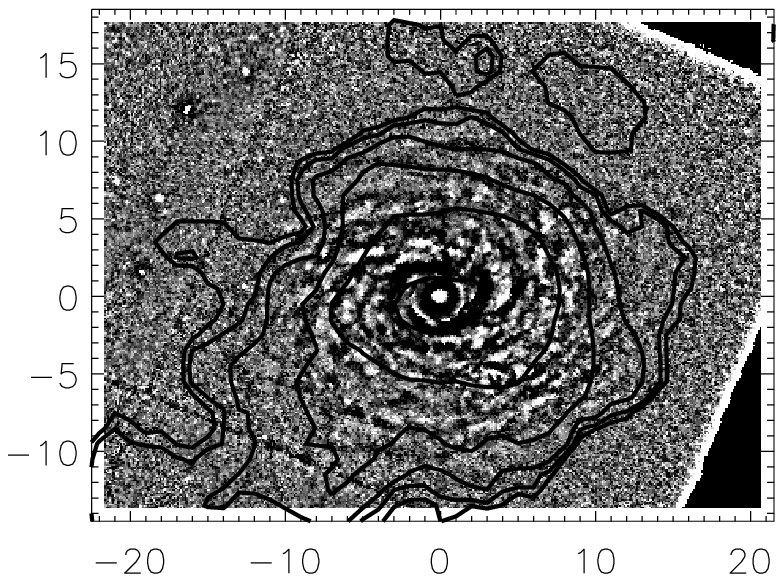}\\
\end{tabular}
\begin{tabular}{cccc}
& NGC~3489 & NGC~4150 & NGC~4459 \\
\begin{sideways}
\phantom{0000000000000}arcsec
\end{sideways} &
\includegraphics[scale=0.68]{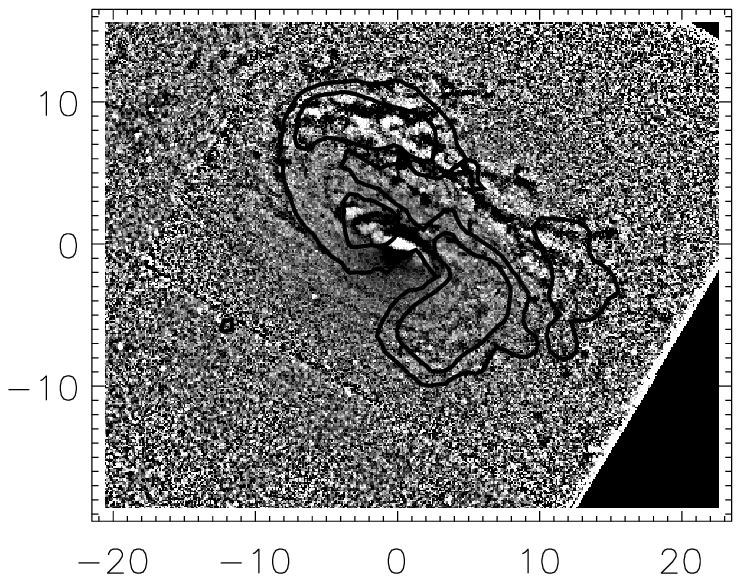}&
\includegraphics[scale=0.68]{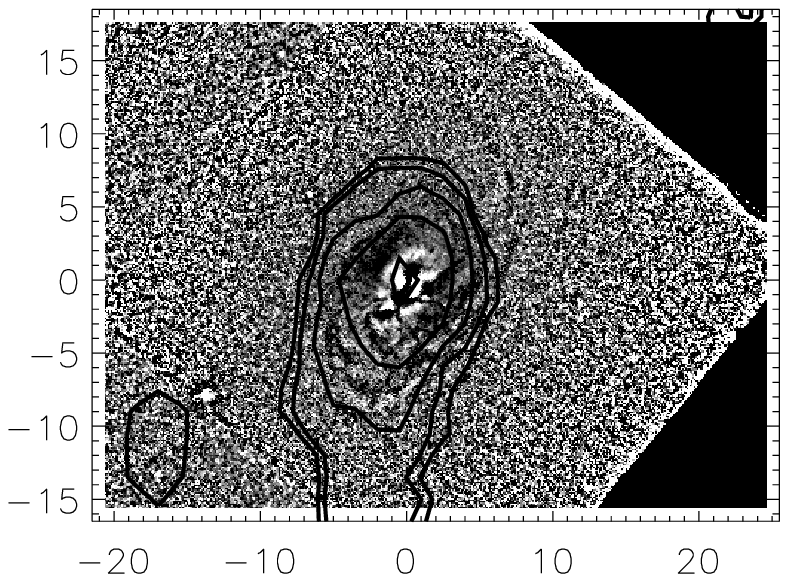}&
\includegraphics[scale=0.68]{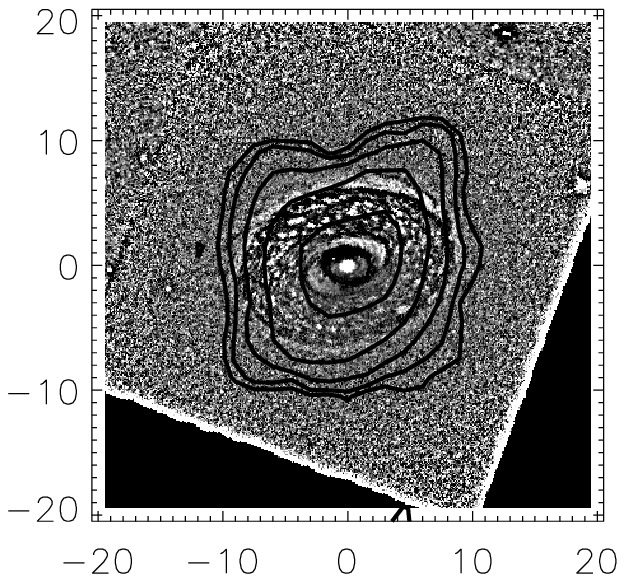}\\
\end{tabular}
\begin{tabular}{cccc}
& NGC~4477 & NGC~4526 & NGC~4550 \\
\begin{sideways} 
\phantom{000000000000}arcsec
\end{sideways} &
\includegraphics[scale=0.68]{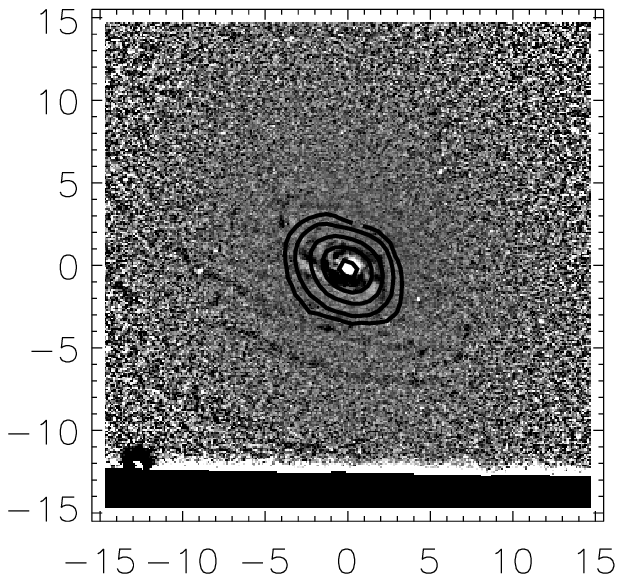}&
\includegraphics[scale=0.68]{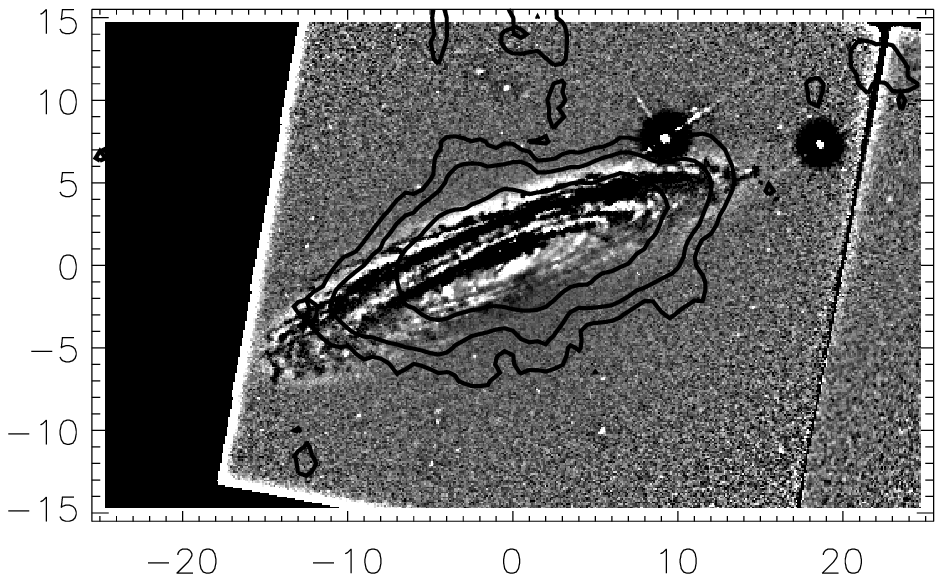}&
\includegraphics[scale=0.68]{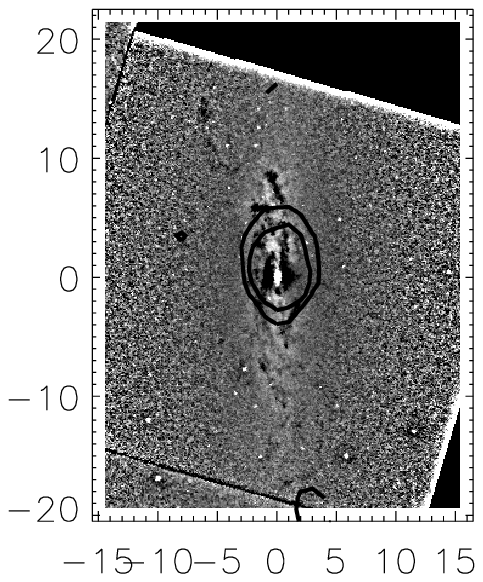}\\
& \phantom{00}arcsec & \phantom{00}arcsec & \phantom{00}arcsec\\
\end{tabular}
\end{center} 
\caption{Unsharp-masked HST WFPC2 F555W images showing 
  optically obscuring dust. CO(1-0) contours are overlaid in black.} 
\label{fig:dustmaps}
\end{figure*}

NGC~524 hosts a 1.1~kpc radius molecular gas disc of $6.7\times10^{7}$
M$_{{\tiny \sun}}$ according to the PdBI
map (see Fig.~\ref{fig:COmaps}). The molecular disc coincides with a tightly wound spiral
of dust seen in a F555W unsharp-masked image from the Hubble Space Telescope
Wide-Field Planetary Camera 2 (HST/WFPC2; see
Fig.~\ref{fig:dustmaps}). The CO velocity map shows an ordered velocity
field with a surprisingly large maximum velocity for such a face-on
configuration. While \citet{silchenko00} interprets the similarly
large velocities she finds for the ionised gas as indicating a
kinematically decoupled gas component, the CO and ionised gas
velocities actually agree well with the circular velocities (maximum
465 km s$^{-1}$) computed by assuming the multi-Gaussian expansion
(MGE) mass distribution, dynamical mass-to-light ratio and
inclination derived in \citet{cappellari06} from a model of NGC~524's
stellar kinematics.  At 6\arcsec, this model predicts a velocity of 150
\kms, which is very close to the 160 \kms we observe.  Thus, the gas
is not kinematically decoupled, NGC~524 is simply a massive galaxy.  

NGC~524 has central 3mm continuum emission, with a flux
density of $4.5\pm0.2\pm0.9$~mJy (random and systematic error). It has
previously been detected in radio continuum at 1.4, 5 and 8.4~GHz \citep{condon98,
  filho02, filho04}. The 5~GHz Very Long Baseline Array (VLBA)
observations show that NGC~524's radio core is compact on milliarcsecond-scales,
clearly indicating an active galactic nucleus (AGN). With a beam of
2\farcs66 $\times$ 2\farcs47, the 8.4~GHz 
VLA observations are very well matched to the resolution of our 3mm
observations. At this resolution, the flux density increases from
1.95~mJy at 8.4~GHz to 4.5~mJy at 3~mm (115~GHz). A maximum of 
1.5~mJy at 3mm can be explained by dust based on the Infrared Astronomical
Satellite (IRAS) 100 and 60 \mm data points, so most of the emission
at 3~mm must originate from self-absorbed synchrotron emission from the AGN \citep[e. g.][]{krips07}.

\subsubsection{NGC~3489}
Unlike any other early-type galaxy, NGC~3489's $2.1 \times 10^{7}$ M$_{{\tiny \sun}}$ of molecular gas is
clearly resolved into an extended spiral-like pattern
(Fig.~\ref{fig:COmaps}). The spiral consists of an arc to 
the north-east ($\approx570$~pc from the centre), a nearly central clump
($\approx170$~pc to the north-east) and an arc to the south-west
($\approx460$~pc). The three blobs (one very faint) to the west are likely
an extension of the northern spiral arm
. The galaxy centre is determined from the
Two Micron All Sky Survey (2MASS),
so it should be 
accurate to 2\arcsec, implying that  the nearly-central clump
is indeed offset and cannot be directly associated with the galaxy
nucleus. This clumpy and not settled distribution suggest that the gas has been recently acquired or recently disturbed. The molecular gas velocity map shows 
general corotation with both the ionised gas and the stars. However,
to the south, negative relative velocities  extend further east
than might be expected, similar to a feature seen in the
ionised gas map. This velocity irregularity also suggests the recent acquisition or disruption of the gas.

\subsubsection{NGC~4477}

\begin{figure}
\begin{center}
\includegraphics[clip=true,width=4.3cm]{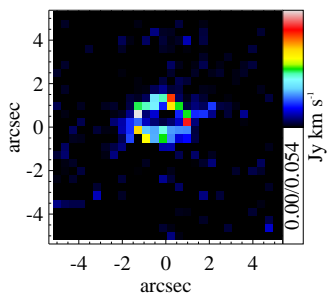}
\end{center}
\caption{Centre of NGC~4477. Integrated clean component map of NGC~4477 in CO(2-1). The
  ring structure is clear, with a radius of 1\farcs3 ($\approx$
  100~pc). }

\label{fig:cct4477}
\end{figure}

The barred S0 galaxy NGC~4477 shows a small nuclear ring of
$2.3\times10^{7}$ M$_{{\tiny \sun}}$ of molecular gas. This H$_{2}$
mass matches the IRAM~30m value within the errors. While the large
CO(1-0) beam makes the gas appear centrally peaked, the CO(2-1)
integrated intensity map shows a hint of 
ring structure (Fig.~\ref{fig:COmaps}), which is then much more visible
in the integrated clean component map (no convolution by the `clean
beam'; Fig.~\ref{fig:cct4477}). The 
size and location of this molecular gas ring agrees well with the
ring-like structure seen in the optically obscuring dust
(Fig. \ref{fig:dustmaps}).

\subsubsection{NGC~4278}

The CO observations of NGC~4278 give an upper limit of $6.9
\times 10^{6}$ M$_{{\tiny \sun}}$ of molecular gas.

The 1~mm continuum observations reveal a central unresolved point source in
both the December 2005 and November 2007 runs. However, the flux of
this source is 34~mJy in December 2005 and 65~mJy in November
2007. Sometime between November 2002 and May 2003, \citet{doi05}
detected 3~mm (96~GHz) continuum at 57~mJy using the Nobeyama Millimetre
Array. As all these observations should have a flux accurate to around
20\%, we think such variation in the mm continuum of
NGC~4278 is real. NGC~4278 has a known AGN,
with a compact X-ray source 
\citep{colbert99} and possible sub-parsec-scale radio jets
\citep{giovannini01}. Periods of variability of the radio emission have
been reported previously, for example \citet{schilizzi83} report a
~25\% increase 
in 6~cm radio emission between February 1980 and April 1981.

\subsubsection{NGC~7457}

The PdBI observations limit the amount of molecular gas in NGC~7457 to $1.8 \times 10^6$ \Msun.
The absence of significant molecular gas also fits well with
the low level of ionised emission and the lack of optically obscuring
dust in this galaxy \citep{sarzi06}. 

\begin{table}
\begin{center}
 \caption{CO fluxes and molecular gas mass estimates from PdBI. } 
 \begin{tabular}{@{}lrrrrr}
  \hline
  Galaxy & Flux & M$_{\mathrm{H}_{2}}$ & H$_{2}$ radius \\
   &  (Jy km s$^{-1}$) & (M$_{{\tiny \sun}}$) & (kpc) & \\
  \hline
  NGC~\phantom{0}524 & 10.5 & $6.7\times10^{7}$ & 1.1  \\
  NGC~3489 & 12.5 & $2.1\times10^{7}$ & 0.8  \\
  NGC~4278 &$<2.4$& $<6.9\times10^{6}$& --      \\ 
  NGC~4477 & 7.3  & $2.3\times10^{7}$ & 0.4  \\
  NGC~7457 &$<0.9$& $<1.8\times10^{6}$& --     \\ 
  \hline
\label{tab:COFlux}
 \end{tabular}
\end{center}

\end{table}

\section{Analysis}

\begin{table*}
\begin{center}
 \caption{CO-based molecular gas properties. } 
\vspace{0.05cm}
 \begin{tabular}{@{}lrrrrrrrrr}

\hline
  Galaxy & Instrument & D\phantom{c)} & M$_{\mathrm{H}_{2}}$\phantom{00)} &
  R$_{\mathrm{H_{2}}}$\phantom{00} &
  $\Sigma_{\mathrm{H}_{2}}$\phantom{000} & H$_{2}$ Morphology 
  & CO Ref. & Dust Map Ref. \\
   &  & (Mpc)\phantom{} & (M$_{{\tiny \sun}}$)\phantom{00} &
  (kpc)\phantom{00} & (M$_{{\tiny \sun}}$) pc$^{-2}$ & \\

\hline
  NGC~\phantom{0}524 & PdBI\phantom{0} & 23.3 & $6.7\times10^{7}$ & 
  1.1\phantom{70}  & 23  & face-on disc & (1) & (1)\\
  NGC~2320 & BIMA\phantom{0}  & 83\phantom{.0} & $4.6\times10^{9}$ &
  4.0\phantom{70}  & 120 & edge-on disc  & (2) & (2) \\
  NGC~2685 & OVRO\phantom{0}  & 15.0 & $2.3\times10^{7}$ &
  2.5\phantom{70}  & --  & extended polar ring & (3) & (3)\\
  NGC~2768 & PdBI\phantom{0}  & 21.8 & $6.3\times10^{7}$ &
  0.6\phantom{70}  & 68  & polar ring/disc & (4) & (1)\\
  NGC~3032 & BIMA\phantom{0}  & 21.4 & $5.0\times10^{8}$ &
  1.5\phantom{70}  & 96  & disc  & (5) & (1)\\
  NGC~3156 & 30m\phantom{0}   & 21.8 & $2.2\times10^{7}$ & 
  --\phantom{70}   & --  &  -- & (6) & (3)\\
  NGC~3489 & PdBI\phantom{0}  & 11.8 & $2.1\times10^{7}$ &
  0.8\phantom{70}  & 14  & spiral  & (1) & (1)\\
  NGC~4150 & BIMA\phantom{0}  & 13.4 & $5.5\times10^{7}$ &
  0.6\phantom{70}  & 70  & ring/disc & (5) & (1)\\
  NGC~4459 & BIMA\phantom{0}  & 16.1 & $1.7\times10^{8}$ &
  1.0\phantom{70}  & 72  & disc & (5) & (1) \\
  NGC~4477 & PdBI\phantom{0}  & 16.5 & $2.3\times10^{7}$ &
  0.4\phantom{70}  & 63  & ring & (1) & (1) \\
  NGC~4526 & BIMA\phantom{0}  & 16.4 & $5.7\times10^{8}$ &
  1.0\phantom{70}  & 230 & disc  & (5) & (1) \\ 
  NGC~4550 & PdBI\phantom{0}  & 15.5 & $7.9\times10^{6}$ &
  0.4\phantom{70}  & 24  & lopsided ring/disc  & (7) & (1) \\
  NGC~5666 & BIMA\phantom{0}  & 35\phantom{.0} & $5.6\times10^{8}$ &
  2.7\phantom{70}  & 33  &  ring  & (8) & (2) \\

\hline 
\label{tab:COdist}
 \end{tabular}
\\

\end{center}
\vspace{0.2cm}
 {\sc CO references:} (1) this paper, (2) \citet{young05}, (3)
\citet{schinnerer02}, (4) \citet{crocker08}, (5) \citet{young08}, (6)
\citet{combes07}, (7) \citet{crocker09}, (8) \citet{young02}.
{\sc Dust map references:} (1) Fig.~\ref{fig:dustmaps}, (2)
\citet{young09}, (3) \citet{sarzi06}. 
{\sc Note:} The molecular gas mass of NGC~2685 was recalculated using our
$I_{\mathrm{CO}}$ factor.

\end{table*}

\subsection{Distribution of molecular gas}
The set of CO maps presented here and in \citet{schinnerer02},
\citet{young02}, \citet{young05}, \citet{young08}, 
\citet{crocker08} and \citet{crocker09} reveal the distribution of
the molecular gas in early-type galaxies. The 
molecular gas is mostly found in central discs or rings (see
Table~\ref{tab:COdist}), although the map resolutions are usually too poor to
distinguish between these structures. The two known
exceptions to the central disc/ring rule are the molecular spiral of
NGC~3489 and the spots of CO emission in the extended \hi polar
ring of NGC~2685 (at a radius of 
34\arcsec = 2.5~kpc).  Two other galaxies, NGC~2768 and 
NGC~4150, also show faint tails of CO emission.  
Other interferometric CO maps in the literature
also suggest that a central disc and/or ring is the most common distribution for 
the molecular gas of early-type galaxies \citep[][]{wiklind97,young02, young05, okuda05}.  

However, there is a bias toward detecting central molecular
structures because almost all interferometric targets
have been selected based on central single-dish observations. For example, the
CO emission in the polar ring of NGC~2685 would 
not have been detected by a central observation
using the IRAM 30m telescope (instead its molecular gas was detected in an
interferometric study of its \hi ring by \citealp{schinnerer02}). \hi and UV observations prove that
outlying molecular gas does exist in a fraction of early-type
galaxies, assuming molecular gas is necessary for 
star formation. For example, two E/S0 galaxies from the {\tt SAURON} sample, NGC~2974 and 
NGC~4262, have substantial \hi rings at radii of $\approx8$ and
$\approx13$~kpc, respectively \citep[][Oosterloo et al. in preparation]{weijmans08}. Within these rings, clumps 
of UV emission are seen, which must be due to knots of young
stars ($<$100 Myr) recently formed from molecular gas \citep{jeong07}. These outer star-forming rings are similar to the case
of ESO 381-47, thoroughly analysed in \citet{donovan09}.

Attempting to quantify the statistics of where molecular gas is found
in E/S0 galaxies requires an unbiased sample. Taking the {\tt SAURON}
representative sample of 48 early-type 
galaxies, 10/48 have  
centrally-detected CO while 3/48 present evidence of outlying
molecular gas (NGC~2685, NGC~2974 and NGC~4262). Although a larger
 and complete sample will solidify these numbers considerably, these
 fractions provide a rough estimate of where molecular gas is found in 
early-type galaxies.  

The resolution limit of the interferometric observations means that
the molecular gas may lie in more detailed structures than the CO maps currently probe. A suggestion of these
structures is indicated by the dust maps shown in
Fig.~\ref{fig:dustmaps} (unsharp-masked optical images). 
 In all cases, the gas and dust appear
spatially linked.
Galaxies with over $10^{8}$ M$_{{\tiny \sun}}$ of molecular gas have clear
flocculent spiral patterns (NGC~3032,
NGC~4459 and NGC~4526), 
while galaxies with less molecular mass have less 
obscuring dust and the dust structures are less ordered. However,
the total molecular mass does not 
alone determine the appearance of the dust. For
example, NGC~524, NGC~2768 and NGC~4150 all have similar molecular gas
masses of about $10^{8}$ \msun, but NGC~524 has an extensive
flocculent, tightly-wound spiral dust disc while NGC~2768 
has a small polar nuclear ring with extended wisps and NGC~4150 has an
irregular, very central dust structure. 

None of our sample galaxies have a
well-ordered two-arm spiral pattern. Even in the case of NGC~3489, with
its molecular two-armed spiral, the dust reveals a multiple-armed more
flocculent distribution. However, the absence of such two-arm structure may be
a selection effect - galaxies with a strong two-arm spiral in dust and
molecular gas are likely to have associated star formation and be
classified as Sa or later type galaxies. A combination of 
the galaxy properties (bar, shape of potential) and the history of
the cold gas (time since accretion or cooling) must affect the cold gas and
dust morphology. In early-type galaxies, this cold material is most
often found in compact, flocculent spiral or nuclear ring
configurations. 

We determine an outer radius ($R_{\mathrm{H_2}}$) for the molecular gas based
on the radius of largest extent in the CO maps (one-half
the largest diameter through the centre; reported in Table~\ref{tab:COdist}). We wish to
characterise the outer radius for the molecular discs or rings, and so
ignore the clear tails of emission seen in some galaxies. However, the
$R_{\mathrm H_2}$ values measured depend critically on the depth
and spatial resolution of the CO observations. 
 Ideally, better resolution and
sensitivity will allow the measurement of a proper characteristic
scale radius for these molecular gas distributions and enable a more
consistent study of their extent.

Nevertheless, the mass measured with an interferometer corresponds to
  the radial limit of the CO observed and we can calculate an average
  molecular gas density for this region, listed in Table~\ref{tab:COdist}. These average molecular gas
  densities range between 14 and 230 \Msun pc$^{-2}$. The 
  lower limit is related to the density at which neutral hydrogen
  becomes predominantly molecular, which seems to be around 10 \Msun
  pc$^{-2}$ \citep{bigiel08}. At the high end, densities are approaching
  the densities of individual molecular clouds.

\begin{figure*}
\begin{center}
   \begin{tabular}[c]{ccccc}
& NGC 524 & NGC2320 & NGC 2768 & 
\begin{rotate}{-90}
\phantom{000000000} V$_{\mathrm{star}}$
\end{rotate}  \\
\begin{sideways}
\phantom{0000000000}arcsec
\end{sideways} &
\rotatebox{0}{\includegraphics[width=3.9cm]
  {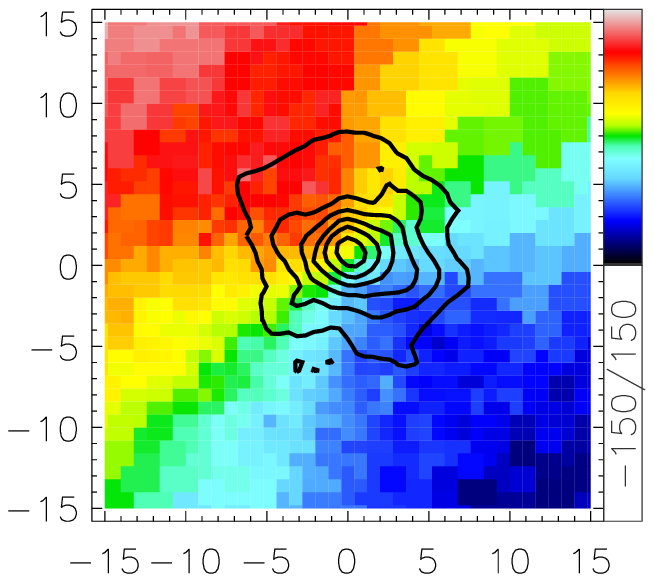}} &
\rotatebox{0}{\includegraphics[width=3.3cm]
  {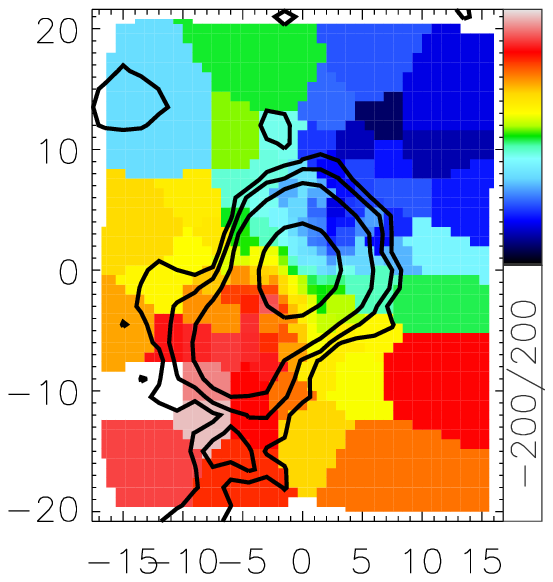}} &
\rotatebox{0}{\includegraphics[width=3.9cm]
  {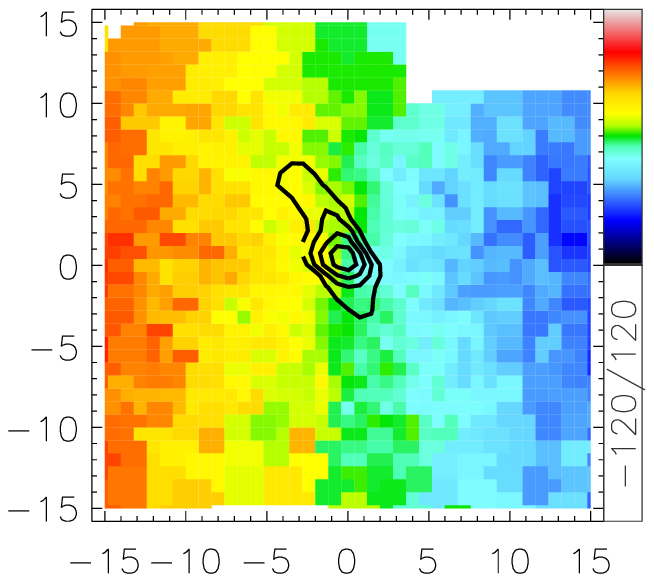}}  &
\begin{rotate}{-90}
\phantom{000000000} $\sigma_{\mathrm{star}}$
\end{rotate}
\\
\begin{sideways}
\phantom{0000000000}arcsec
\end{sideways} &
\rotatebox{0}{\includegraphics[width=3.9cm]
  {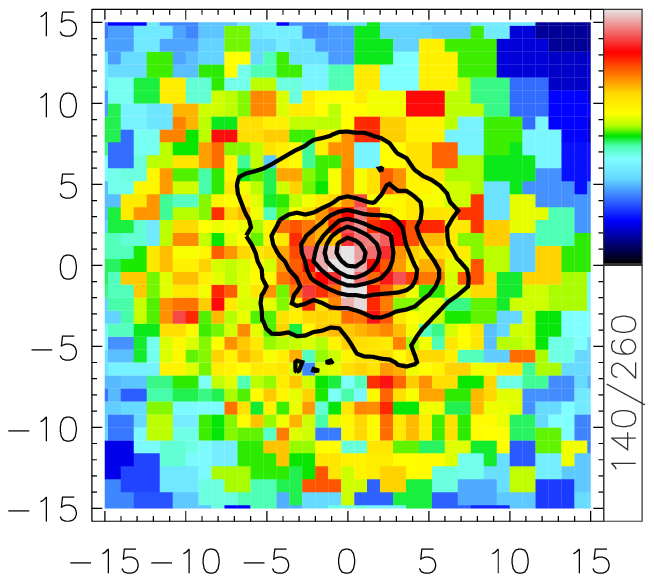}} &
\rotatebox{0}{\includegraphics[width=3.3cm]
  {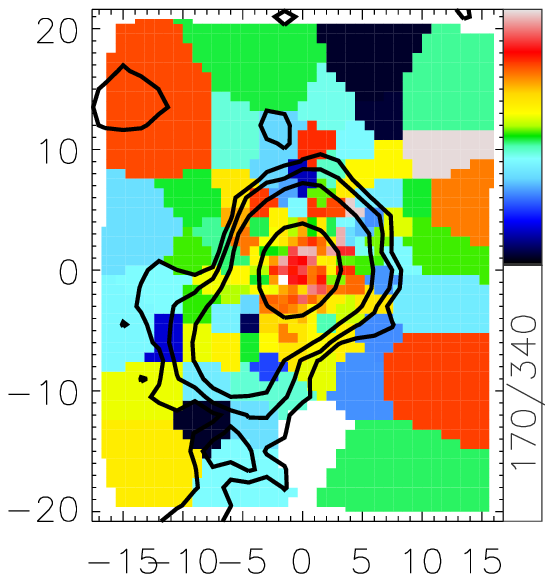}} &
\rotatebox{0}{\includegraphics[width=3.9cm]
  {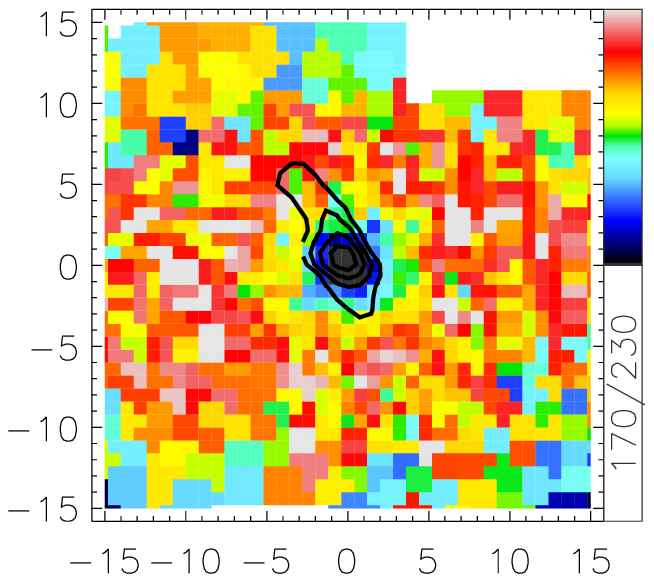}} &
\begin{rotate}{-90}
\phantom{0000000} log(F$_{\mathrm{H\beta}}$)
\end{rotate}
\\
\begin{sideways}
\phantom{0000000000}arcsec
\end{sideways} &
\rotatebox{0}{\includegraphics[width=3.9cm]
  {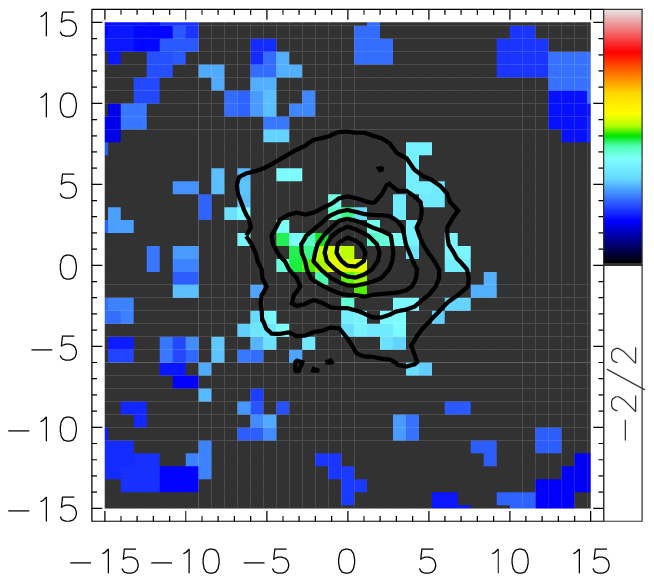}} &
\rotatebox{0}{\includegraphics[width=3.3cm]
  {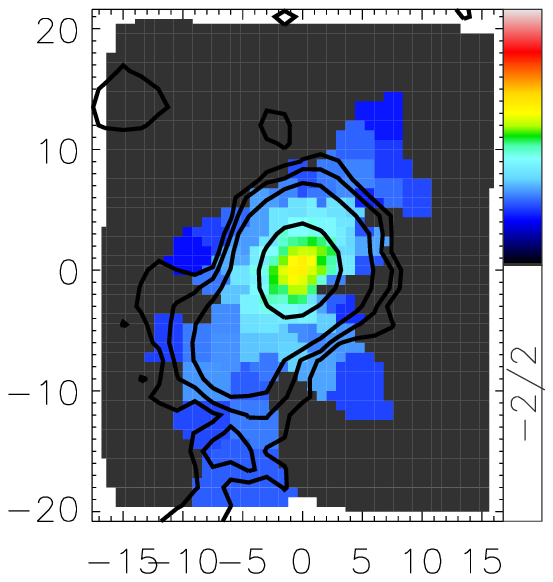}} &
\rotatebox{0}{\includegraphics[width=3.9cm]
  {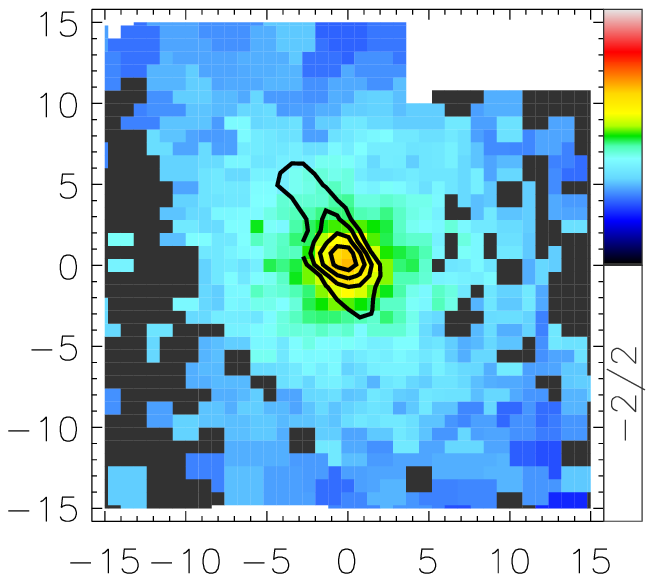}}   &
\begin{rotate}{-90}
\phantom{00000} log(EW$_{\mathrm{[OIII]}}$)
\end{rotate}
\\
\begin{sideways}
\phantom{0000000000}arcsec
\end{sideways} &
\rotatebox{0}{\includegraphics[width=3.9cm]
  {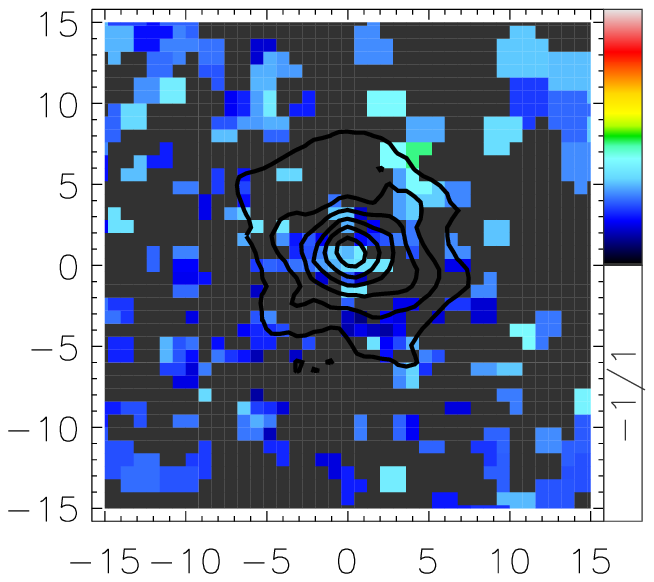}} &
\rotatebox{0}{\includegraphics[width=3.3cm]
  {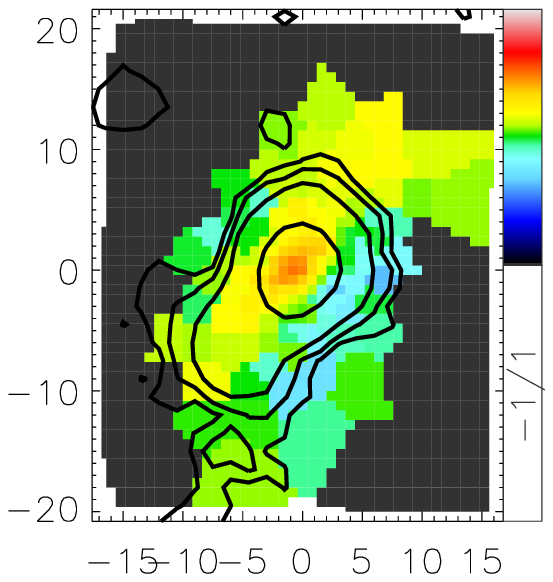}} &
\rotatebox{0}{\includegraphics[width=3.9cm]
  {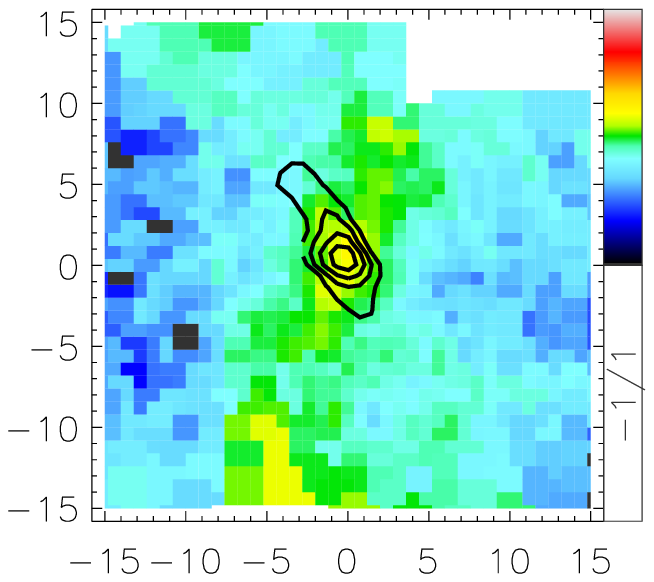}}  &
\begin{rotate}{-90}
\phantom{0000} log(\otohb)
\end{rotate}
\\
\begin{sideways}
\phantom{0000000000}arcsec
\end{sideways} &
\rotatebox{0}{\includegraphics[width=3.9cm]
  {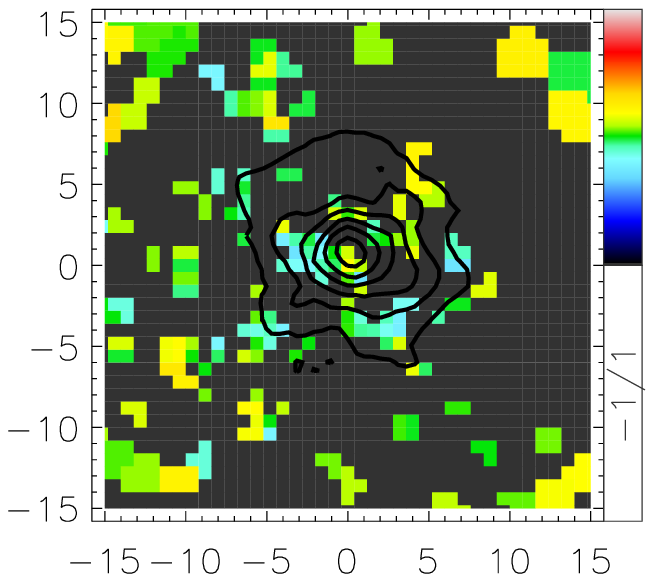}} &
\rotatebox{0}{\includegraphics[width=3.3cm]
  {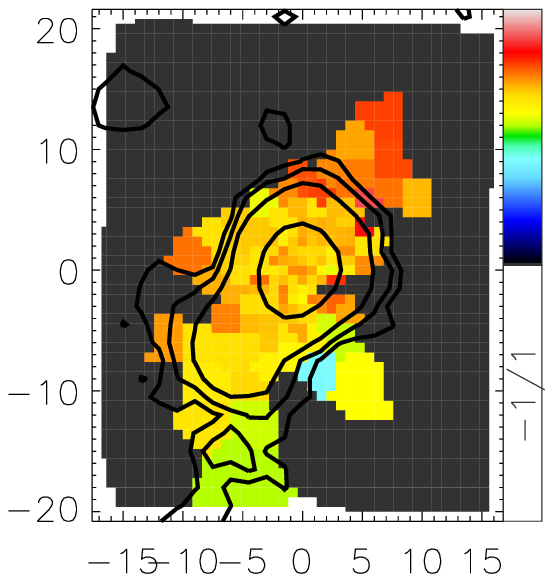}} &
\rotatebox{0}{\includegraphics[width=3.9cm]
  {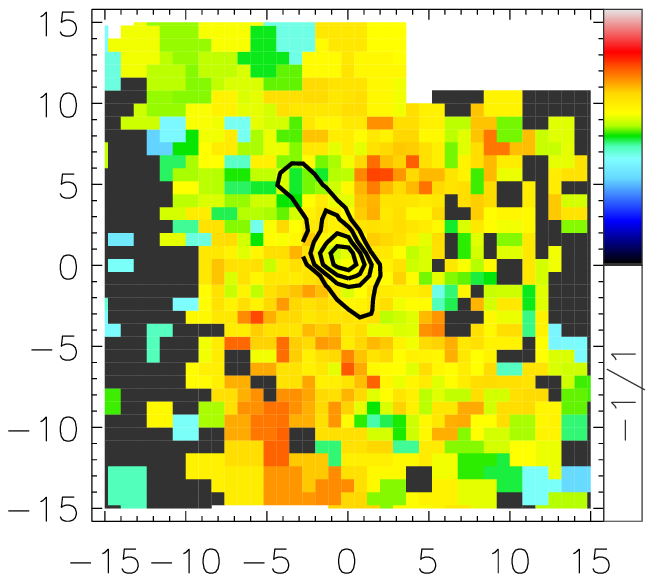}}  &
\begin{rotate}{-90}
\phantom{000000000}Age
\end{rotate}
\\
\begin{sideways}
\phantom{0000000000}arcsec
\end{sideways} &
\rotatebox{0}{\includegraphics[width=3.9cm]
  {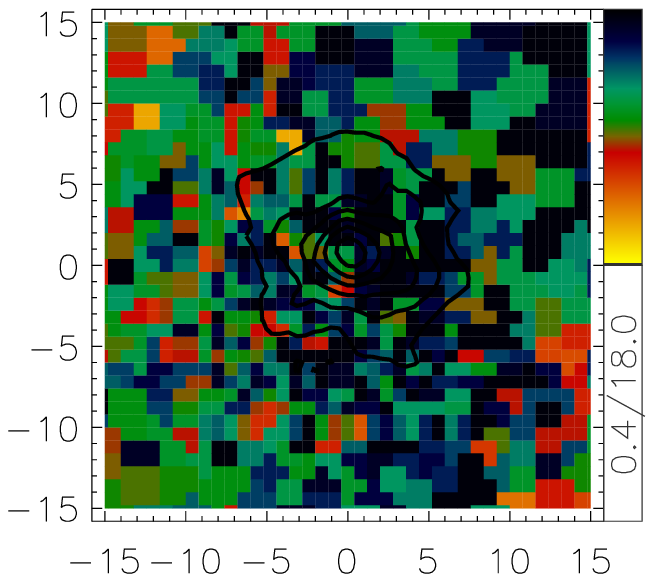}} &
\rotatebox{0}{\includegraphics[width=3.3cm]
  {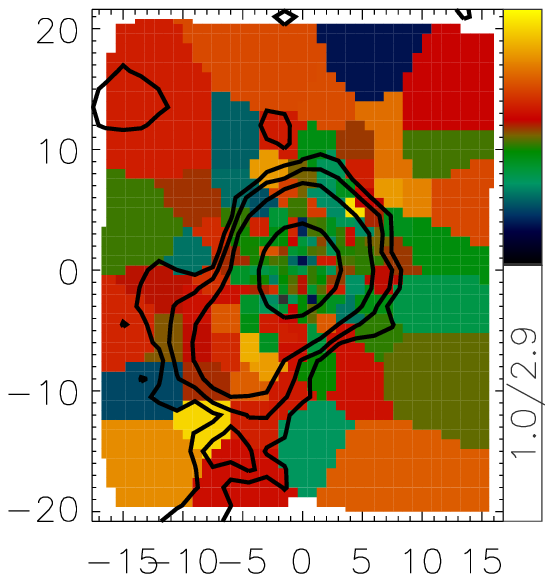}} &
\rotatebox{0}{\includegraphics[width=3.9cm]
  {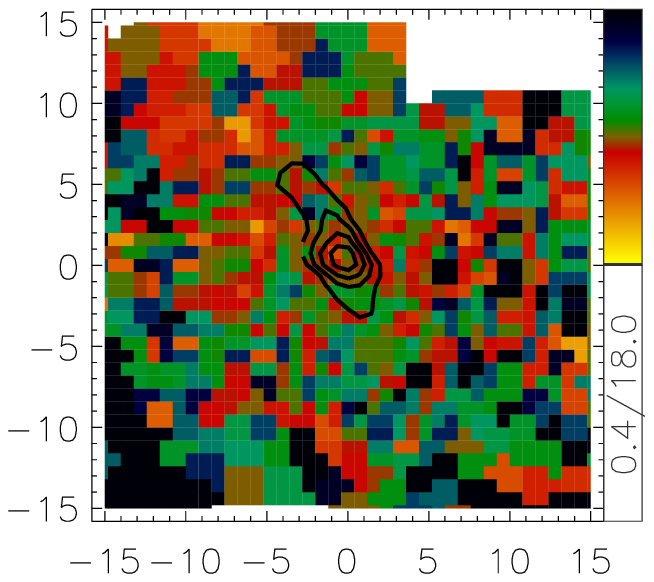}} &
\\
 & \phantom{00}arcsec & \phantom{00}arcsec & \phantom{00}arcsec\\
   \end{tabular}
\end{center}
\caption{CO contours overlaid on IFU maps. {\em Top to bottom:} mean stellar
  velocity (km s$^{-1}$), stellar velocity dispersion (km s$^{-1}$),
  logarithmic  H$\beta$ emission line flux (10$^{-16}$ erg s$^{-1}$ cm$^{-2}$
  arcsec$^{-2}$), logarithmic equivalent width of
  [O$\:${\small III}] (\AA), logarithmic [O$\:${\small III}]/H$\beta$
  emission line ratio, SSP-equivalent age (Gyr; except for NGC~2320
  and NGC~5666 where the H$\beta$ absorption linestrength is shown in \AA).}
  
\label{fig:overlays}
\end{figure*}

\begin{figure*}
\begin{center}
   \begin{tabular}[c]{ccccc}
& NGC 3032 & NGC 3489 & NGC4150  &
\begin{rotate}{-90}
\phantom{000000000} V$_{\mathrm{star}}$
\end{rotate}
\\
\begin{sideways}
\phantom{0000000000}arcsec
\end{sideways} &
\rotatebox{0}{\includegraphics[width=4.9cm]
  {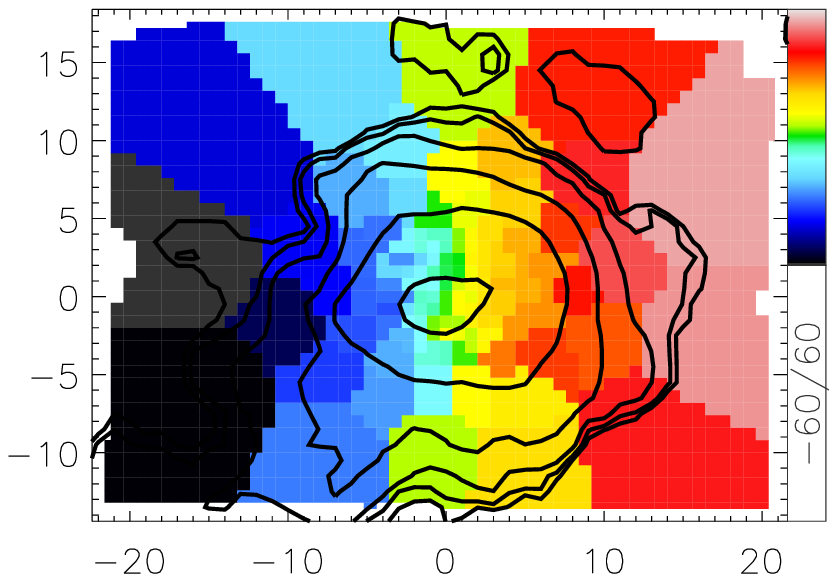}} &
\rotatebox{0}{\includegraphics[width=4.7cm]
  {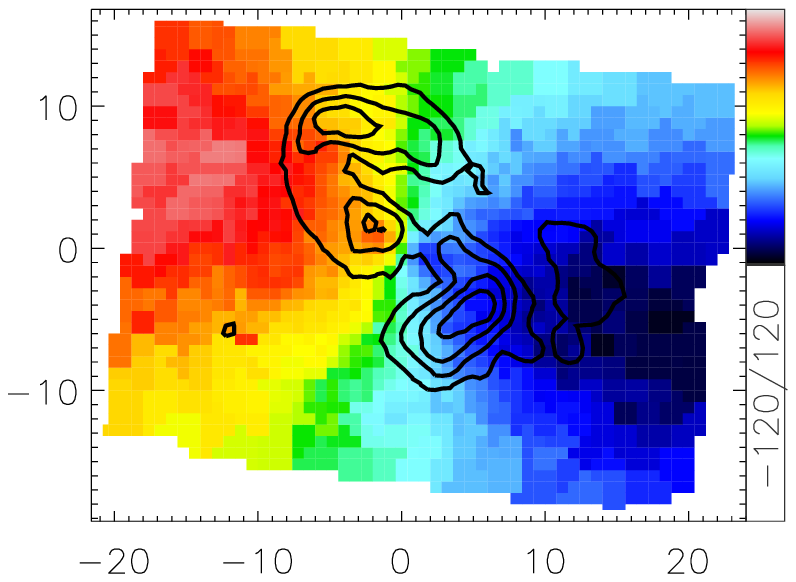}} &
\rotatebox{0}{\includegraphics[width=4.9cm]
  {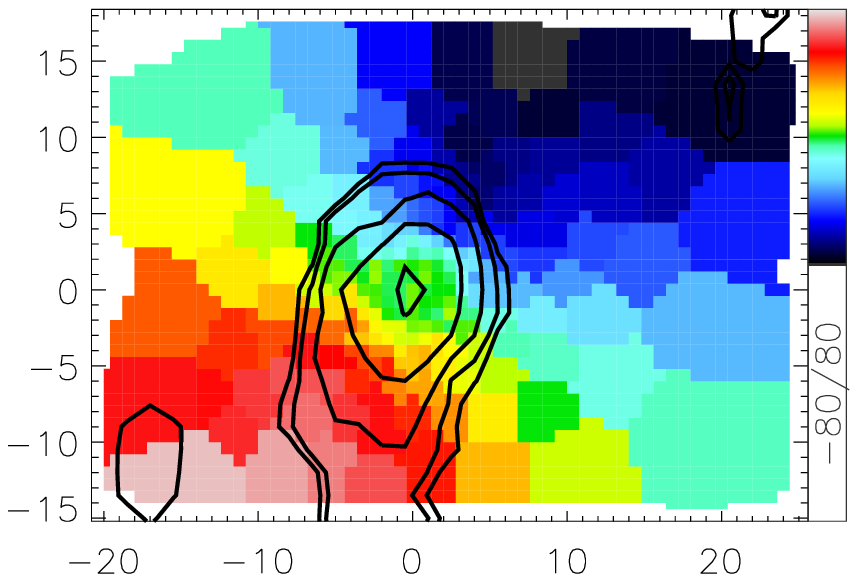}} &
\begin{rotate}{-90}
\phantom{000000000} $\sigma_{\mathrm{star}}$
\end{rotate}
\\
\begin{sideways}
\phantom{0000000000}arcsec
\end{sideways} &
\rotatebox{0}{\includegraphics[width=4.9cm]
  {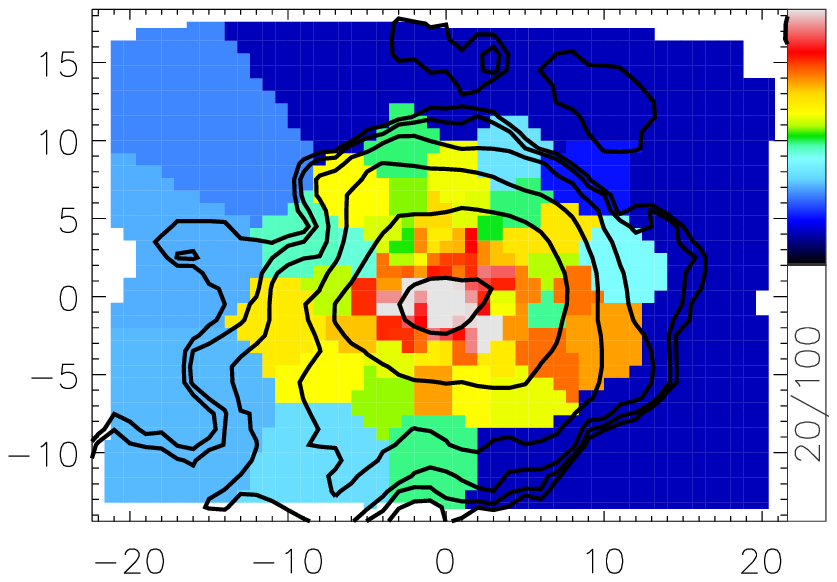}} &
\rotatebox{0}{\includegraphics[width=4.7cm]
  {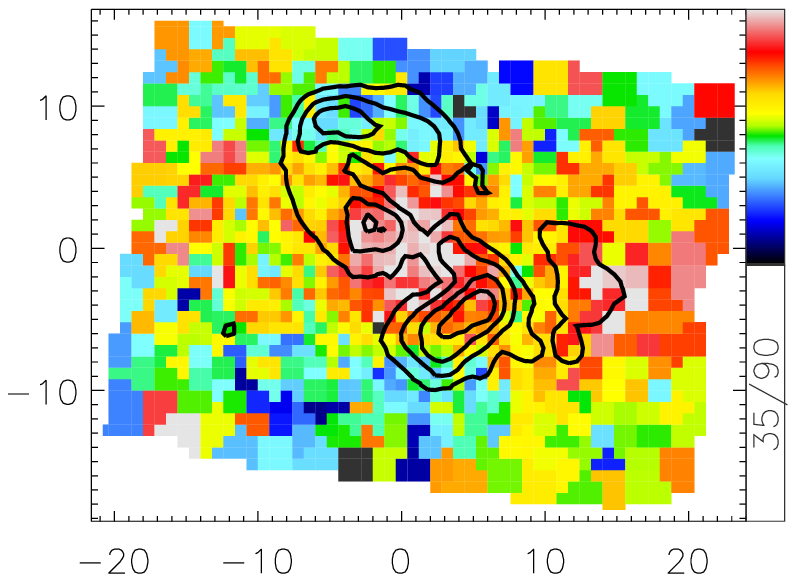}} &
\rotatebox{0}{\includegraphics[width=4.9cm]
  {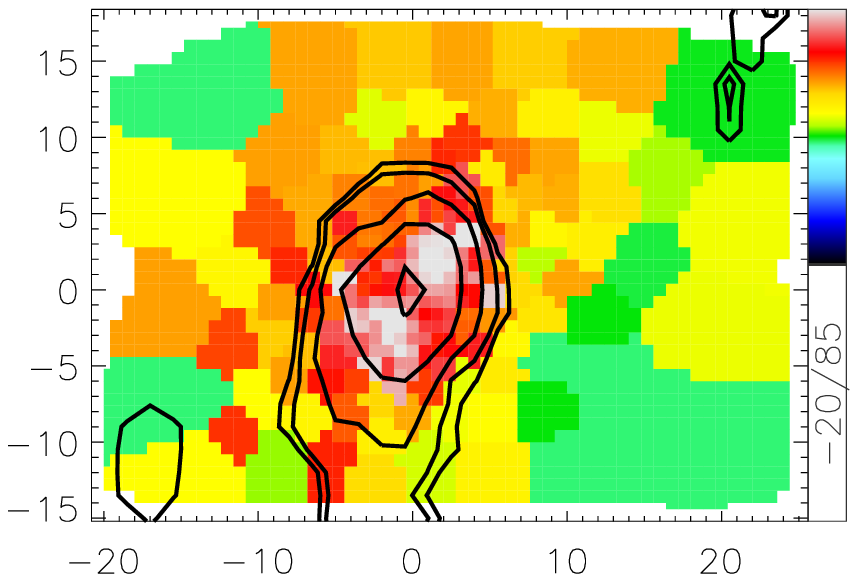}}  &
\begin{rotate}{-90}
\phantom{0000000} log(F$_{\mathrm{H\beta}}$)
\end{rotate}
\\
\begin{sideways}
\phantom{0000000000}arcsec
\end{sideways} &
\rotatebox{0}{\includegraphics[width=4.9cm]
  {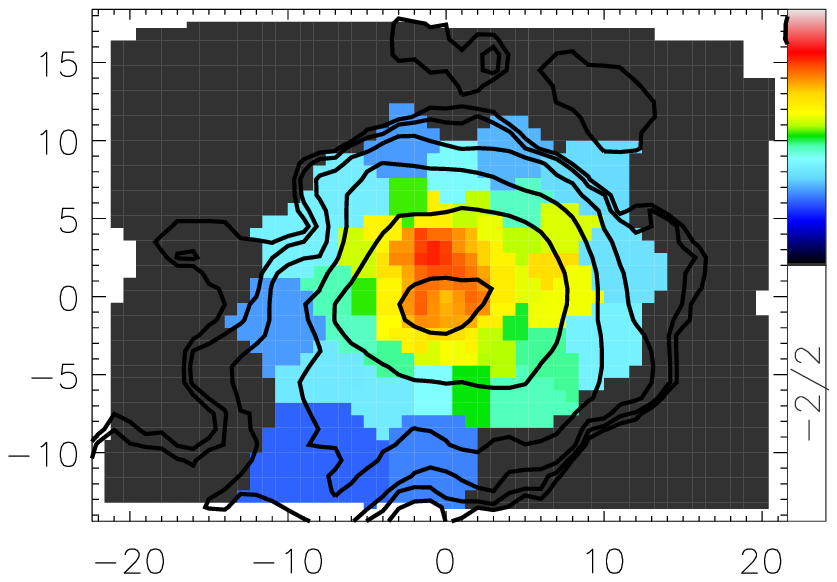}} &
\rotatebox{0}{\includegraphics[width=4.7cm]
  {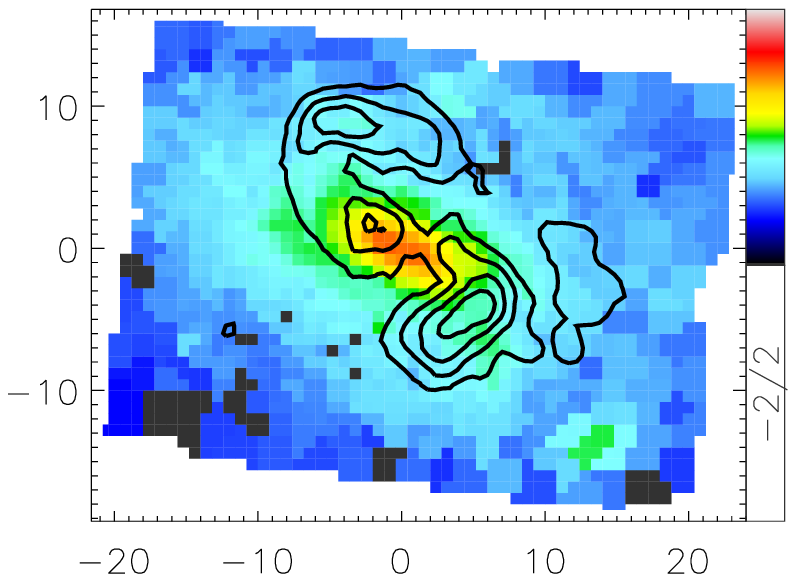}} &
\rotatebox{0}{\includegraphics[width=4.9cm]
  {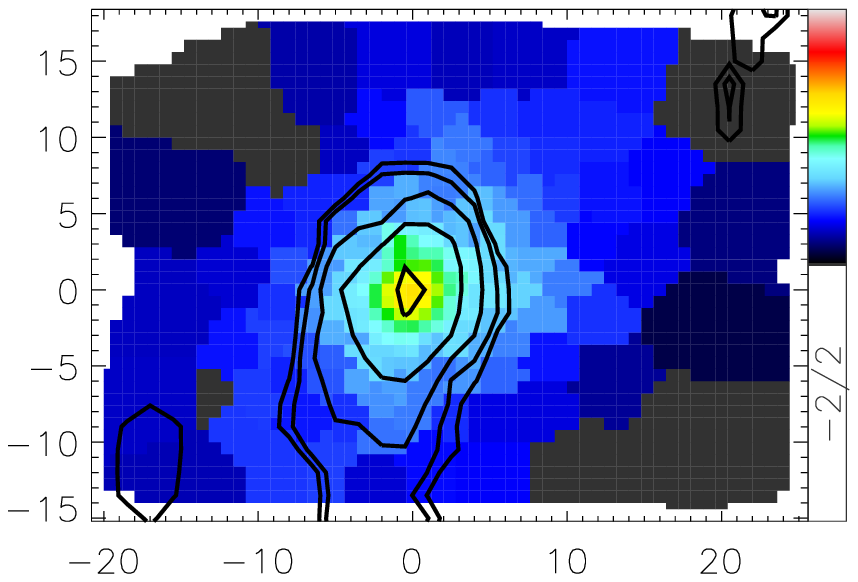}}  &
\begin{rotate}{-90}
\phantom{00000} log(EW$_{\mathrm{[OIII]}}$)
\end{rotate}
\\
\begin{sideways}
\phantom{0000000000}arcsec
\end{sideways} &
\rotatebox{0}{\includegraphics[width=4.9cm]
  {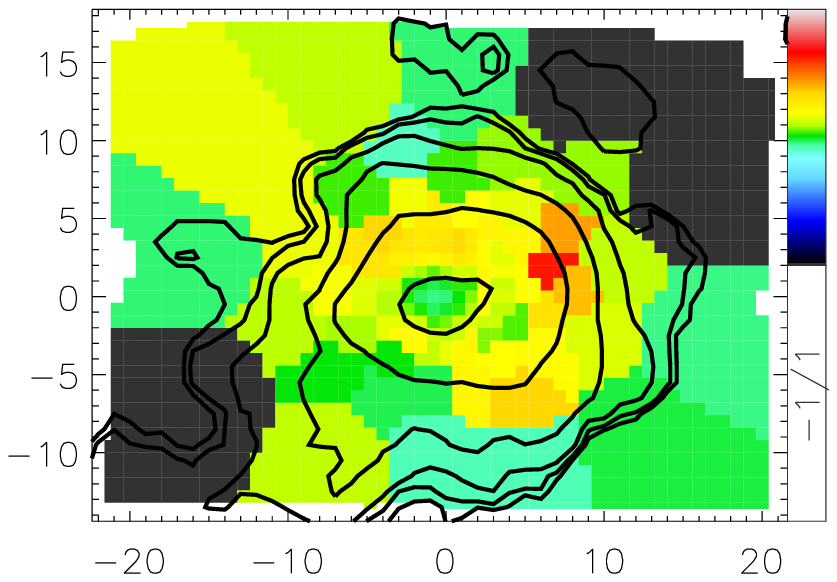}} &
\rotatebox{0}{\includegraphics[width=4.7cm]
  {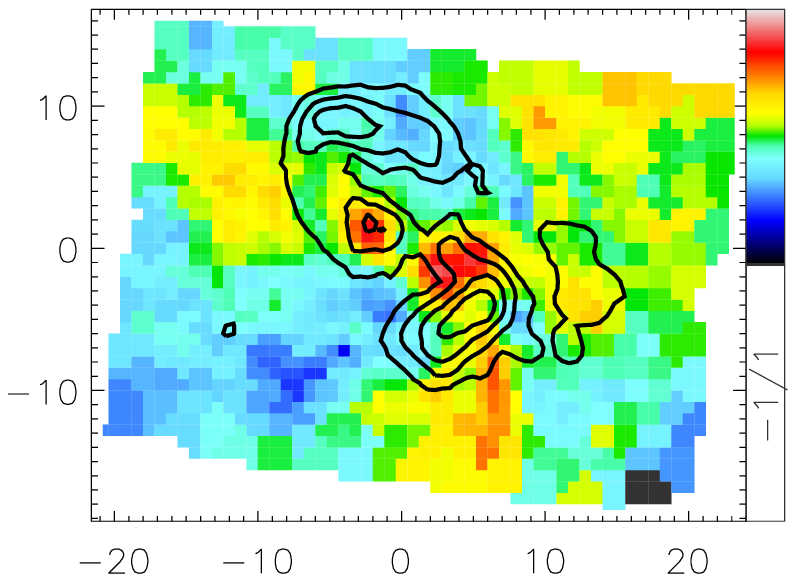}} &
\rotatebox{0}{\includegraphics[width=4.9cm]
  {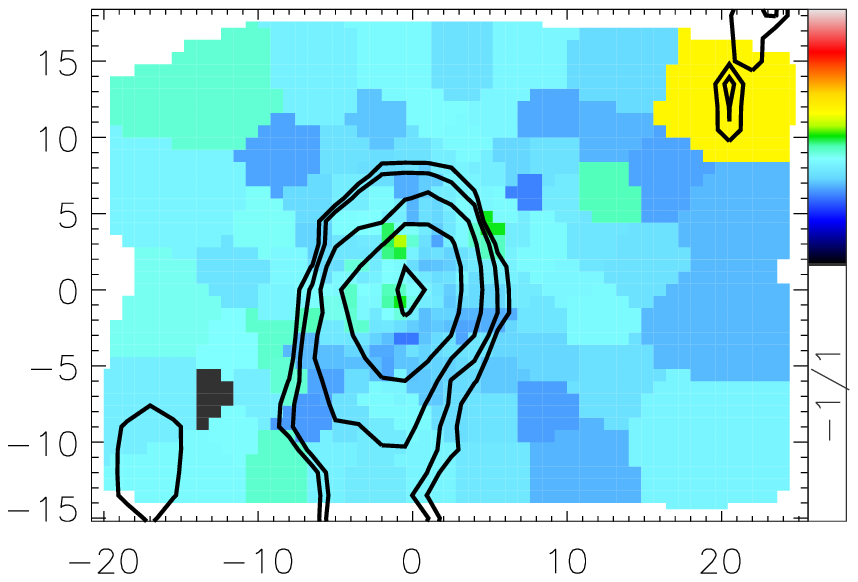}}  &
\begin{rotate}{-90}
\phantom{0000} log(\otohb)
\end{rotate}
\\
\begin{sideways}
\phantom{0000000000}arcsec
\end{sideways} &
\rotatebox{0}{\includegraphics[width=4.9cm]
  {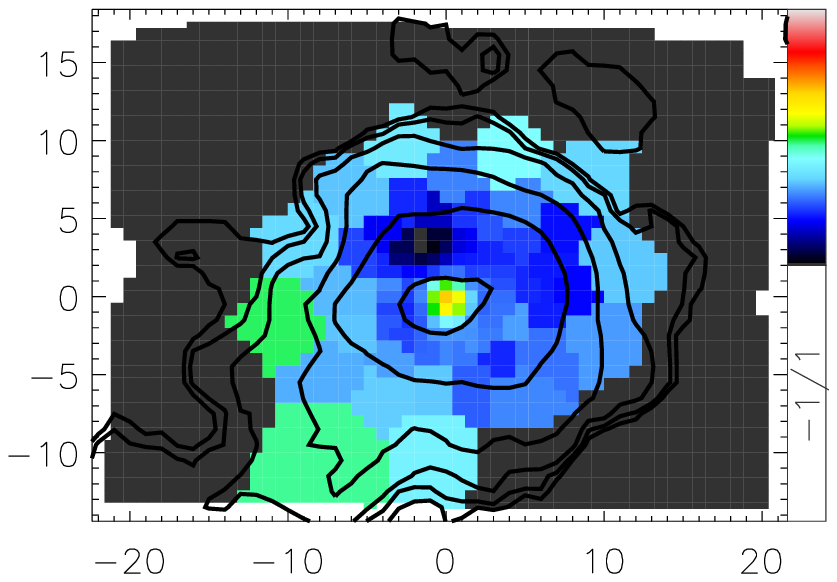}} &
\rotatebox{0}{\includegraphics[width=4.7cm]
  {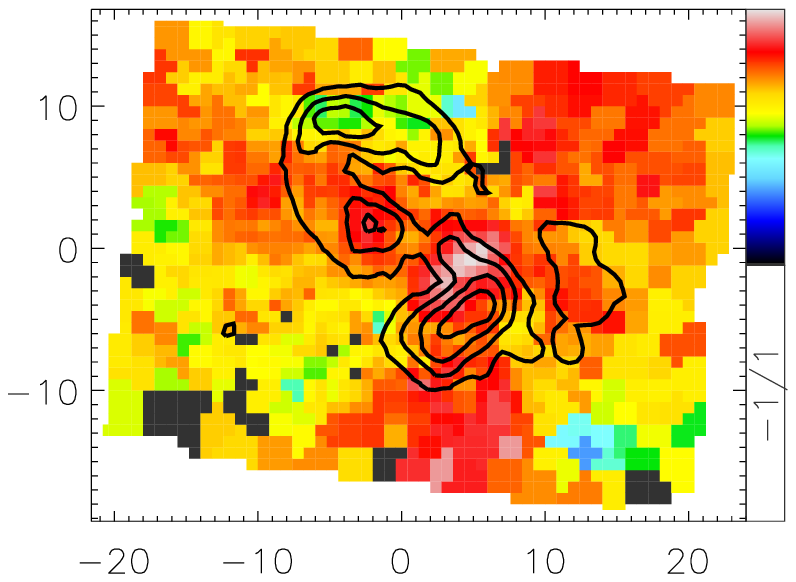}} &
\rotatebox{0}{\includegraphics[width=4.9cm]
  {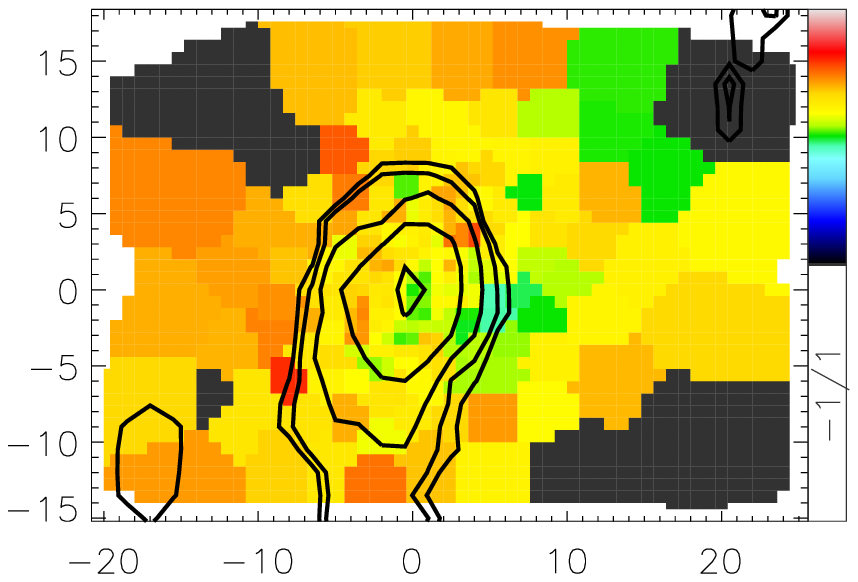}}  &
\begin{rotate}{-90}
\phantom{000000000}Age
\end{rotate}
\\
\begin{sideways}
\phantom{0000000000}arcsec
\end{sideways} &
\rotatebox{0}{\includegraphics[width=4.9cm]
  {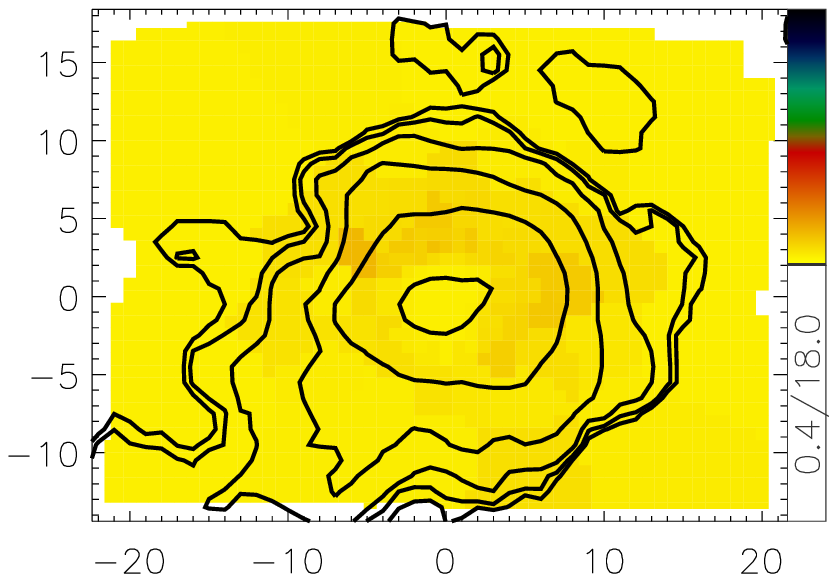}} &
\rotatebox{0}{\includegraphics[width=4.7cm]
  {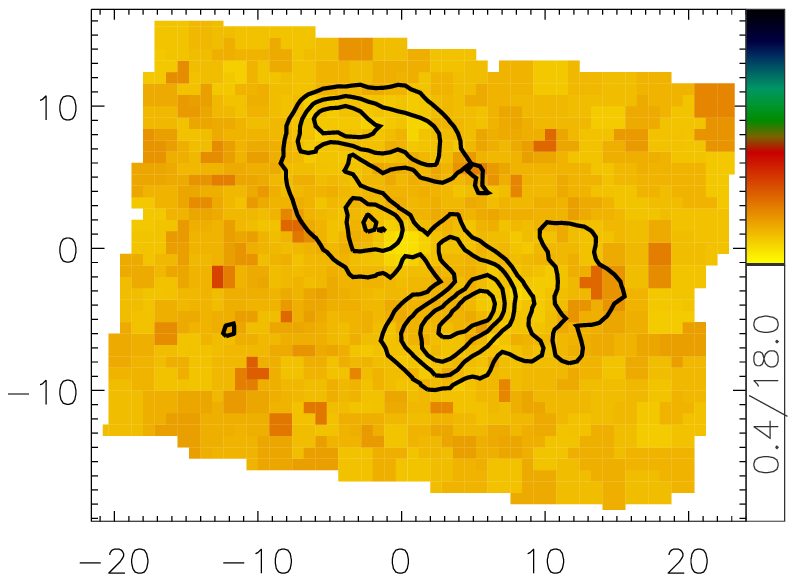}} &
\rotatebox{0}{\includegraphics[width=4.9cm]
  {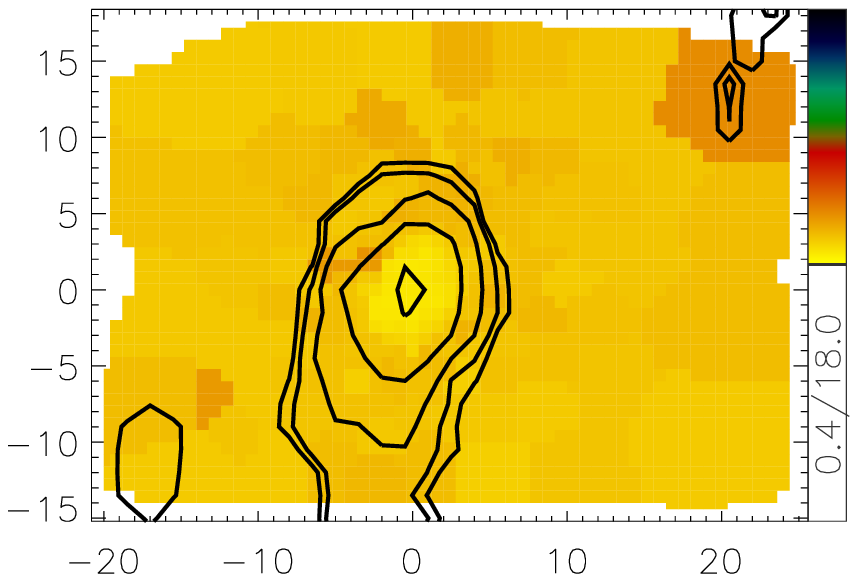}} &
\\
 & \phantom{00}arcsec & \phantom{00}arcsec & \phantom{00}arcsec\\
   \end{tabular}
\end{center}
\contcaption{}
\end{figure*}

\begin{figure*}
\begin{center}
   \begin{tabular}[c]{ccccc}
 & NGC 4459 & NGC 4477 & NGC 4526 &
\begin{rotate}{-90}
\phantom{000000000} V$_{\mathrm{star}}$
\end{rotate}
\\
\begin{sideways}
\phantom{0000000000}arcsec
\end{sideways} &
\rotatebox{0}{\includegraphics[width=3.9cm]
  {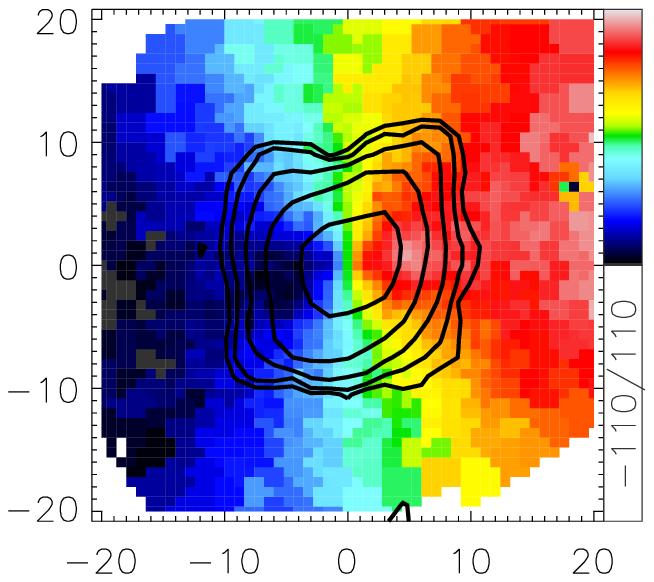}} &
\rotatebox{0}{\includegraphics[width=3.9cm]
  {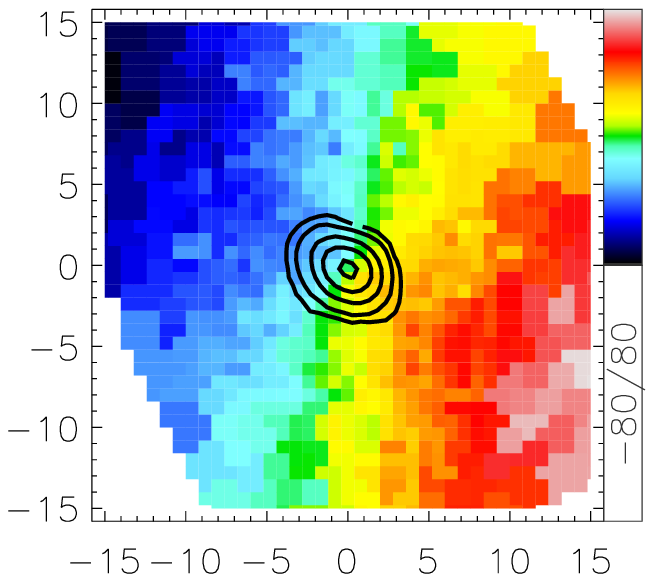}} &
\rotatebox{0}{\includegraphics[width=5.7cm]
  {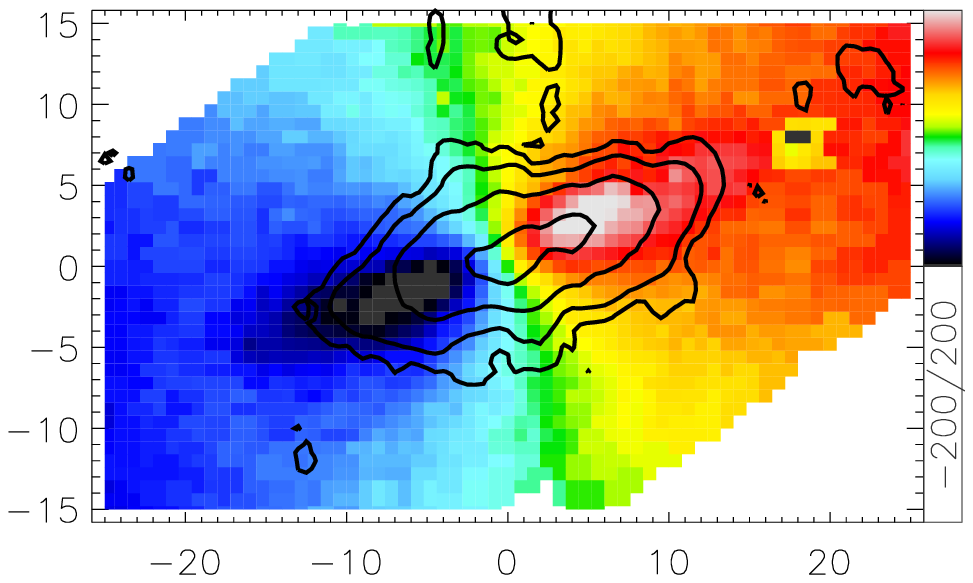}} &
\begin{rotate}{-90}
\phantom{000000000} $\sigma_{\mathrm{star}}$
\end{rotate}
\\
\begin{sideways}
\phantom{0000000000}arcsec
\end{sideways} &
\rotatebox{0}{\includegraphics[width=3.9cm]
  {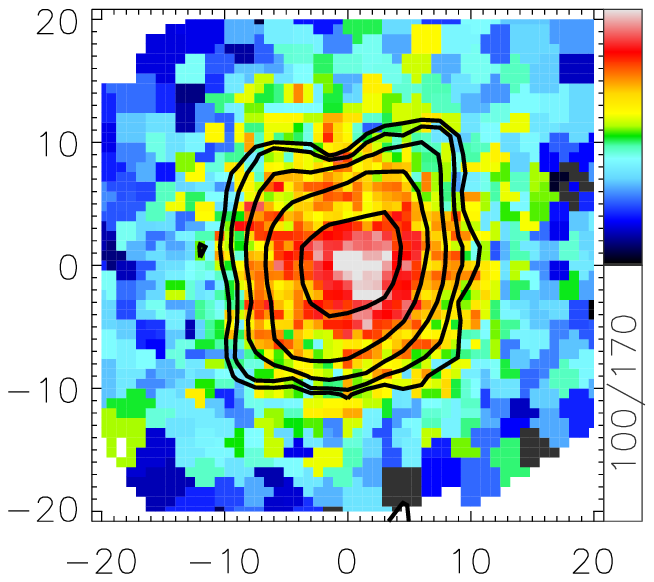}} &
\rotatebox{0}{\includegraphics[width=3.9cm]
  {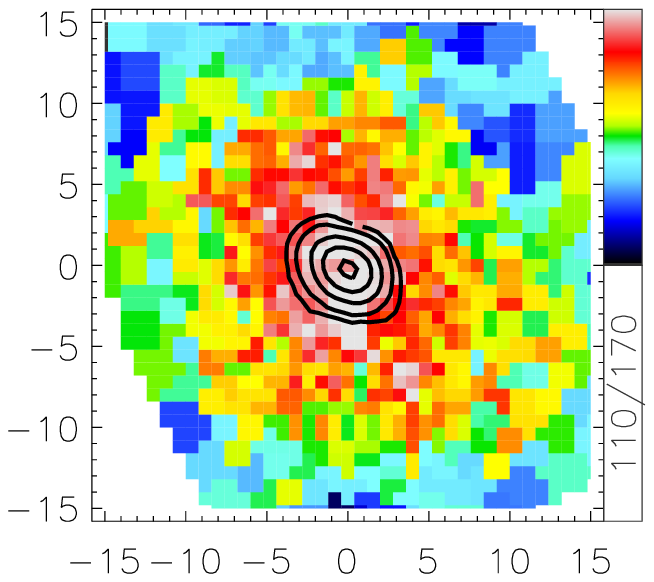}} &
\rotatebox{0}{\includegraphics[width=5.7cm]
  {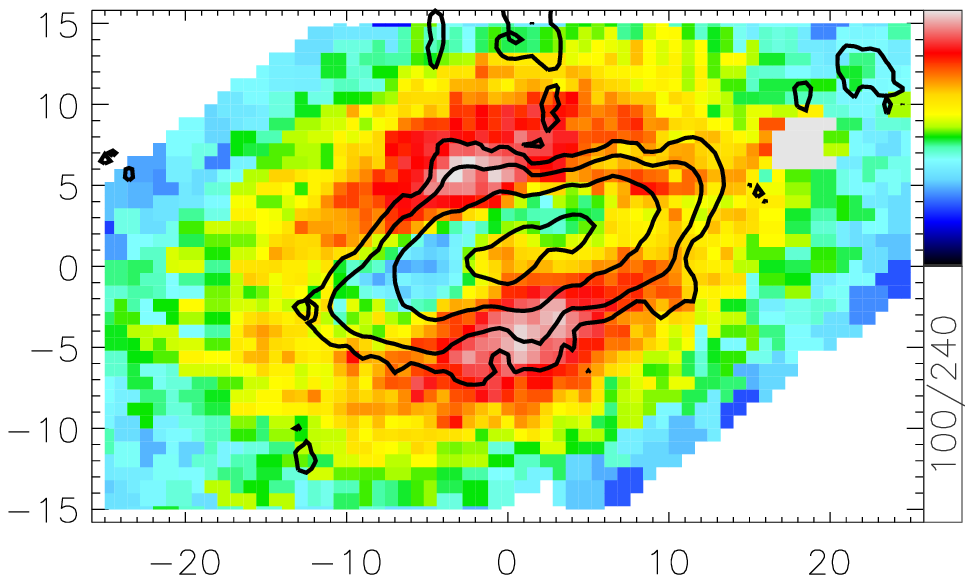}}  &
\begin{rotate}{-90}
\phantom{0000000} log(F$_{\mathrm{H\beta}}$)
\end{rotate}
\\
\begin{sideways}
\phantom{0000000000}arcsec
\end{sideways} &
\rotatebox{0}{\includegraphics[width=3.9cm]
  {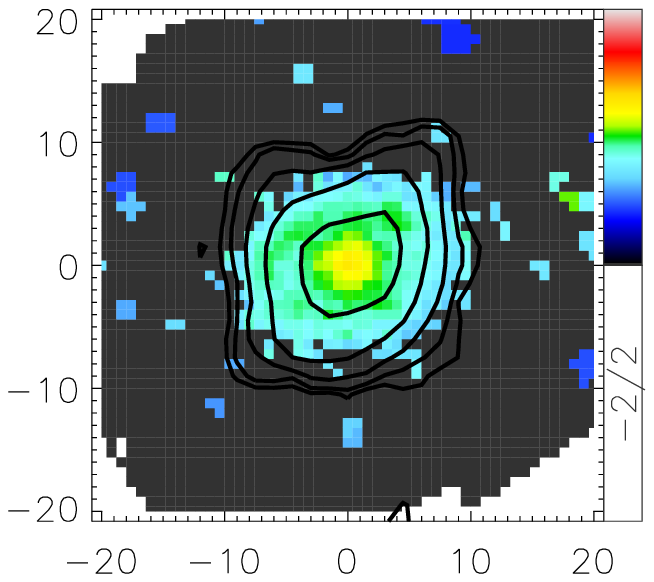}} &
 \rotatebox{0}{\includegraphics[width=3.9cm]
  {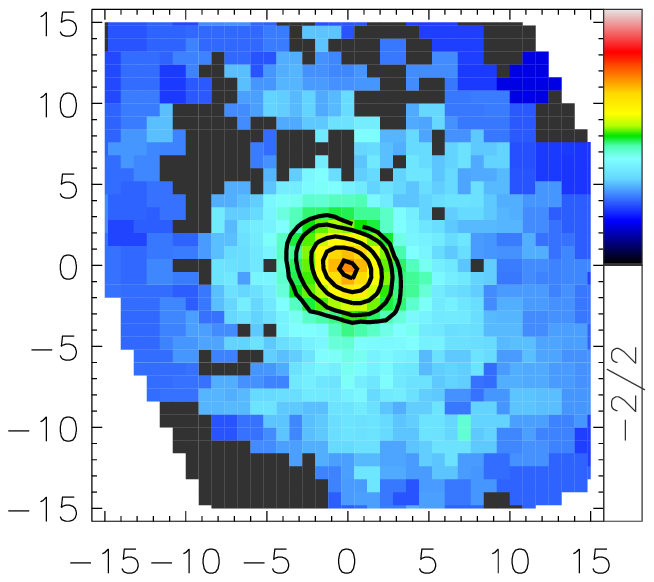}} &
\rotatebox{0}{\includegraphics[width=5.7cm]
  {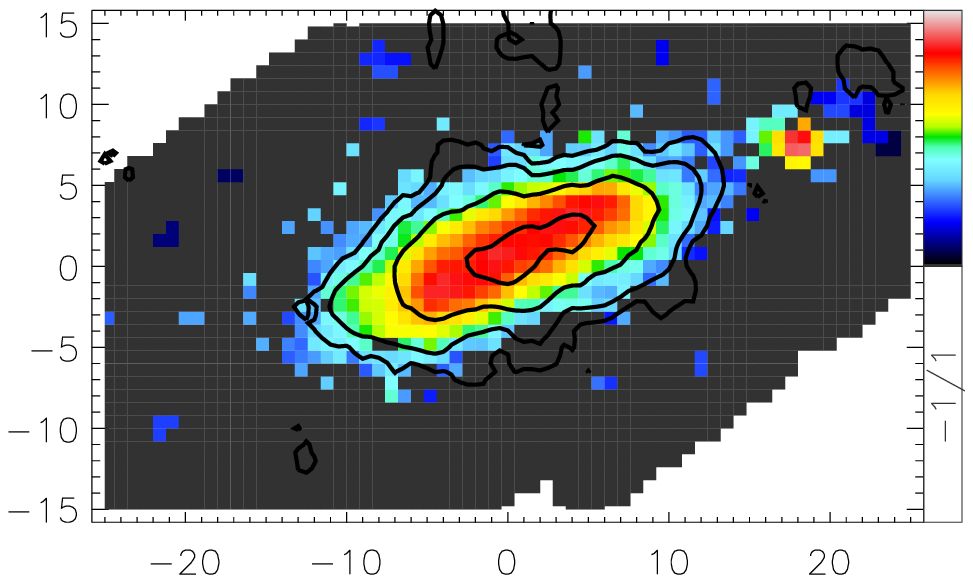}} &
\begin{rotate}{-90}
\phantom{00000} log(EW$_{\mathrm{[OIII]}}$)
\end{rotate}
\\
\begin{sideways}
\phantom{0000000000}arcsec
\end{sideways} &
\rotatebox{0}{\includegraphics[width=3.9cm]
  {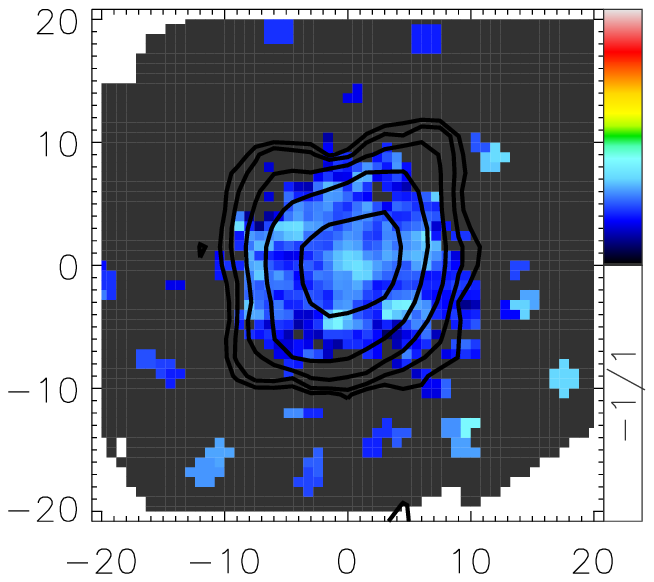}} &
 \rotatebox{0}{\includegraphics[width=3.9cm]
  {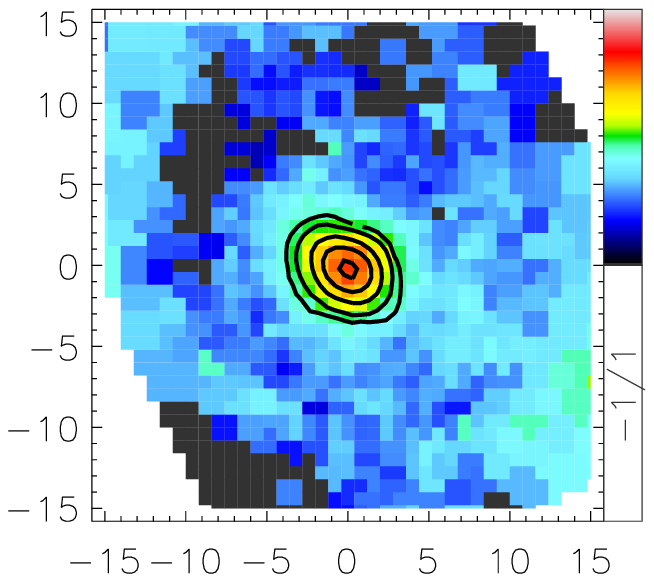}} &
\rotatebox{0}{\includegraphics[width=5.7cm]
  {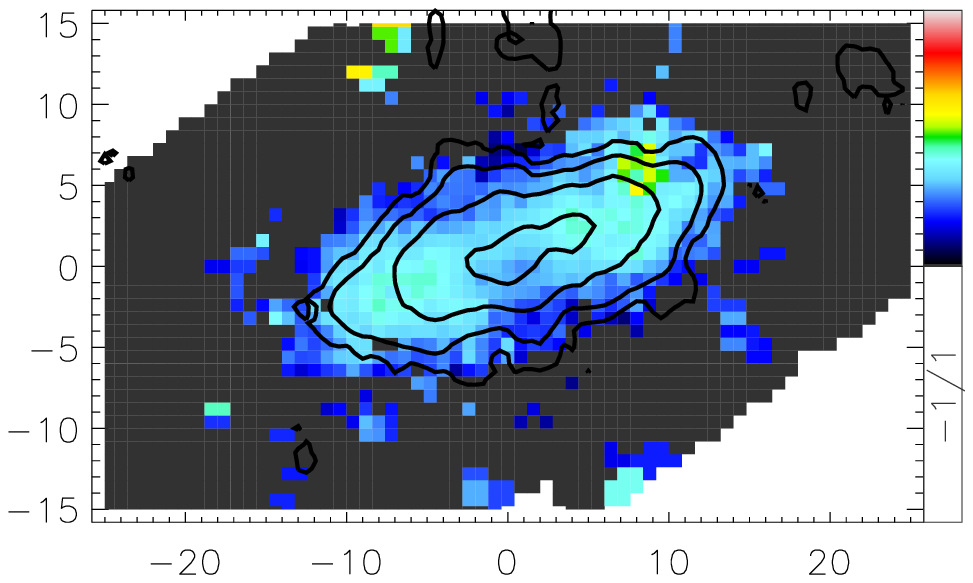}} &
\begin{rotate}{-90}
\phantom{0000} log(\otohb)
\end{rotate}
\\
\begin{sideways}
\phantom{0000000000}arcsec
\end{sideways} &
\rotatebox{0}{\includegraphics[width=3.9cm]
  {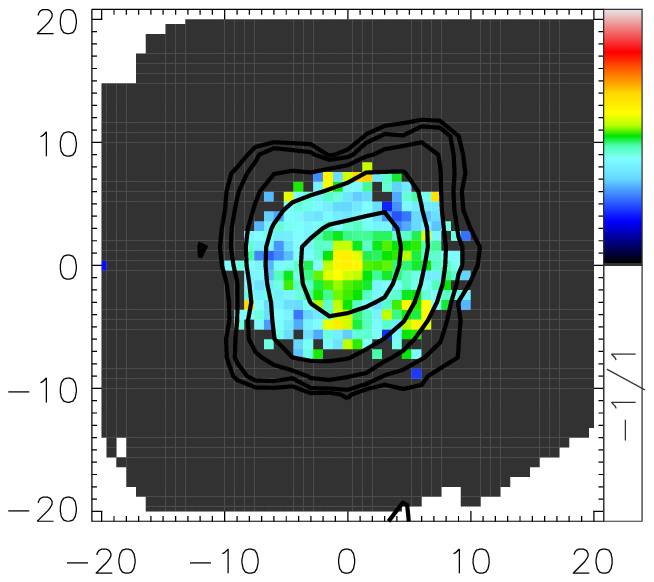}} &
\rotatebox{0}{\includegraphics[width=3.9cm]
  {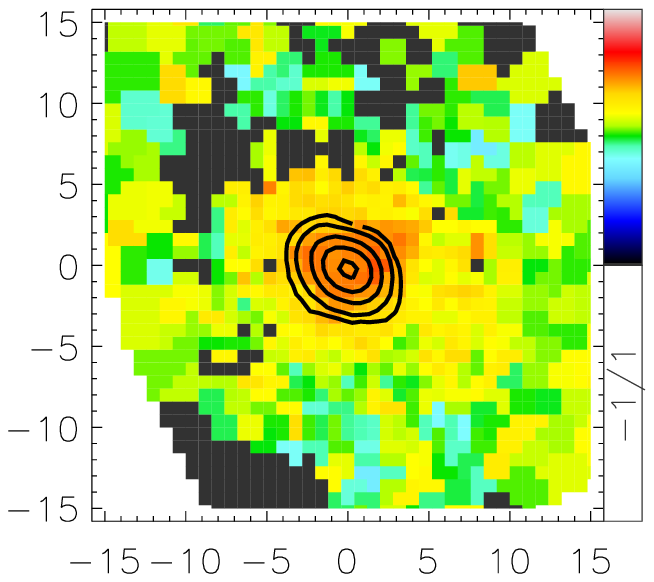}} &
\rotatebox{0}{\includegraphics[width=5.7cm]
  {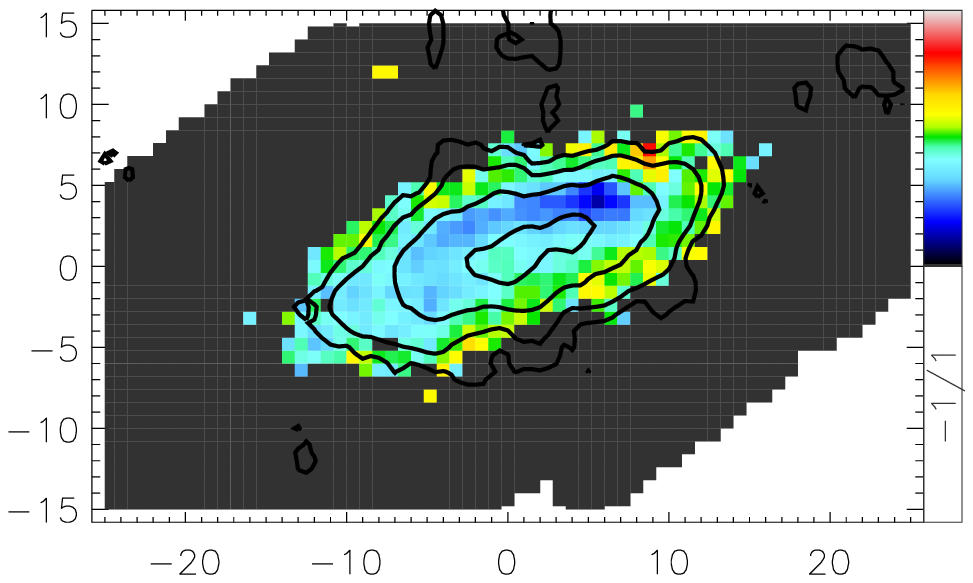}} &
\begin{rotate}{-90}
\phantom{000000000}Age
\end{rotate}
\\
\begin{sideways}
\phantom{0000000000}arcsec
\end{sideways} &
\rotatebox{0}{\includegraphics[width=3.9cm]
  {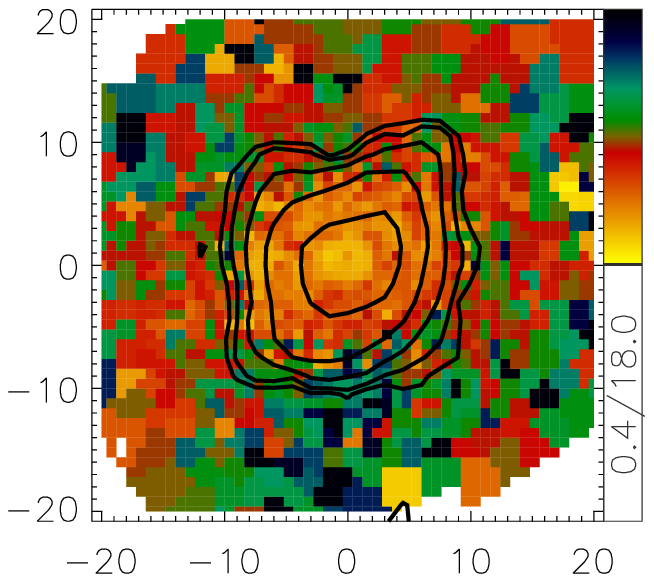}} &
\rotatebox{0}{\includegraphics[width=3.9cm]
  {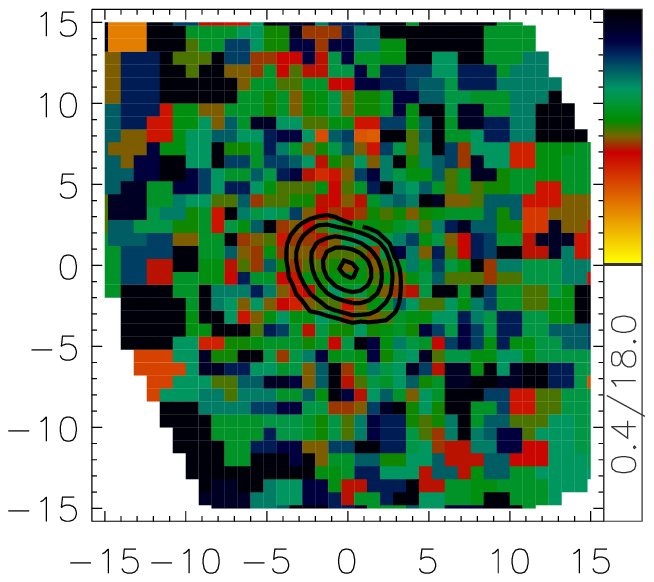}} &
\rotatebox{0}{\includegraphics[width=5.7cm]
  {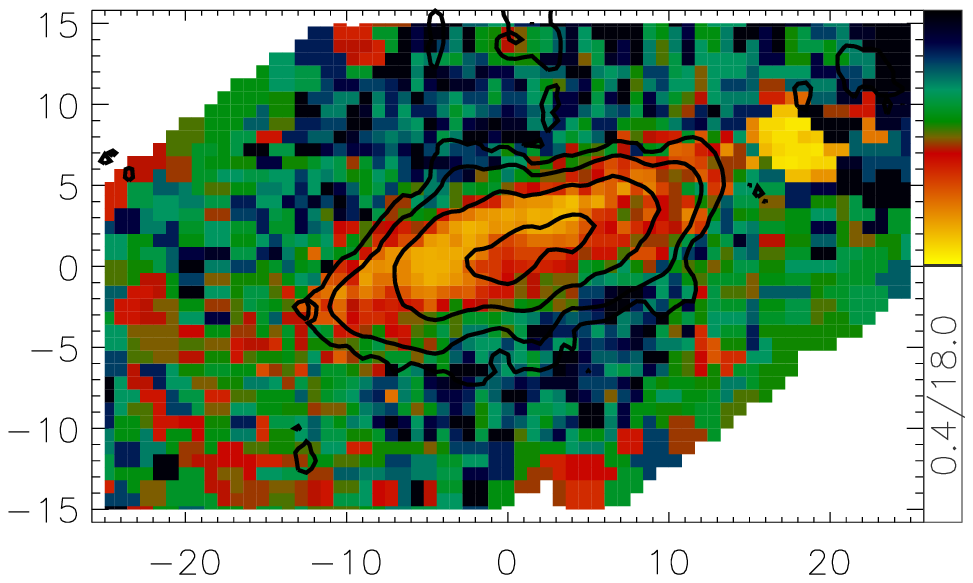}}&
\\
 & \phantom{00}arcsec & \phantom{00}arcsec & \phantom{00}arcsec\\
   \end{tabular}
\end{center}
\contcaption{}
\end{figure*}

\begin{figure*}
\begin{center}
   \begin{tabular}[c]{cccc}
 & NGC 4550 & NGC 5666 & 
\begin{rotate}{-90}
\phantom{000000000} V$_{\mathrm{star}}$
\end{rotate}
\\
\begin{sideways}
\phantom{0000000000}arcsec
\end{sideways} &
\rotatebox{0}{\includegraphics[width=3.9cm]
  {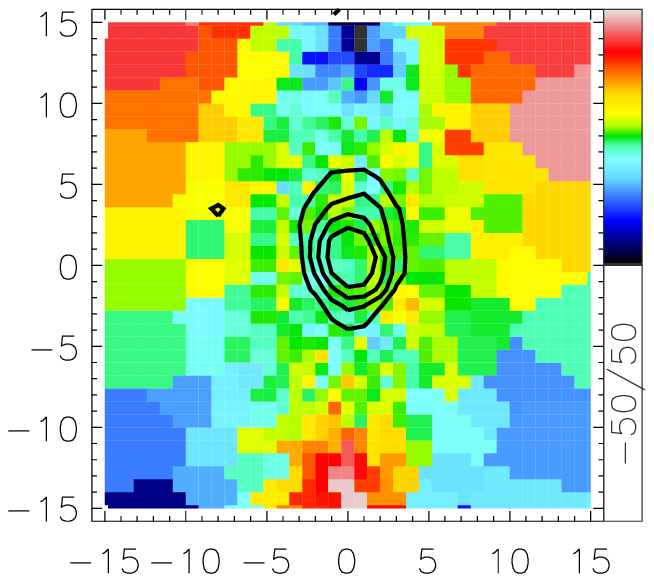}} &
\rotatebox{0}{\includegraphics[width=3.7cm]
  {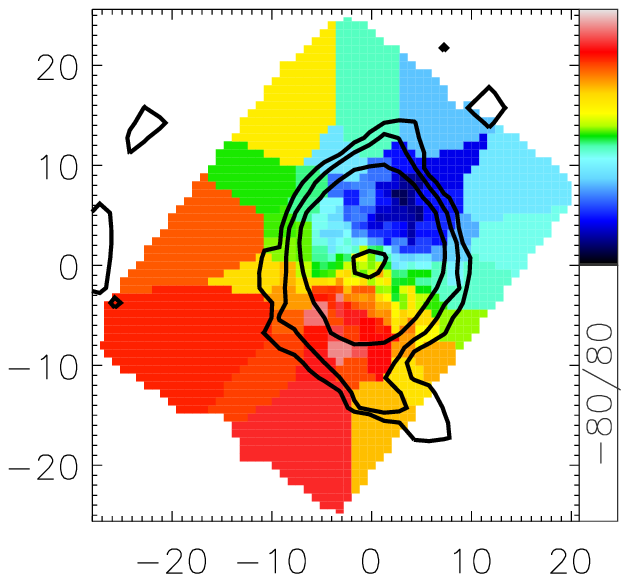}}  &
\begin{rotate}{-90}
\phantom{000000000} $\sigma_{\mathrm{star}}$
\end{rotate}
\\
\begin{sideways}
\phantom{0000000000}arcsec
\end{sideways} &
\rotatebox{0}{\includegraphics[width=3.9cm]
  {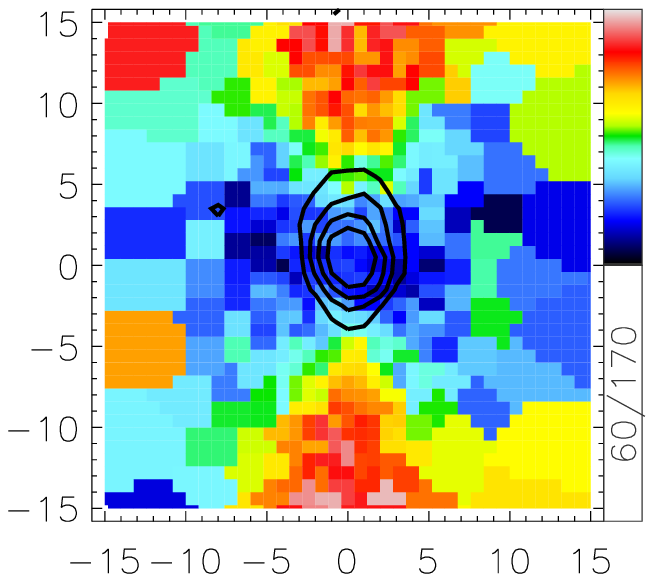}} &
\rotatebox{0}{\includegraphics[width=3.7cm]
  {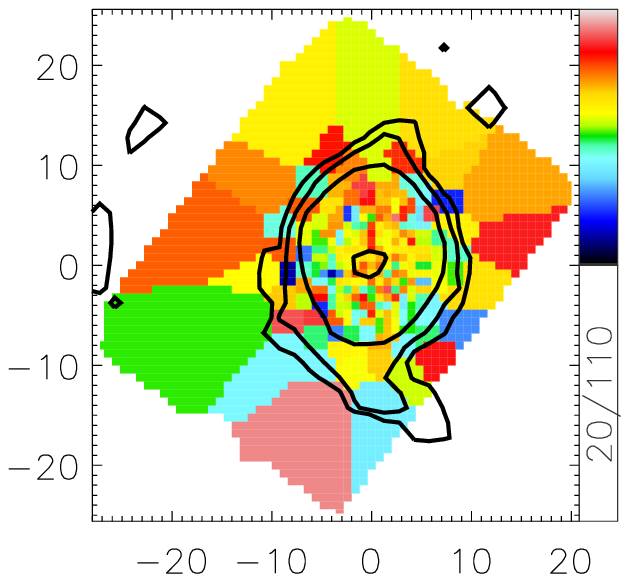}} &
\begin{rotate}{-90}
\phantom{0000000} log(F$_{\mathrm{H\beta}}$)
\end{rotate}
\\
\begin{sideways}
\phantom{0000000000}arcsec
\end{sideways} &
\rotatebox{0}{\includegraphics[width=3.9cm]
  {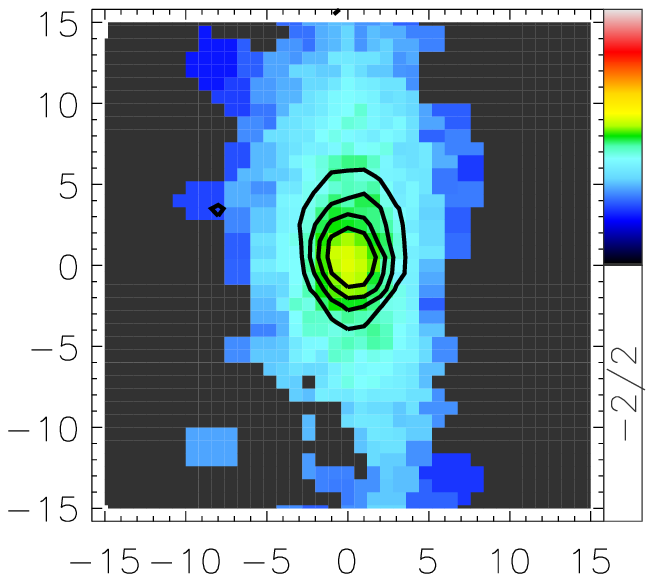}} &
\rotatebox{0}{\includegraphics[width=3.7cm]
  {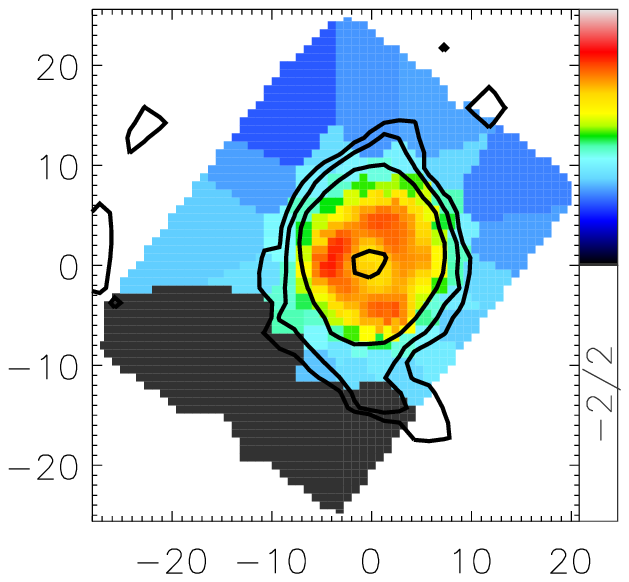}}  &
\begin{rotate}{-90}
\phantom{00000} log(EW$_{\mathrm{[OIII]}}$)
\end{rotate}
\\
\begin{sideways}
\phantom{0000000000}arcsec
\end{sideways} &
\rotatebox{0}{\includegraphics[width=3.9cm]
  {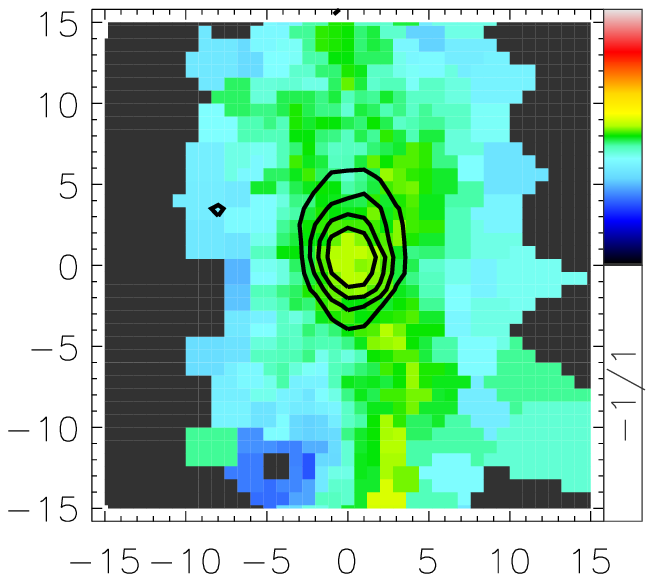}} &
\rotatebox{0}{\includegraphics[width=3.7cm]
  {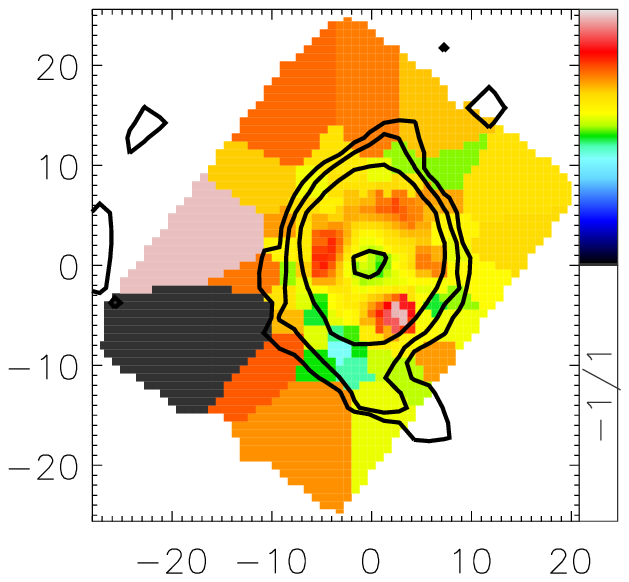}} &
\begin{rotate}{-90}
\phantom{0000} log(\otohb)
\end{rotate}
\\
\begin{sideways}
\phantom{0000000000}arcsec
\end{sideways} &
\rotatebox{0}{\includegraphics[width=3.9cm]
  {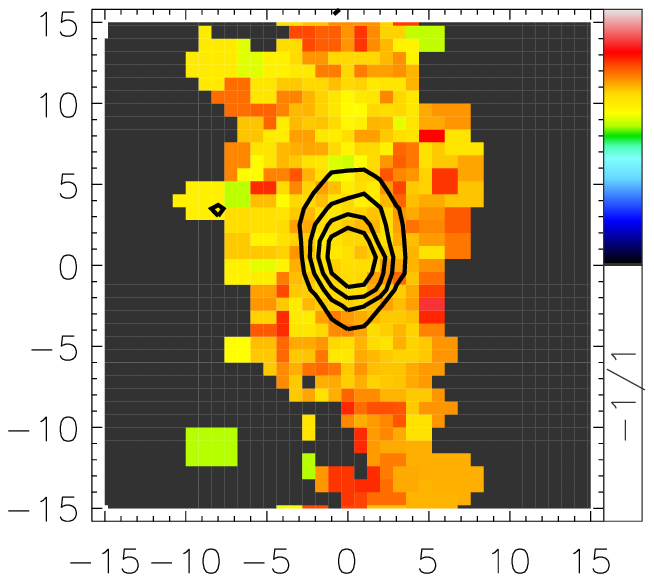}} &
\rotatebox{0}{\includegraphics[width=3.7cm]
  {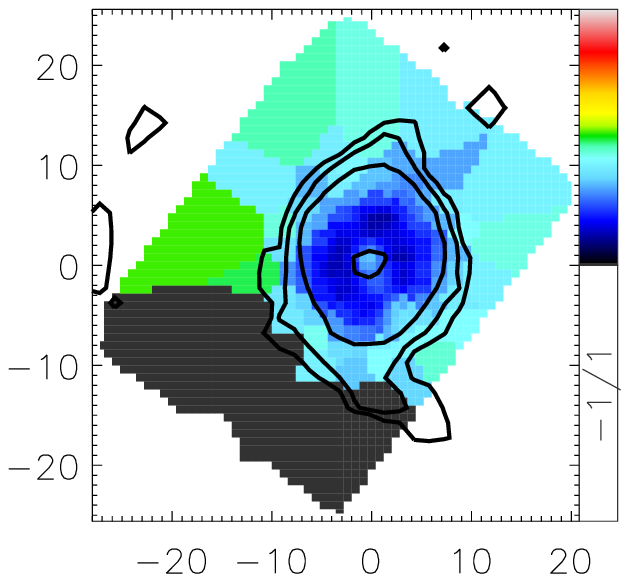}} &
\begin{rotate}{-90}
\phantom{000000000}Age
\end{rotate}
\\
\begin{sideways}
\phantom{0000000000}arcsec
\end{sideways} &
\rotatebox{0}{\includegraphics[width=3.9cm]
  {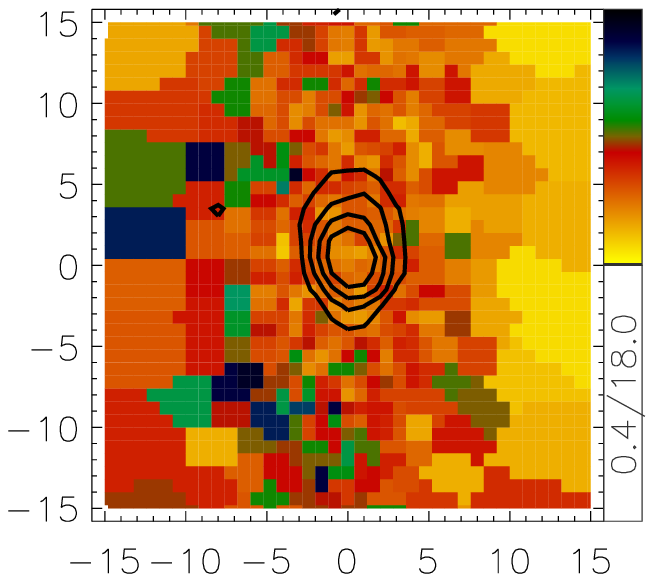}} &
\rotatebox{0}{\includegraphics[width=3.7cm]
  {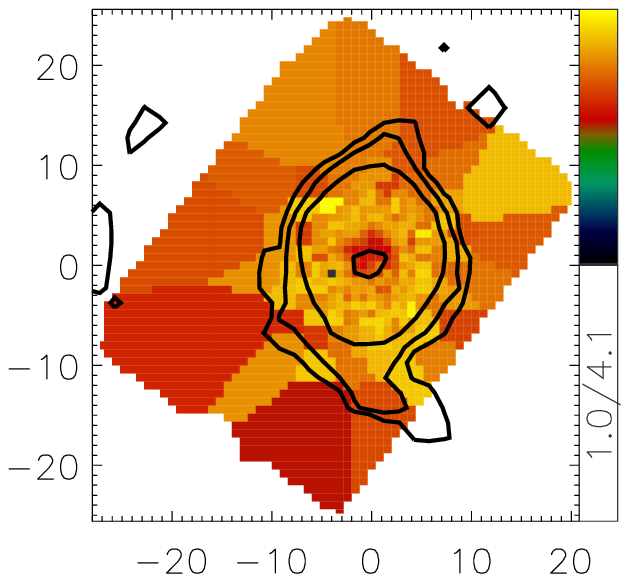}} &
\\
& \phantom{00}arcsec & \phantom{00}arcsec \\
   \end{tabular}
\end{center}
\contcaption{}
\end{figure*}


\begin{table*}
\begin{center}
 \caption{Ionisation and stellar population properties} 
\vspace{0.05cm}
 \begin{tabular}{@{}lrrrrrrrrr}

\hline
  Galaxy & Ion. Source & \otohb  & \bve & Age$_{\mathrm{NUV}-V}$ & H$\beta$ (R$_{\mathrm e}$/8)   \\

\hline
  NGC~\phantom{0}524 & star-formation(?) & 0.00 & 1.07  & old &1.34 \\
  NGC~2320 & young pAGBs, AGN & 0.36 & 0.87 & --- & 1.80 \\
  NGC~2685 & young pAGBs & 0.28 &  0.94 & --- & 1.99 \\
  NGC~2768 & old pAGBs, AGN & 0.26 & 0.96 & old & 1.68 \\ 
  NGC~3032 & star formation &  -0.41 & 0.63 & young & 4.77 \\
  NGC~3156 & young pAGBs & 0.39 & 0.77 & --- & 4.44 \\
  NGC~3489 & young pAGBs &  0.57 & 0.85 & --- & 2.81 \\
  NGC~4150 & young pAGBs & 0.19 & 0.83 & young & 3.53 \\
  NGC~4459 & star-formation & -0.07 & 0.97 & young & 2.14 \\
  NGC~4477 & old pAGBs, AGN & 0.34 & 0.97 & old & 1.62 \\
  NGC~4526 & star formation & -0.20 & 0.98 & young & 1.84\\
  NGC~4550 & young pAGBs & 0.36 & 0.89 & young & 2.04 \\
  NGC~5666 & star-formation & -0.63 & 0.86 & --- & 3.47\\
  \hline

\label{tab:organize}
 \end{tabular}
 \\
 {\sc Notes for columns:} (2) based on \citet{sarzi06}, (3) integrated total \otohb ratio, (4) from HyperLeda database, (5) based on \citet{jeong09}, (5) from \citet{kuntschner09}.

\end{center}

\end{table*}

\subsection{Molecular gas and ionised gas}

All of our CO-detected sample galaxies are also detected in the \hbeta and \oiii optical emission lines. Contours of the molecular gas distributions over the \hbeta flux maps from the {\tt SAURON} IFU \citep{sarzi06} reveal that ionised gas is detected everywhere there is molecular gas (see Fig.~\ref{fig:overlays}; except in NGC~524 where the IFU data have lower signal-to-noise ratios due to bad weather). However, in approximately half of the galaxies, the ionised gas  extends to larger radii than the molecular gas distribution. For these galaxies, cool gas in the form of atomic hydrogen may extend to these radii and provide the material to be ionised. The detection of \hi in nearly all of these galaxies (see Table~\ref{tab:origin}) strengthens this hypothesis. The kinematics of the molecular and ionised gas are always shared (not shown), indicating a strong link between the two gas phases. 

Emission-line equivalent width (EW) features are seen in both the \oiii and \hbeta lines of many of our sample galaxies. These include central peaks attributable to AGN, ring features associated with star formation, and spiral or filamentary enhancements of unclear origin \citep{sarzi09}. However, only the strongly star-forming rings in NGC~3032 and NGC~5666 match the CO distribution.  Filaments or spirals of increased EW([O$\:${\small III}]) are seen in NGC~2320, NGC~2768, NGC~3489 and NGC~4550, but the molecular gas does not obviously relate to these structures.  Furthermore, similar structures are seen in many of the {\tt SAURON} sample galaxies without molecular gas. Aside from the two strongly star-forming galaxies NGC~3032 and NGC~5666, the EW features are not directly tied to the molecular gas, suggesting it is the source(s) of the ionisation and not the underlying gas density that mediates these structures.

The sources of ionisation of all the {\tt SAURON} galaxies are discussed at length in \citet{sarzi09}. The basic conclusions of this work are that 1) AGN cannot dominate the ionisation outside of the central few hundred parsecs (central few arcsec), 2) the shock speeds required for significant shock ionisation are too high given the galaxies' potentials, 3) interaction with the hot gas phase gives a distinct morphology and low-excitation to the ionisation, but is only relevant for the most massive galaxies, 4) low \oiii/\hbeta ratios point to ongoing star formation in a few galaxies, 5) young pAGB stars are likely dominant in a few  galaxies with a significant young stellar component, and 6) generally, old pAGB stars seem the most likely source of much of the non-star forming ionisation in early-type galaxies. We tabulate the suggested dominant ionisation source of our sample galaxies in Table~\ref{tab:organize}, classifying the two additional special galaxies according to the same principles. 

\begin{figure}
\begin{center}
\rotatebox{0}{\includegraphics[clip=true, width=7cm]
   {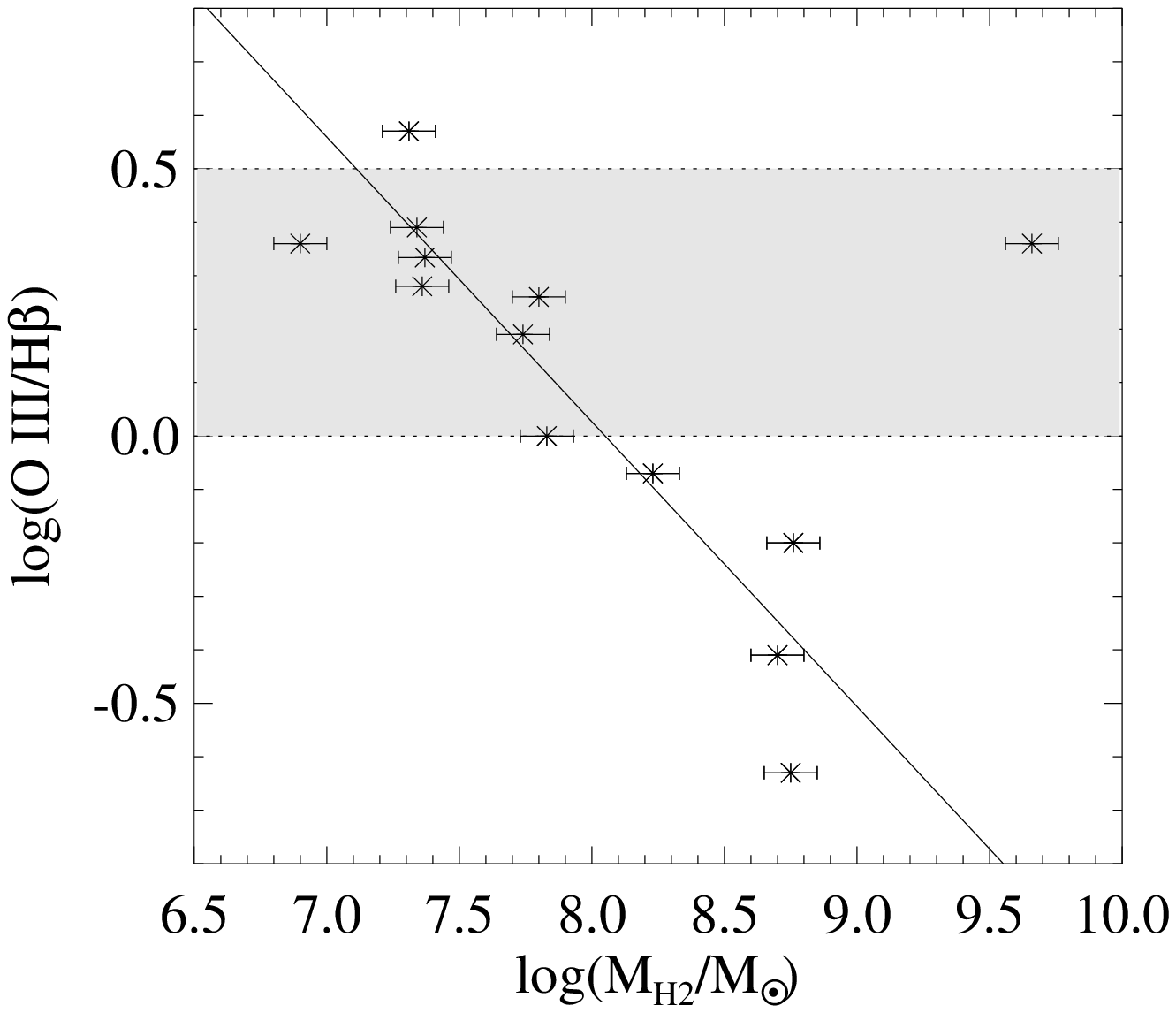}}
\caption{Total integrated \otohb ratios from \citet{sarzi09} data as a function of molecular gas mass. Galaxies with more molecular gas have correspondingly lower
  \otohb ratios, with higher ratios for galaxies with less
  molecular gas. The one exception is NGC~2320, which has a very high
  \otohb ratio despite having the most molecular gas of our E/S0
  sample. The solid line represents a regression to all the galaxies
  except NGC~2320. The shaded region represents the area with \otohb
  values found for {\tt SAURON} galaxies without any molecular gas.}
\label{fig:otohb}
\end{center}
\end{figure}

Surprisingly few galaxies of our sample contain gas definitely ionised by star formation (5/13), while the gas in a similar number is ionised by young post-AGB stars (6/13) and the gas in the remaining few is ionised by an AGN or old pAGB stars (2/13; see Table~\ref{tab:organize}). The division between star formation and other types of ionisation is made based on the \otohb ratio, since this ratio is generally the only ratio available and is lower for star formation (unless the star-forming gas is of low metallicity, when the ratio remains high). We note that in several of the galaxies ionised by young pAGB stars, the ionised gas is more extended than the CO distribution, further suggesting a source other than star formation is responsible for their ionisation. Total integrated log(\otohb) ratios of our sample galaxies range from -0.63 to 0.57 (listed in Table~\ref{tab:organize}), while for {\tt SAURON} galaxies without molecular gas this range is 0.03 to 0.47. For our sample galaxies, the total \otohb ratio anti-correlates with the total molecular gas mass (except for NGC~2320; see Fig.~\ref{fig:otohb}). Galaxies with less molecular mass have higher \otohb ratios similar to those found in galaxies without any detected molecular gas, indicating that star formation does not dominate over other ionisation sources at these lower gas contents. 

We note that a possible alternative cause for this trend would be both a higher \otohb and a higher X$_{\mathrm{CO}}$ factor (thus H$_{2}$ content currently underpredicted) at low gas-phase metallicities. Early-type galaxies might contain such low-metallicity gas if they have accreted it from a low-mass satellite or other unenriched source. Directly determining the gas phase metallicities would be ideal to investigate this scenario, but this is not possible with the current data. However, the work of \citet{schawinski07} suggests that low-metallicity star-forming gas is rare in E/S0s, as does the relation between IR and H$\alpha$ star-formation indicators discussed in Section~7.2.4 for our sample. Thus, the anti-correlation observed is more likely explained by star formation dominating the ionisation only when high star formation rates (associated with large molecular gas reservoirs) are reached. 

NGC~2320 has the most molecular gas of our sample, but also displays a uniformly high \otohb ratio. The balance between ionisation from star formation and other ionisation sources must be different in this galaxy. Either star formation is weaker than expected given its molecular gas content (supported by the lack of extended radio continuum and 24\mum emission; \citealp{young09}) or other ionisation processes are particularly strong, possibly its AGN. Further data are needed to distinguish these scenarios, particularly to investigate whether such a larger reservoir of molecular gas could have a very low star formation efficiency. 

In summary, we see ionised gas wherever molecular gas is detected and the two gas phases are kinematically aligned in our sample of E/S0s with molecular gas. However, this ionised gas is only ionised by star formation in 5 galaxies. In the other galaxies (mostly with less molecular gas), other ionisation sources dominate, such as young and/or old pAGB stars and AGN.

\subsection{Molecular gas and stellar populations}

Because they contain molecular gas, we expect our sample galaxies to host young stellar populations. Here we investigate the presence of such young populations using optical and UV colours and absorption linestrengths.  

\begin{figure}
\begin{center}
\rotatebox{270}{\includegraphics[clip=true, width=6cm]
   {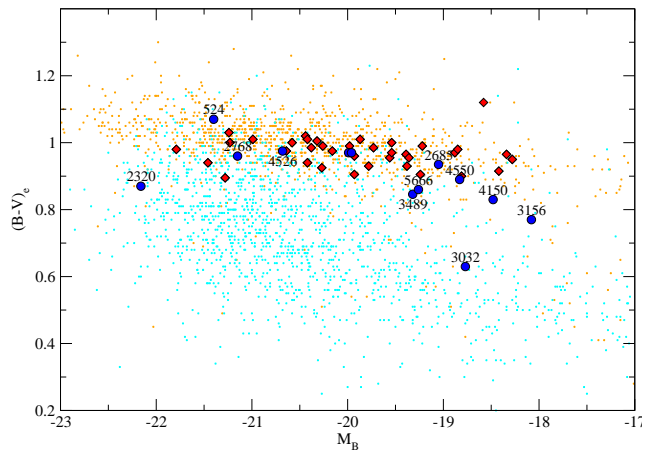}}
\caption[Colour-magnitude diagram]{Colour-magnitude diagram. Blue circles represent the early-type 
galaxies with molecular gas in our sample. Red
  diamonds are the {\tt SAURON} sample E/S0s without molecular
  gas.  Small cyan and orange
  circles represent late-type and early-type galaxies from HyperLEDA,
  respectively. From left to right, the two
unlabelled galaxies are NGC~4459 and NGC~4477. }
\label{fig:cm}
\end{center}
\end{figure}

A first check for young stars can be
made with optical colours. To do this, we plot a
colour-magnitude diagram using \bve\footnote{\bve is the \bv colour
  within the aperture that contains half of the total $B$-band
flux. Values are from the HyperLEDA database (http://leda.univ-lyon1.fr) and are listed in Table~\ref{tab:organize}.} colour against absolute 
blue magnitude (Fig.~\ref{fig:cm}). Our sample galaxies are shown as blue circles, while {\tt SAURON} galaxies without molecular gas are shown as red diamonds. Other
late-type (t $>$ 1) and 
early-type (t $<$ -1) galaxies are shown as small
cyan and orange points, respectively. The {\tt SAURON}
galaxies without molecular gas lie on the colour-magnitude relation
characteristic of most early-type galaxies, i.e. the red
sequence. While some of our sample galaxies also lie on this red sequence, about half are bluer, a property presumably attributable to their hosting a significant fraction of young stars.

However, optical colours are not very sensitive to young stellar populations and the UV provides a distinct improvement \citep[e.g.][]{schawinski07b}. Based on \citet{jeong09}, 5/8 of our sample galaxies with GALEX (Galaxy Evolution Explorer) UV imaging data contain blue NUV$-V$ colours indicative of a young stellar population (see Table~\ref{tab:organize}, while this fraction is only 4/26 for the rest of the {\tt SAURON} sample. Of the five sample galaxies with young NUV$-V$ colours, two are on the optical red-sequence, confirming the greater sensitivity of the UV to young stars. However, the other three sample galaxies with GALEX data show no sign of a young stellar population in the UV. As for the four galaxies with blue NUV$-V$ colours from the non-CO detected {\tt SAURON} sample, three have such blue colours only at large radii, outside of the central area searched for CO. Either there is molecular gas and star formation at these large radii, or young blue stars have been recently accreted from a satellite. The fourth galaxy is NGC~7457, for which we have documented a firm CO upper limit. It seems to be a post-starburst galaxy that has evacuated all its molecular gas. Thus there is a strong, but not one-to-one, link between central blue NUV$-V$ colours and the presence of molecular gas in early-type galaxies.  

Absorption linestrengths obtained from the IFU data can also be
used to estimate the ages of the stellar populations and helpfully provide information about the spatial distribution of the young stars. The measured linestrengths (see Table~\ref{tab:organize})
are used in combination with theoretical models for
SSPs to determine the age, metallicity and
[$\alpha$/Fe] ratio of a galaxy. Fits to these parameters for the {\tt SAURON} galaxies are presented in 
\citet{kuntschner09}. We show CO distributions over age maps in 
Fig.~\ref{fig:overlays}, except for the two galaxies
outside of the {\tt SAURON} sample, NGC~2320 and NGC~5666. For these
two galaxies, we show the CO distributions over the H$\beta$
linestrength maps instead, as H$\beta$ is the most age-sensitive linestrength measured.

The CO contours over the age maps reveal various relative distributions of
young stars and molecular gas. Four galaxies have globally
young ($<$ 3~Gyr) SSP ages, extending beyond the region of detected
molecular gas, while two galaxies have extended intermediate (3-7~Gyr) SSP ages.
In two other galaxies, a distinct change from intermediate to old SSP age occurs exactly at the boundary of the detected molecular gas. Because the SSP fit tries to account for both a young and old population in a single age, the age derived partly reflects the age of the young population and partly the fraction of young to old stars. Since all of these galaxies are likely forming some stars from their molecular gas, the intermediate ages found in some of the galaxies probably reflect lower mass fractions of truly young stars. However, three sample galaxies also have purely old populations based on their SSP fits. These are the same galaxies that have purely red NUV$-V$ colours. Using a central ($R_{\mathrm e}$/8) age of 5~Gyr to divide young and old galaxies, around 55\% of galaxies with molecular gas have young SSPs and only 11\% of those without molecular gas.  If a SSP age of 2~Gyr is used instead, these percentages change to 36 and 0\%, respectively.

The optical colours, UV-optical colours and the SSP-equivalent age maps described above trigger two interesting
questions. Firstly, why do some galaxies have no sign of a young
stellar population despite having molecular gas?
Secondly, why are the young stellar populations in four galaxies so
much more extended than the molecular gas we observe? 

Three sample galaxies have predominantly old stellar populations:
NGC~524, NGC~2768 and NGC~4477. One possibility is that the young populations in
these galaxies are simply too small relative to the old population to
be detected.  To test this hypothesis, we have calculated the
stellar mass contained within the region of detected molecular
gas for each sample galaxy, using I-band images from the
MDM observatory (Falc\'on-Barroso et al., in prep). This mass is
listed in Table~\ref{tab:mol2star}, along with the ratio of molecular gas
to stellar mass and the expected percentage of young stars given an
extrapolation of the current SFR over the past 1~Gyr (SFRs taken from
Table~\ref{tab:sfrs} with the following order of preference:
H$\alpha$+PAH, H$\alpha$+24\mum, H$\alpha$+TIR). 
We also note the mass-to-light ratios used, which are based
upon the Jeans models of \citet{cappellari06} and \citet{scott09}. For
galaxies not in either of these papers, we use an average
mass-to-light ratio from 
galaxies with similar absorption linestrengths, which should capture
the first-order effects of different stellar populations. 

This calculation shows that the galaxies without detected young stars 
have some of the lowest molecular gas
to stellar mass ratios and predicted young star mass fractions. It is difficult
for both UV colours and absorption linestrengths to detect young
stellar populations of less than about 0.1\% in mass. Our three purely old galaxies
are all below this limit. However, several galaxies with detected young stars have only slightly higher fractions. Whether a galaxy has a detectable fraction of young stars may thus be due more to peculiarities in its recent star formation history -  either star formation is
just starting or has been less efficient than assumed in the purely old
galaxies, or star formation
has recently been more vigorous in the others. This calculation of the molecular gas to stellar mass ratio shows that the old ages found for some galaxies indicate very low ratios of young to old stellar populations, but they do not
completely rule out the presence of young stars. 

\begin{table}
\begin{center}
 \caption{Molecular to stellar mass ratios}
 \begin{tabular}{@{}lccccc}
  \hline
  SSP Age & NGC & $(\frac{M}{L})_{\mathrm {Jeans}}$ & log($M_{\mathrm
  {*}}$) & log($\frac{M_{\mathrm H_{2}}}{M_{\mathrm {*}}}$) &
  $f_{\mathrm{YC}}$ \\[1mm]
  & & ($\frac{M_{\odot}}{L_{\odot,I}}$) &(M$_{\odot}$) & & (\%) \\
  \hline
  Old & \phantom{0}524 & 5.04 & 11.2 & -3.4 & 0.02 \\
       & 2768 & 5.32 & 10.7 & -2.9 & 0.05 \\
       & 4477 & 3.22 & 10.4 & -3.1 & 0.09 \\
\hline
  Young & 3032 & 1.40 & 10.3 & -1.5 & 2.88 \\
         & 3489 & 0.99 & 10.2 & -2.9 & 0.21 \\
         &4150 & 1.56 & 10.0 & -2.2 & 0.32 \\
           & 4459 & 2.76 & 10.8 & -2.6 & 0.13 \\
	  & 4526 & 3.51 & 10.3 & -2.4 & 0.18 \\
  	& 4550 & 2.81 & 10.0 & -3.1 & 0.06 \\
         &5666 & 1.40 & 10.5 & -1.8 & 4.71 \\ 
  \hline
 \end{tabular}
\label{tab:mol2star}
\end{center}
{\sc note:} $(\frac{M}{L})_{\mathrm {Jeans}}$ is the
mass-to-light ratio derived from the Jeans models of
\citet{cappellari06} or \citet{scott09}. $M_{\mathrm {*}}$ is the stellar mass within the
region with detected CO. $\frac{M_{\mathrm H_{2}}}{M_{\mathrm {*}}}$
is the ratio of molecular gas to stellar mass within the region with
detected CO. $f_{\mathrm{YC}}$ is the expected mass fraction of
young to old stars based on the current SFR
extrapolated over the past 1~Gyr.

\end{table}

Four sample galaxies have young stellar populations ($<$ 3~Gyr) more spatially
extended than their molecular gas (NGC~3032, NGC~3489, NGC~4150, NGC~5666). They do not show a distinct change in stellar population at the molecular gas boundary; instead each has a smooth age gradient, with younger stellar populations toward the centre where the molecular gas lies. The age gradients combined with the spatially limited molecular gas suggests that star formation has ended at larger radii first and is ceasing in an outside-in manner, a possibility first suggested by \citet{shapiro09} for these same galaxies based on the spatial distribution of polycyclic aromatic hydrocarbon (PAH) containing dust. The exhaustion of the gas supply through star formation could explain such a radial trend, as could the enhanced removal of outer gas from ram-pressure or dynamical stripping. AGN feedback cannot be invoked for these galaxies, as an AGN would presumably disrupt the inner gas first. Additionally, no clear sign of AGN activity is seen. 
These four galaxies with extended young stellar populations all lie below the optical red sequence  at the low-mass end of our sample distribution and should join the red sequence soon as their star formation continues to decline. However, we note that the same process, if present in more massive galaxies, may be hidden by their dominant old stellar populations. 

In short, the comparison of the molecular gas content and extent shows that young populations and molecular gas are closely related, but not uniquely. Some galaxies with molecular gas do not show signs of a young stellar population (likely because the old population dominates) and a small proportion of galaxies without any detected molecular gas do have young stars. In a handful of our sample galaxies, we see evidence that star formation ceased at large radii before the inner radii where the molecular gas is currently located.

\subsection{Molecular gas and stellar kinematics}

Stars recently formed from a molecular gas disc or ring will have discy
kinematics, dominated by high rotation and low velocity
dispersion. If these disc stars contribute enough to a galaxy's
light, we should see this contribution in the mean stellar velocity and
velocity dispersion fields, with some dependence on inclination. Maps
of these quantities with overlays of 
molecular gas contours are shown in Fig.~\ref{fig:overlays} for all of
our sample galaxies.

 Four galaxies (NGC~3489, NGC~4459, NGC~4526 and NGC~5666) show discy stellar components related to the CO. Additionally, in one galaxy (NGC~3032), such a component is detected in a higher-resolution stellar velocity map from \citet{mcdermid06a}. The remaining galaxies have no stellar kinematic feature directly attributable to the molecular gas via star formation, although peculiar features which most likely indicate past interactions are visible in three galaxies (see \citealp{crocker08, crocker09, young08} for discussions). 
In general, it is thus the galaxies with clear young
stellar populations that show signs of a related cold kinematic component, while
galaxies with intermediate and old SSP ages do not. This bimodality is as expected since both properties depend on the light from young stars dominating that from old stars.

\begin{figure}
\begin{center}
\rotatebox{0}{\includegraphics[clip=true, width=7.2cm]
   {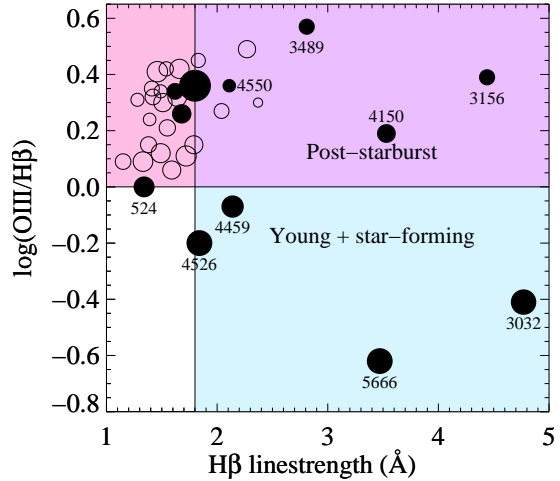}}
\caption{A loose classification for E/S0s with molecular gas based on the total integrated \otohb emission and the H$\beta$ absorption linestrength
within $R_{\mathrm 
  e}/8$. Galaxies detected in CO are shown as filled black circles while those undetected in CO are shown hollow. Three regions are inhabited by our sample galaxies. In the blue
region are galaxies whose gas ionisation is currently dominated by
star formation, also showing 
either locally or globally young stellar populations (only galaxies with CO). In purple are probable
post-starburst galaxies, which have young stellar populations but
whose ionisation is not currently dominated by star formation (both with and without CO). In pink are galaxies
whose gas is not ionised by star formation, nor do they display young
stellar populations (mostly galaxies without CO). The size of the symbols logarithmically scales
with the molecular gas mass (or mass upper limit) in each galaxy.}
\label{fig:classify}
\end{center}
\end{figure}

\subsection{A classification}
Some early-type galaxies have detectable young stellar populations, while others do not. Similarly, some have their ionisation dominated by star formation while others do not. Fig.~\ref{fig:classify} illustrates where galaxies lie with regards to their H$\beta$ absorption linestrength within $R_{\mathrm e}$/8 from \citet{kuntschner09}
and their total integrated log(\otohb) in emission from data from \citet{sarzi09}. The sample CO-detected galaxies are shown as filled circles whose size is proportional to the logarithm of their molecular gas mass. {\tt SAURON} sample galaxies not detected in CO are shown as hollow circles with size proportional to the logarithm of their molecular mass upper limit. (Some galaxies without CO are also undetected in ionised gas and are not shown at all.)

The galaxies without detected molecular gas inhabit a relatively limited region of this diagram. All have a log(\otohb) ratio above 0.0 and most have old populations indicated by a H$\beta$ linestrength value less than 1.8 \AA. These boundaries are motivated by the expectation of ionisation only from star formation at log(\otohb) $<0.0$ and the definite presence of a young stellar population when the H$\beta$ linestrength $>1.8$ \AA\xspace (this choice of limit also agrees with the NUV$-V$ criterion in \citealp{jeong09}). Only around 14\% of the galaxies with ionised gas but no detected molecular gas have young stellar populations. These galaxies have probably undergone an episode of star formation in the recent past but have little or no remaining molecular gas. 

Galaxies with molecular gas are spread fairly equally between 3 different categories: those that are young and strongly star-forming, those that are young but whose ionisation is not currently dominated by star formation, and those that appear old and not star-forming, similar to the majority of galaxies without molecular gas. 
The galaxies in the young and star-forming region all have over $10^8$  \Msun of molecular gas and have the highest star formation rates (see Table~\ref{tab:sfrs}).
Galaxies in the second category have young stellar populations but high \otohb ratios probably due to young pAGB stars. All have less than $10^8$ \Msun of H$_{2}$, including some that have not been detected with CO. The apparently old and not star-forming galaxies with molecular gas have received attention previously in Section~5.3. These galaxies may be star-forming, but the amount of star formation relative to their old stellar mass is not sufficient to make the young stars detectable.  They thus appear similar to galaxies that do not contain molecular gas.

In general, we expect galaxies to cycle through this diagram as they acquire, use and then exhaust a supply of cold gas. Consider a quiescent early-type galaxy that accretes a significant amount of gas.  First, we expect it to move down from the old region as star formation starts to dominate the ionisation and the \otohb ratio thus drops. When the young stars become a significant mass fraction of the old stellar population, the H$\beta$ linestrength will increase, moving the galaxy into the young and star-forming region in Fig.~\ref{fig:classify}. The galaxy may remain there star-forming for some time, but eventually it will start to exhaust its gas supply and its SFR will decline. Then young pAGB stars will start to dominate the ionisation and the galaxy will move up to the `post-starburst' region of the diagram, even if it is still moderately star-forming. As the SFR continues to decline and then stops, the young stars will be less important relative to the old stars and the H$\beta$ linestrength will decrease, eventually reaching values consistent with a purely old population.  

\section{Discussion}

\subsection{Origin of the gas}

 \begin{table}
 \caption{Accretion signs}
 \begin{tabular}{@{}lrrrr}
  \hline
  NGC & PA$_{\mathrm{g}}$-PA$_{\mathrm{\star}}$ &log($M_{\mathrm {H {\tiny I}}})$ & H$\:${\scriptsize I} Morphology &
  H$\:${\scriptsize I} Ref.  \\
   & (deg) & ($M_{\sun}$) & \\
  \hline
    \phantom{0}524 & 1           & 6.41    & distant cloud & 1\\
    2320 & $\approx 0$ & $<7.75$ &  --- & 2 \\
    2685 & 73 & 9.26 & warped polar disc & 3\\
    2768 & -95         & 8.23    & central cloud; tail& 3\\
    3032 & -151        & 7.98    & central disc & 1\\
    3489 & -6          & 6.76    & central disc; tail & 1\\
    4150 & 21          & 6.40    & central disc; tail & 3\\
    4459 & -1          & $<6.59$ & --- & 1\\
    4477 & -28         & $<6.61$ & --- & 1\\
    4526 & 6           & $<7.88$ & --- & 4\\
    4550 & 0           & $<6.57$ & --- & 1\\
    5666 & $\approx 0$ & 9.23    & extended disc & 1\\ 
  \hline
 \end{tabular}
\label{tab:origin}
{\sc Notes:} PA$_{\mathrm{g}}$-PA$_{\mathrm{\star}}$ is the difference between the
ionised gas and stellar kinematic position angle from \citet{sarzi06}, except
for NGC~2320 and NGC~5666 where it is estimated by eye to be close to
zero (definitely co-rotating). H$\:${\scriptsize I} references: (1)
 Oosterloo et al. (in preparation), (2) \citet{lucero08}, (3) \citet{morganti06}, (4)
\citet{diseregoalighieri07}.  
\end{table}

The cold gas in early-type galaxies is probably either accreted from an
external source or results from internal stellar mass loss. For some galaxies, cold gas may
also remain from before the morphological transition to early-type (i.e. from their presumed spiral progenitors). These scenarios are
difficult to distinguish, but the kinematic misalignment of gas and
stars or a disturbed atomic hydrogen distribution are telltale signs of
external accretion.  

As ionised gas in our sample galaxies always seems kinematically linked with the
molecular gas and its kinematic position angle is more reliably determined, we
use here the difference in kinematic position angles between the ionised
gas and stars from \citet{sarzi06}, listed in
Table~\ref{tab:origin}. In 8/12 galaxies, the gas rotates in the same
direction as the stars and thus may be contributed by either internal
or external processes. However, the polar rotating gas in NGC~2685 and NGC~2768 and the
counter-rotating gas in NGC~3032 must have been contributed by
external sources. In NGC~4550, the gas most likely remains from a
merger of two disc galaxies \citep{crocker09}.  

About half the sample galaxies are also detected in H$\:${\small
  I}. Out of the six firmly-detected galaxies, three show signs of
  possible ongoing accretion. The case is most 
clear for NGC~2768, that has a small amount of H$\:${\small I} at the galaxy
centre, but a distinct and massive tail of H$\:${\small I} extending toward the
northeast \citep{morganti06}. \citet{crocker08} discuss possible accretion scenarios for
this galaxy in detail, suggesting that a nearby spiral galaxy is the
  most likely donor of the gas. NGC~3489 and NGC~4150 both have very low column
density tails of atomic hydrogen (Oosterloo et al. in preparation) that indicate
ongoing accretion. However, neither of these galaxies has much additional
  cold  gas to accrete. 

Thus out of our sample, three galaxies must have acquired their gas
externally, as indicated by gas misalignment, and 
two more show hints of ongoing accretion. All of these galaxies with
recent or ongoing accretion are located in the field, hinting at the difficulty of gas exchange when a hot intracluster medium is present. These five galaxies are a lower limit to the incidence of accretion; in others, the obvious signs of accretion may have already
disappeared. In NGC~4550, gas appears to have been contributed along with stars in a major merger, potentially the merger that transformed it to an early-type galaxy. But internal stellar
mass loss may also be important for the other galaxies, especially those in
a cluster environment.

\subsection{Star formation tracers}
Early-type galaxies provide unique conditions to study star
formation, with their deep potential wells, hot gaseous halos,
relatively little cold gas, frequent AGN and only weak, if present,
stellar discs. Considering
these physical differences, we are interested in whether the molecular
gas in early-type galaxies forms stars in the same manner as 
in spirals. To do this, we must reliably estimate the star formation 
rate. Multiple potential tracers exist: radio continuum
(synchrotron emission from supernova remnants and thermal emission from H$\:${\small II}
regions), mid-infrared (MIR) and FIR emission (produced by young stars heating 
their dusty environs, i.e. obscured star formation), Balmer line emission
(by young stars ionising surrounding gas) and UV radiation (unobscured young O
and B stars). Each of these tracers is used as a SFR measure in the
literature (see \citealp{kennicutt98b} for a review). 

All of these tracer-SFR conversions, however, rely on star formation
and directly related processes being the dominant
source of the observed emission. In early-type galaxies, this may not always
be the case. For example, MIR and FIR emission is present in early-type
galaxies that are clearly quiescent \citep{temi07}, as is extended
H$\beta$ emission \citep{sarzi06}. Radio AGN can
contribute significantly to radio emission and some old stars emit in the UV
(the UV-upturn phenomenon; see e. g. \citealp{oconnell99} for a
review). By comparing their fluxes, we will investigate here the
reliability of  radio, FIR and H$\beta$ emission as star formation tracers in
early-type galaxies. If two of these
measures accurately trace star formation, a linear relation should
hold between them. The use of 8 and 24\mum emission as 
star formation tracers for the {\tt SAURON} sample is discussed in
\citet{shapiro09} and \citet{temi09}, respectively. The UV is treated in
\citet{jeong09}, but traces young stellar populations better 
than it does the current star formation rate.

\begin{figure}
\begin{center}
\rotatebox{0}{\includegraphics[clip=true, width=7.2cm]
{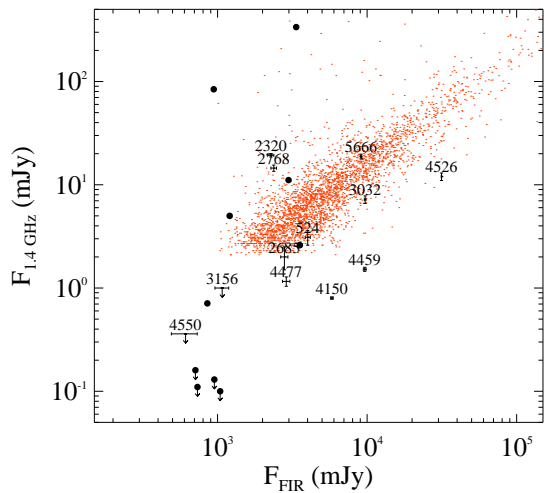}} 
\caption{Radio continuum versus FIR flux. The UGC sample
  from \citet{condon02} is shown as red 
  points. Our sample galaxies are
  shown with black error bars, while {\tt SAURON} sample E/S0s without detected molecular gas are filled black circles.
  }
\label{fig:q}
\end{center}
\end{figure}

\subsubsection{Radio-FIR correlation}

The tight relation of radio continuum and FIR fluxes in star-forming
galaxies holds over five orders of magnitude \citep{price92}. This
relation is attributed to both the FIR and radio emission tracing the
current star formation rate. Star formation-associated FIR emission is
produced when hot young stars heat their dusty environments. Radio
emission is linked to the SFR through the release of cosmic rays by
supernovae. These cosmic rays are then accelerated in the magnetic field of the
galaxy, producing synchrotron radiation.  However, the relation is slightly
non-linear below $L_{\mathrm{FIR}}<10^{9}$ L$_{{\tiny \sun}}$. Weaker
FIR galaxies have either radio deficits or FIR excesses with respect
to higher FIR galaxies \citep{yun01}. 

To compare the radio and FIR emission we plot the radio flux against
the FIR flux in Fig.~\ref{fig:q}. The small red
points represent the \citet{condon02} sample of NRAO VLA Sky Survey
(NVSS) and IRAS detected galaxies from the Uppsala General Catalogue
(UGC).  The UGC catalogue contains both spiral and early-type galaxies, although many early-type galaxies are not detected and most that are detected are radio AGN.   Our sample of CO-detected E/S0 galaxies is overplotted with error bars and marked with each galaxy's NGC
number.  We also plot the few (10)  {\tt SAURON} sample galaxies without CO that are detected by IRAS (filled circles). Fig.~\ref{fig:q} shows that early-type galaxies do not follow the star-forming relation, even those with molecular gas. The E/S0s with molecular gas are primarily below (FIR-excess or radio deficient) the star-forming relation, aside from two galaxies hosting radio AGN (NGC~2320 and NGC~2768) and NGC~5666. 

The mismatch between early-type galaxies and the radio-FIR correlation was first thoroughly presented in \citet{walsh89}, although they had no external way to separate galaxies likely to be star-forming. \citet{lucero07} discuss the radio-FIR correlation for a sample of 6 early-types with molecular gas (thus likely star-formers) and find that most lie close to the radio-FIR relation, although half are below, similar to what we find.  Results on
spirals from \citet{condon91} also indicate that more FIR or less radio emission is
found for earlier types (Sa-type spirals). They offer a simple model to explain this change of ratio, based on lower-mass stars (which will not produce
supernovae) contributing significantly to the dust heating in these galaxies. Our work in Section~6.2.4 also shows a FIR-excess with respect to the 8 and 24\mum emission, further supporting the low-mass star heating explanation. However, a handful of galaxies are extremely deviant, lying a factor 10 below the relation (NGC~4150, NGC~4459 and NGC~4550). The FIR excess with respect to the mid-IR is not strong enough to explain this large offset, and it is possible another cause (more efficient cosmic ray escape or weak magnetic fields) is also at work. Four galaxies without CO, but with IRAS detections, also lie far below the radio-FIR relation. These are possibly false IRAS detections, as they are right at the detection limit. Otherwise, they may host a small quantity of dust, heated by old stars, without any recent supernovae to provide the cosmic rays necessary to produce the radio synchrotron emission. 

The poor relation between the FIR and radio emission in early-type galaxies with molecular gas warns of a flaw in one or both of these tracers as a star formation rate indicator.

\begin{figure*}
\begin{center}
\rotatebox{0}{\includegraphics[clip=true,width=7cm]
{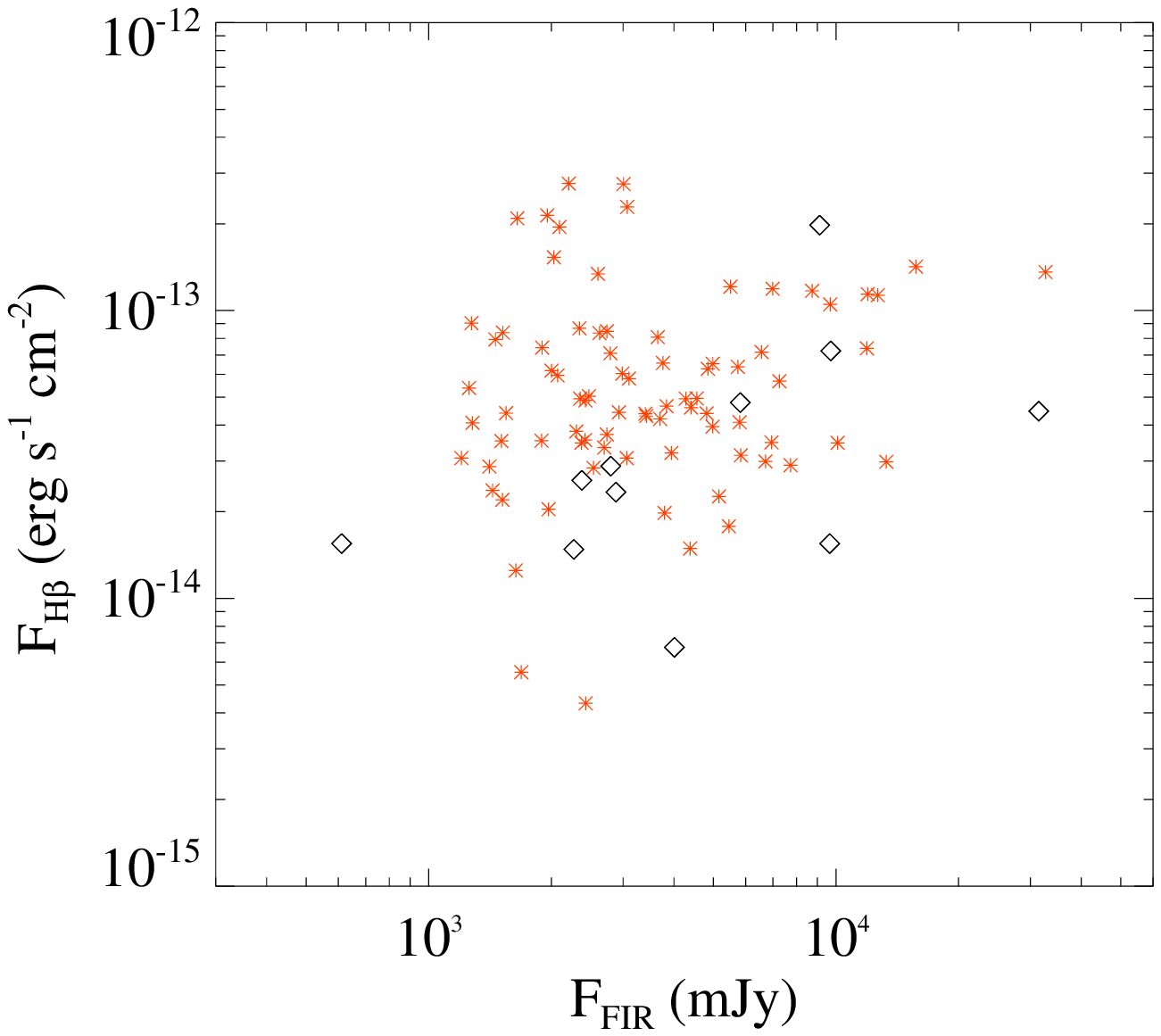}} \hspace{0.8cm}
\rotatebox{0}{\includegraphics[clip=true, width=7cm]
{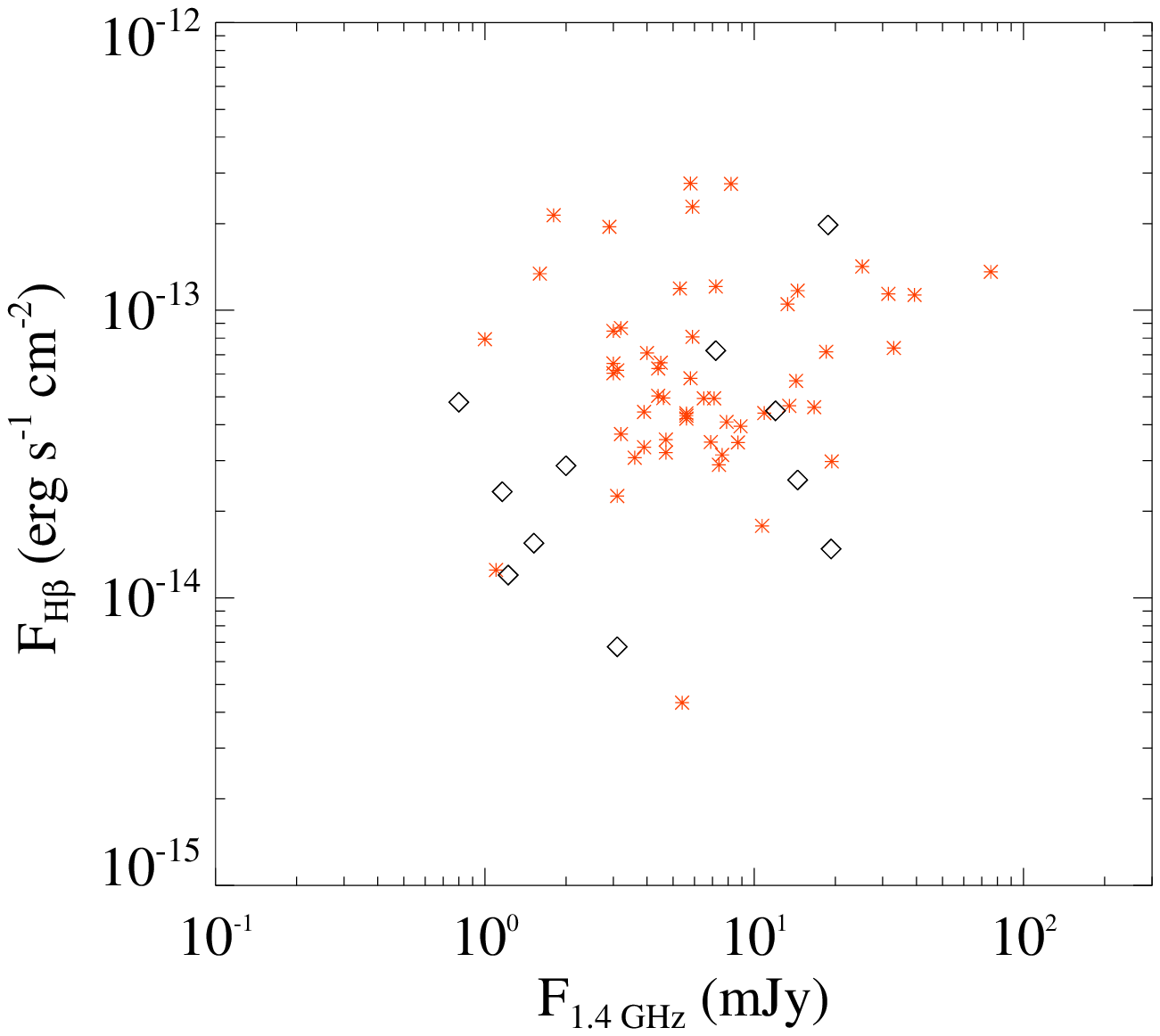}}\\
\rotatebox{0}{\includegraphics[clip=true,width=7cm]
{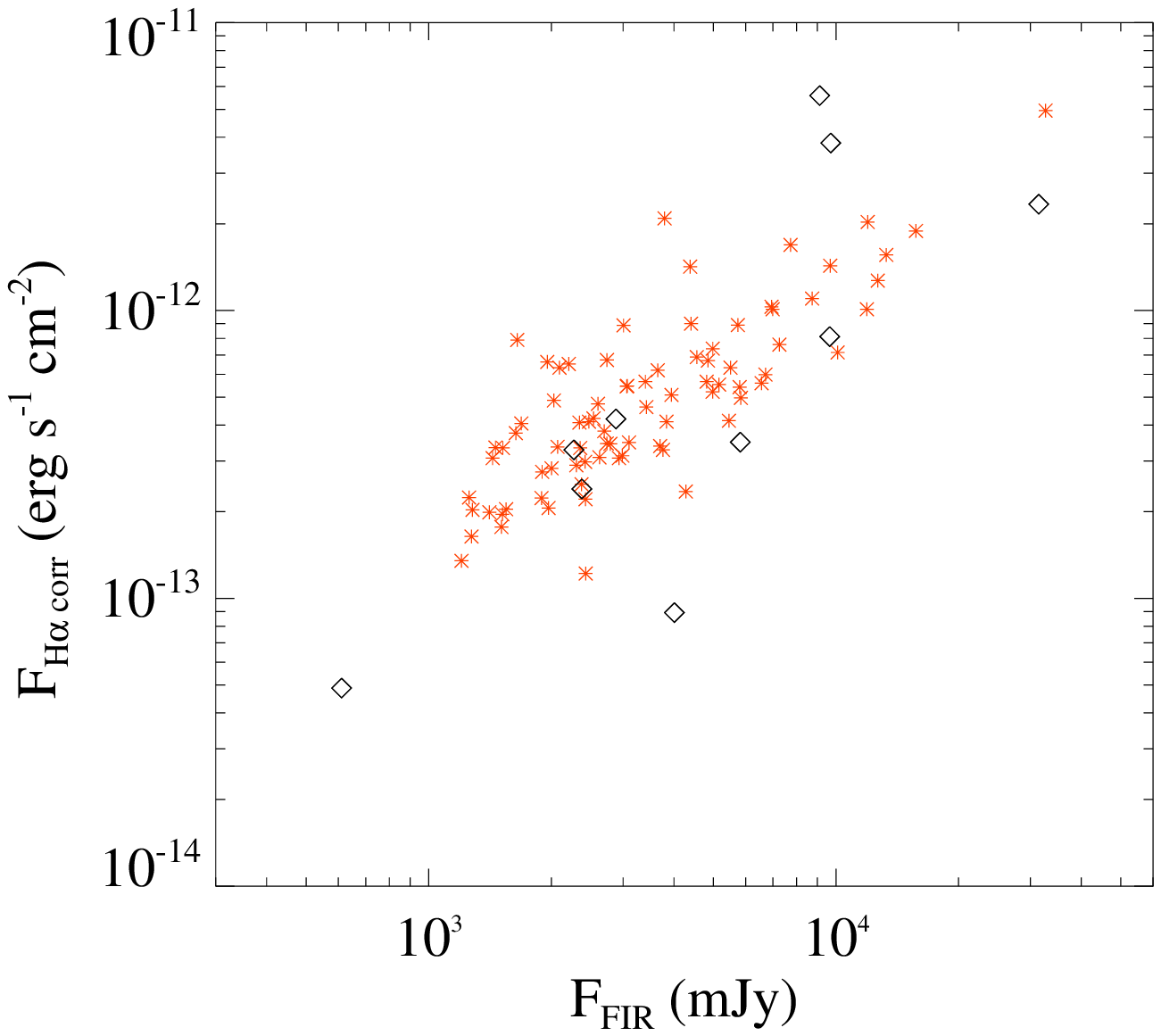}} \hspace{0.8cm}
\rotatebox{0}{\includegraphics[clip=true, width=7cm]
{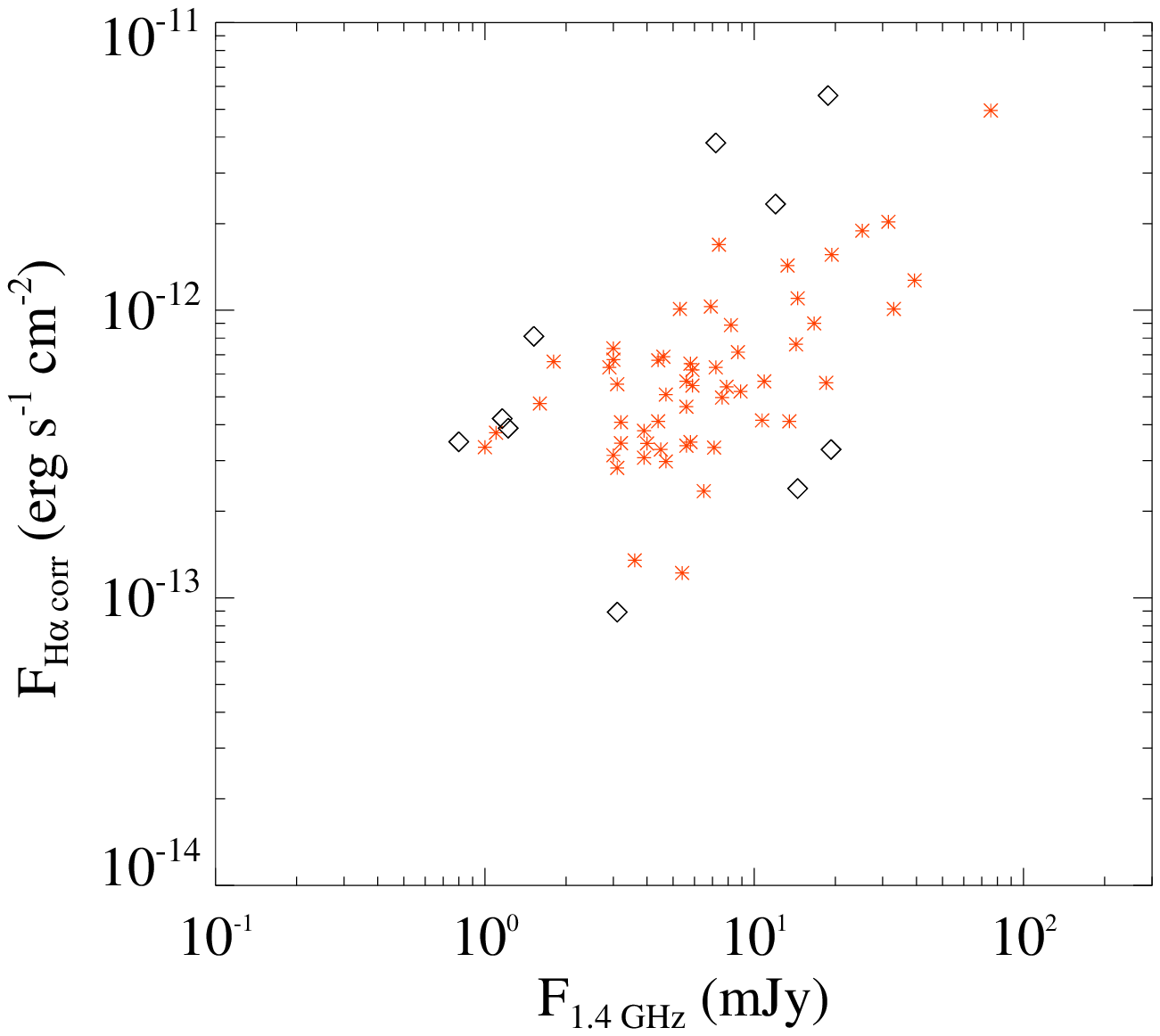}}
\caption{{\em Top-left:} H$\beta$ flux versus FIR flux. {\em Top-right:}
  H$\beta$ flux versus 1.4~GHz continuum flux. {\em Bottom-left:}
  Extinction-corrected H$\alpha$ flux versus FIR flux. {\em
  Bottom-right:} Extinction-corrected H$\alpha$ flux versus 1.4~GHz
  continuum 
  flux. Red asterisks represent galaxies from the NFGS sample, while
  our E/S0s with molecular gas are shown as black diamonds. Correlations are
  not seen with the H$\beta$ emission either in our sample or the NFGS
  galaxies, while correlations are clearly observed with extinction-corrected
  H$\alpha$ in the NFGS and probably for our sample E/S0s as well.}
\label{fig:hbeta}
\end{center}
\end{figure*}

\subsubsection{H$\beta$ as a star formation tracer}

If H$\alpha$ emission is corrected for dust extinction, the FIR and H$\alpha$
emission are proportional over four orders of magnitude in
star-forming galaxies \citep{kewley02}. Unfortunately, the {\tt SAURON}
wavelength range only includes the weaker and more dust-extincted
H$\beta$ emission line. Even in 
clearly star-forming galaxies, H$\beta$ does not trace star
formation, as can be seen in the top two plots of
Fig.~\ref{fig:hbeta}. These plots show no correlation between H$\beta$
and either the FIR or radio flux
for the galaxies of the Near Field Galaxy Survey (NFGS) sample, a local
field galaxy-only 
sample that is comprised mostly of
star-forming galaxies \citep{jansen00}. However, when these same
galaxies are extinction-corrected based on the Balmer decrement between H$\alpha$
and H$\beta$, a strong correlation between extinction-corrected
H$\alpha$ and FIR is present, and a slightly weaker correlation
is observed between the corrected H$\alpha$ and radio emission
(bottom of Fig.~\ref{fig:hbeta}; \citealp{kewley02}). 

Without observed H$\alpha$ emission, we cannot perform the same
simple extinction correction for the {\tt SAURON} galaxies. Instead,
we assume an 
extinction to hydrogen column density ratio and use our molecular gas
maps along with HI data to estimate the extinction in each pixel of the H$\beta$ flux
maps. Summing over the region with detected molecular gas then  gives
us the corrected H$\beta$ flux associated with the region where star
formation is expected. The details of this process are as follows. 
First, we calculate the H$_{2}$ column density using the same
X$_{\mathrm{CO}}$ factor, $3.0 \times 10^{20}$ cm$^{-2}$ (K km s$^{-1}$)$^{-1}$,
used to calculate the total molecular masses. As the \hi
maps are at much lower spatial resolutions, we calculate the average central
\hi to H$_{2}$ mass ratio for each galaxy and multiply the H$_{2}$ column density by
this factor to estimate the column density of \hi. If no \hi is detected for a galaxy, we assume no \hi is present. We then add the
H$_{2}$ and \hi column densities and divide by the mass of a hydrogen
atom to get the number column density of all hydrogen atoms in
cm$^{-2}$.  We note
that including the \hi is a 10\% or less addition for all galaxies (except 
NGC~5666, where it is 30\%). However, only a portion of the gas and dust
is actually in front of the line-emitting gas. A first order
approximation is that half of the dust is in front (see Appendix 1 of \citealp{calzetti07}). The colour excess is 
based on the Milky Way-derived relation: $E_{{B-V}} =
1.7\times10^{-22}$ cm$^{2}$ $n(\mathrm{H})$ \citep{bohlin78}. We also assume a
Milky Way total-to-selective extinction ratio of $R_{{V}}$=3.1
and then use the \citet{cardelli89} extinction curve to calculate an
extinction-corrected H$\beta$ flux.  

In some galaxies, the local estimated $V$-band extinction, A$_{V}$,
reaches extremely high values (over 20). Any emission actually
detected in H$\beta$ is unlikely to have originated in  
regions with such high extinctions, instead what we observe must come from
outer layers with less extinction. Using such high A$_V$
values produces unrealistically high estimates for the corrected H$\alpha$
emission (calculated assuming a Balmer decrement of 2.86 between
H$\alpha$ and H$\beta$, as expected in the case of no reddening and Case B
recombination from \citealp{osterbrock89}). Instead of using these
very high values, we set an upper 
limit of $A_{V}=2.5$. This choice is motivated by attempting to match
the few galaxies for which we have observed H$\alpha$ fluxes. There
are four such galaxies: NGC~3032, NGC~3489, NGC~4459 and NGC~4526. The
reported H$\alpha$ flux for NGC~3489 is only $9.1 \times 10^{-14}$ erg
s$^{-1}$ cm$^{-2}$ \citep{kennicutt08}, which is less than 2.86
times the observed H$\beta$ flux of  $4.8 \times 10^{-14}$ erg
s$^{-1}$ cm$^{-2}$. We thus do not use this 
observation. For NGC~3032 and NGC~4459, the choice of
A$_{\mathrm{V}}$=2.5 gives expected uncorrected H$\alpha$ fluxes of
$1.3 \times 10^{-12}$ and $2.4 \times 10^{-13}$ erg s$^{-1}$ cm$^{-2}$,
while the actual observed values are $7.7 \times 10^{-13}$ and $4.1 \times
10^{-13}$, respectively. Thus, for these two galaxies, our estimate is
1.9 and 0.6 times the observed value, a good match considering all the
assumptions and the differences in the observations (errors
associated with this method are likely to be at least 50\%). For the very
edge-on galaxy NGC~4526, our method predicts a much lower H$\alpha$
flux than is actually observed \citep{young96}. An $A_{V}$ of 10.9 is
implied by the difference between the observed H$\alpha$ and H$\beta$
fluxes. As this is an unreasonably high extinction (it leads to an
H$\alpha$-estimated SFR of over 3000 \Msun yr$^{-1}$), we will use the
H$\alpha$ value we estimate from the H$\beta$ flux, despite the large
mismatch with the observed H$\alpha$ flux.

Using these corrected H$\alpha$ fluxes, the E/S0s with molecular gas
seem to generally follow the H$\alpha$-FIR correlation delineated by the NFGS,
although they show increased scatter (Fig.~\ref{fig:hbeta}). Any
correlation between 
H$\alpha$ and the radio emission is weaker, although our E/S0s are not
much offset from the NFGS relation, which also shows a larger
scatter than the FIR-based relation. One curious feature
of the E/S0 galaxies revealed by these plots is that the galaxies
predominantly ionised by a mechanism other than star formation (i.e. the high
\otohb ratio galaxies) do not lie clearly off the H$\alpha$-FIR plot, as might
be expected. This suggests that whatever 
additional sources ionise the gas also heat the dust, in a manner similar
to hot young stars. To better explore these trends, direct measurements of
H$\alpha$ in more E/S0 galaxies are required.

\subsubsection{MIR star formation tracers from the literature}

Using Spitzer InfraRed Array Camera (IRAC) data, \citet{shapiro09}
 identify star-forming 
 galaxies among the {\tt SAURON} sample and derive SFRs using
stellar continuum-subtracted 8\mum PAH emission and the conversion from
\citet{wu05}. They identify 8 galaxies with clear signatures of star
formation; 7/8 of these are in our sample (the one galaxy outside our
sample is NGC~2974 that also shows evidence for outer star
formation in the UV, as discussed in Section~6.1). They
 also identify another 5 
galaxies that are likely to host star formation, including three of
our sample galaxies (NGC~524, NGC~4477 and NGC~4550). These galaxies
have less prominent 8\mum PAH distributions and consist of galaxies in or close
to the region of our `old' category in Fig.~\ref{fig:classify}, where
 the IFU data do not conclusively indicate young stars or ongoing
 star formation. Another galaxy from 
 this `old' category, NGC~2768, is classified by \citet{shapiro09} as unlikely
to be star-forming, with its PAH emission instead excited by its
general interstellar radiation field. The SFRs they derive range from 0.006 to 0.4 \msun
 yr$^{-1}$ across the {\tt SAURON} sample. 

The 24\mum emission traces hot dust, either the ejected circumstellar material
around hot old stars or the birth material of hot young stars. A study of
the 24\mum emission in the {\tt SAURON} E/S0s was performed by
\citet{temi09} using Multiband Imaging Photometer for Spitzer (MIPS)
data. They note that subtracting the old-star contribution from the 24\mum is necessary before its use
as a SFR indicator, and they base this subtraction on each galaxy's $K$-band
luminosity. They then use the \citet{calzetti07} calibration to estimate
24\mum SFRs for the {\tt SAURON} E/S0s, finding SFRs between 0.02 and 0.2 \msun
yr$^{-1}$. Out of the 8 of our sample galaxies they consider, they
identify 6 as star-forming. The two not identified as star-forming are
NGC~2768 and NGC~4477, again two galaxies in our `old' category that must have very low specific star formation rates.

\begin{table*}
\begin{center}
 \caption{Star formation rates.} 
\vspace{0.05cm}
 \begin{tabular}{lrrrrrrrrrr}
\hline
 Galaxy & \multicolumn{5}{c}{Partial SFRs (\msun yr$^{-1}$)} &
 \multicolumn{4}{c}{Total SFRs (\msun yr$^{-1}$)} \\
 & TIR & 24 \mum & PAH & 1.4~GHz & H$\alpha$ & TIR+H$\alpha$ &
24\mum+H$\alpha$ & PAH+H$\alpha$ & 1.4~GHz+H$\alpha$ \\
\hline
\phantom{0}524 
     & --     & --     & 0.027  & 0.043  & 0.008  & --    & -- 	  & 0.035  &	0.051 \\   
2320 & --     & --     & --     & 3.412  & 0.495  & --    & --    & --     &    3.908 \\
2685 & --     & --     & 0.045  & 0.012  & --     & --	  & --    & --     &	--    \\	
2768 & 0.041  & 0.006  & --     & 0.177  & 0.018  & 0.059 & 0.025 & --	   &	0.195 \\ 
3032 & 0.151  &	0.084  & 0.112  & 0.085  & 0.384  & 0.535 & 0.468 & 0.496  &	0.469 \\	
3156 & 0.020  &	0.008  & 0.010  & --     & --     & --	  & --	  & --	   &   	--    \\
3489 & --     &	0.012  & 0.017  & 0.004  & 0.016  & --    & 0.028 & 0.034  &	0.020 \\	
4150 & 0.032  &	0.014  & 0.017  & 0.004  & 0.013  & 0.045 & 0.027 & 0.030  &	0.017 \\	
4459 & 0.080  &	0.026  & 0.050  & 0.010  & 0.041  & 0.121 & 0.067 & 0.091  &	0.051 \\	
4477 & --     &	--     & 0.003  & 0.008  & 0.024  & --	  & --    & 0.026  &	0.032 \\	
4526 & 0.294  &	0.070  & 0.102  & 0.083  & 0.147  & 0.441 & 0.216 & 0.249  &	0.229 \\	
4550 & --     & --     & 0.002  & --     & 0.004  & --	  & --    & 0.006  &	--    \\
5666 & 0.352  & --     & --     & 0.591  & 1.311  & 1.663 & --	  & --	   &	1.902 \\	
\hline 	        
\label{tab:sfrs}
\end{tabular}
\end{center}
\end{table*}

\subsubsection{Star formation rate comparison}

For each of the SFR tracers discussed (1.4~GHz, FIR,
24\mum,  8\mum PAH, and H$\alpha$), there are ample different
conversions to SFR in the literature. Consistency between the
tracers is only guaranteed if the conversions are calibrated against
the same set of sample galaxies. We thus turn to the calibrations
found in \citet[][hereafter K09]{kennicutt09}, which linearly combine the observed H$\alpha$ emission
(i.e. unobscured SF) with a measure of the dust emission
(i.e. obscured SF): SFR=pSFR(H$\alpha$)+pSFR(IR) (see their eqn. 16
and table 4; here pSFR denotes a ``partial'' SFR). These calibrations
provide consistent SFRs based on the Spitzer Infrared Nearby Galaxies
Survey (SINGS) sample of star-forming galaxies.  

We first compare the pSFRs for the four obscured SF tracers: 1.4~GHz, total
infrared (TIR), 24\mum and 8\mum PAH (stellar
continuum-subtracted) emission. The radio is justified as an obscured
SF tracer based on the tight radio-FIR correlation. We use 1.4~GHz
fluxes from the literature and 8\mum stellar continuum-subtracted
fluxes from \citet{shapiro09}. For the 24\mum emission, the stellar
contribution is negligible for the 
K09 sample while it is significant for the {\tt SAURON} E/S0
galaxies, as noted above. Thus we use 
the stellar continuum-subtracted 24\mum fluxes from \citet{temi09} as
the input. 

\begin{figure}
\begin{center}
\includegraphics[clip=true,width=8.1cm]{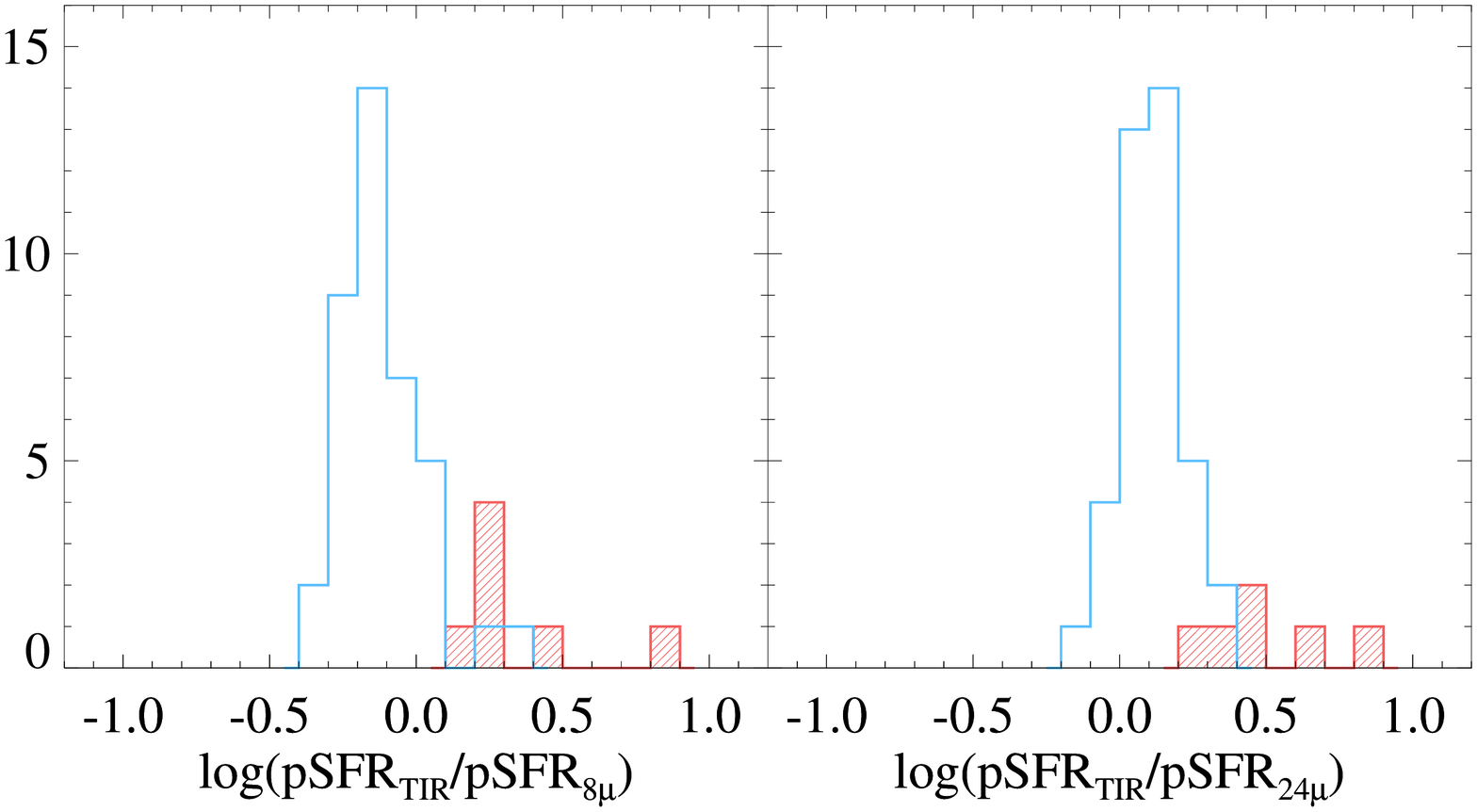}
\includegraphics[clip=true,width=8.1cm]{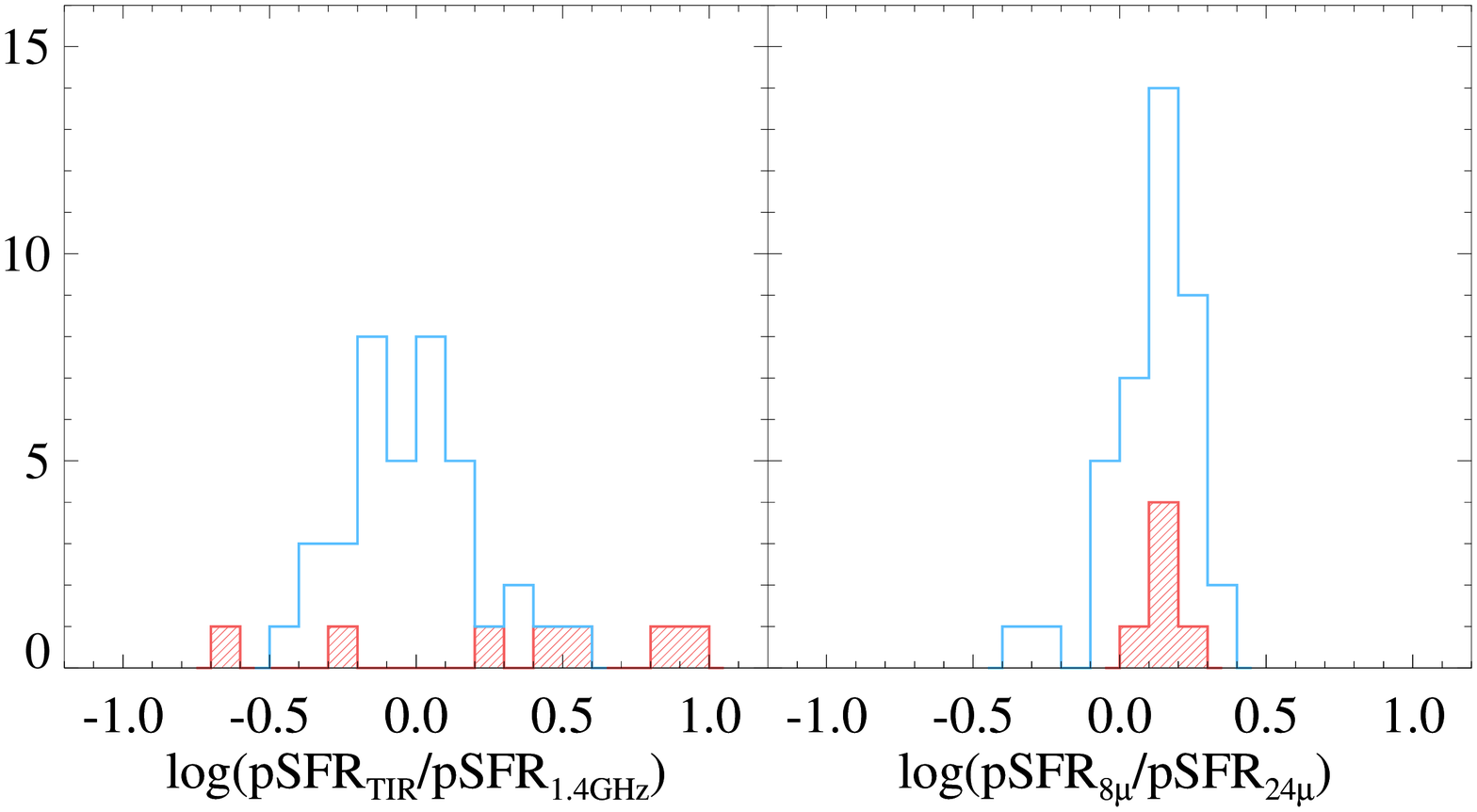}
\includegraphics[clip=true,width=8.1cm]{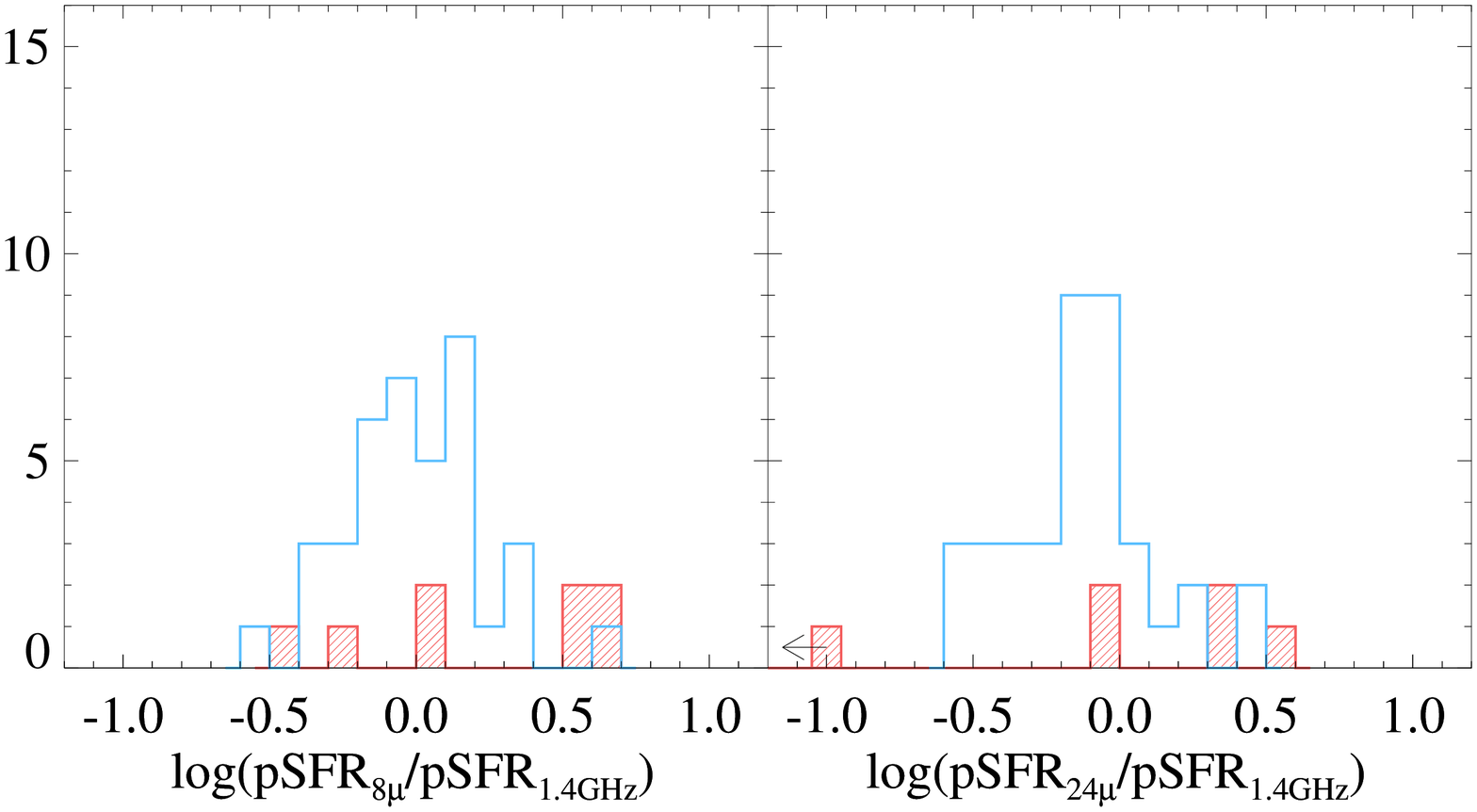}
\caption{A comparison of obscured pSFRs derived via different
  tracers. The spirals from the SINGS sample are plotted in hollow
  histograms. The logarithmic ratio between the spiral pSFRs from
  different tracers peaks around 0, and the small offsets in some
  plots are consistent with the errors in the conversion factors in K09. E
  and S0 galaxies from our sample are plotted in  shaded
  histograms. These plots show that the TIR consistently overestimates
  the SFR in E/S0s when compared to the two mid-IR tracers. The 8 and
  24\mum tracers are in very good agreement. The plots with radio as a
  tracer show the widest spread in both the spirals and E/S0s. The bin
  with an arrow in the 24\mum versus radio histogram represents
  NGC~2768 which has a logarithmic ratio of -1.4.}
\label{fig:comparesfr}
\end{center}
\end{figure}

The TIR luminosity measures the total luminosity from
3 to 1000\mum and is different from the FIR luminosity we used thus
far (based on only IRAS 60 and 100\mum fluxes). For the TIR used in their SFR
calibrations, K09 use the formula from \citet{dale02} that estimates
the TIR flux based on IRAS 25, 60 and  100~\mum data. Unfortunately, only
four of our sample galaxies were detected at 25\mum with IRAS, and
these are all close to the detection limit. However, seven
have 24\mum fluxes from MIPS as reported in \citet{temi09b}.  For
IR-bright SINGS galaxies, $f$(MIPS 24)/$f$(IRAS
 25$)=0.98\pm0.06$, so we use this conversion factor to estimate the IRAS 25\mum flux, adding the 0.06
error in quadrature. We note that the 24\mum flux for
NGC~5666 in \citet{temi09b} was unrealistically large and we have
substituted the value reported in \citet{young09}. We also note a
possible systematic offset of around 20\% in the \citet{temi09b}
fluxes compared to IRAS 25\mum fluxes for galaxies well above the
detection limit in IRAS 25\mum. This possible offset is likely
attributable to differences in background subtraction or aperture
correction. We thus use the SINGS
conversion factor after subtracting the expected stellar contribution
to the 24\mum flux (see \citealp{temi09}), but remain aware that the values
we use for 25\mum fluxes may be about 20\% too low.  We then apply the
\citet{dale02} IRAS-based TIR formula.  
 
We list the consistently calculated obscured pSFRs in
Table~\ref{tab:sfrs}. Values are generally between 0.01 and 0.5 \msun
yr$^{-1}$. Next, we compare each obscured tracer to the others by looking at
the logarithmic ratio between the two derived pSFRs. Plots showing the distribution of this ratio for
our E/S0 galaxies are shown in Fig.~\ref{fig:comparesfr} (shaded
histograms). Spiral galaxies from the SINGS sample are plotted for comparison (hollow
histograms). In E/S0s, the TIR consistently overestimates the SFR
relative to both the 8 and 24\mum tracers (top two plots of
Fig.~\ref{fig:comparesfr}). This
overluminosity in TIR implies a higher fraction of cold dust in
early-type galaxies than in spirals. The two-component model
of \citet{lonsdale87}, in which the far-IR consists of both a
star-formation related component and a component heated by old stars,
suggests that the TIR will be larger relative to the SFR in galaxies with
lower specific SFRs. Here we see this effect in the direct comparison of mid-IR (warm dust) and far-IR (cold dust) emission. 

The ratios of 8\mum PAH pSFRs to 24\mum pSFRs
are very similar between the SINGS spirals and the 6 E/S0s that we are able
to compare (the offset of both distributions is within the error of
the calibration values `a' listed in table~4 of K09). The agreement
of these tracers suggests they are the best tools for deriving
SFRs for early-type galaxies. Compared to these tracers, SFRs directly
derived from TIR or FIR alone may be overestimated by a factor
7 in the worst case, with an average overestimation factor of around 2.8. 

The radio-derived pSFRs show a wide variation relative to all three
IR-derived pSFRs. The FIR-radio has already been extensively discussed
in Section~7.2.1, with one possibility for the FIR-excess or radio deficiency being an additional FIR component not associated with star formation. However, the radio pSFR is also much lower than the 24\mum
or 8\mum derived pSFRs in 3-4 galaxies, indicating that the radio
emission associated with star formation may really be deficient in
these galaxies (NGC~2685, NGC~3489, NGC~4150, NGC~4459).

\begin{figure}
\begin{center}
\rotatebox{0}{\includegraphics[clip=true, width=8cm]
{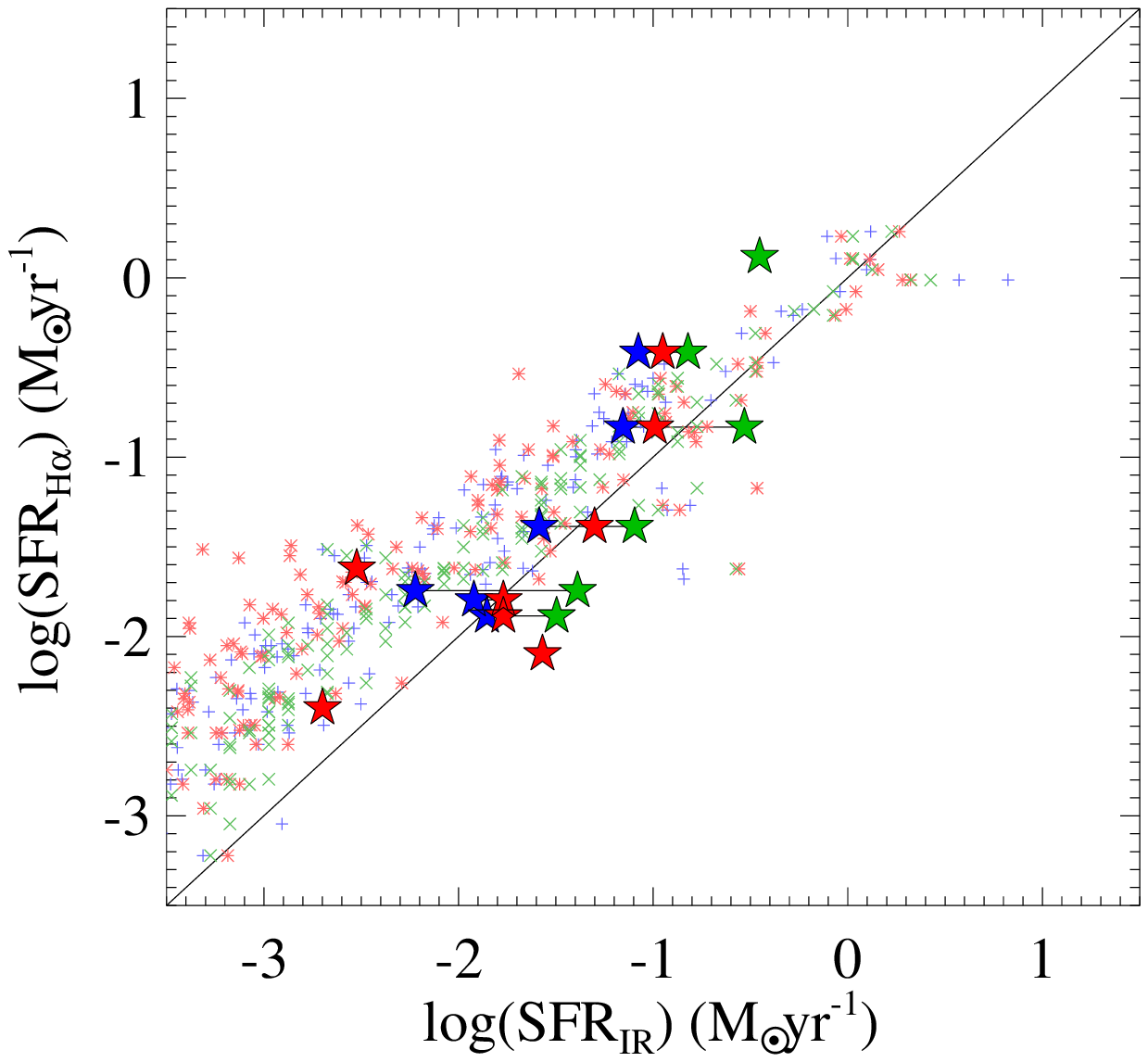}} 
\caption{Unobscured pSFR (traced by H$\alpha$) compared to
  the obscured pSFRs (traced by 8\mum PAHs in red, 24\mum in blue and TIR in
  green). Our sample E/S0s are plotted as large filled stars, with the
  LVL sample 
  plotted as unfilled symbols for comparison.  A
  one-to-one line is plotted indicating equal contributions from unobscured and
  obscured SF.}
\label{fig:irvsha}
\end{center}
\end{figure}

Next we consider the unobscured contribution (measured with H$\alpha$)
to the total SFR. The pSFR from the H$\alpha$ emission is
similar in value to that from the IR and radio tracers (see
Table~\ref{tab:sfrs}). Fig.~\ref{fig:irvsha} illustrates this rough
equality - our E/S0 sample galaxies lie close to the 1-1 line when
pSFR(H$\alpha$) is plotted against pSFR(IR). Note that this line is not 
a fit to either our or the comparison data. The different coloured
points in Fig.~\ref{fig:irvsha} represent the different IR tracers
(PAHs, 24\mum and TIR) and  we have also plotted galaxies
from  the Local Volume Legacy sample (LVL; \citealp{dale09,lee09}) 
for comparison. Many of the E/S0 galaxies with low SFRs deviate from
the trend seen in the LVL galaxies. (The strong H$\alpha$ and weak PAH
emission of the Seyfert NGC~4477 make it the one exception.) The LVL
galaxies at these low star formation rates are primarily dwarf
galaxies with low metallicity and low dust content. These galaxies
thus emit more in H$\alpha$ (low obscuration) and less in the IR (less
dust to heat). The gas phase metallicity in our E/S0s is presumably higher and
thus a different trend is seen at low SFRs. 

Overall, we conclude that the 8 and 24\mum fluxes are the best obscured star formation tracers for early-type galaxies. The TIR leads to an overestimate because of additional dust heating by old stellar populations, and the radio continuum is easily influenced by AGN and may occasionally be deficient for as yet unknown reasons. The ratio between unobscured and obscured star formation is similar to that found in spirals, but is less than that found for galaxies of similar star formation rates from the LVL sample. This difference is probably explained by the difference in metallicity -  our sample galaxies have high metallicities despite low star formation rates. 
 
\subsection{Star formation laws}

\begin{figure*}
\begin{center}
\rotatebox{0}{\includegraphics[clip=true,width=7.5cm]
{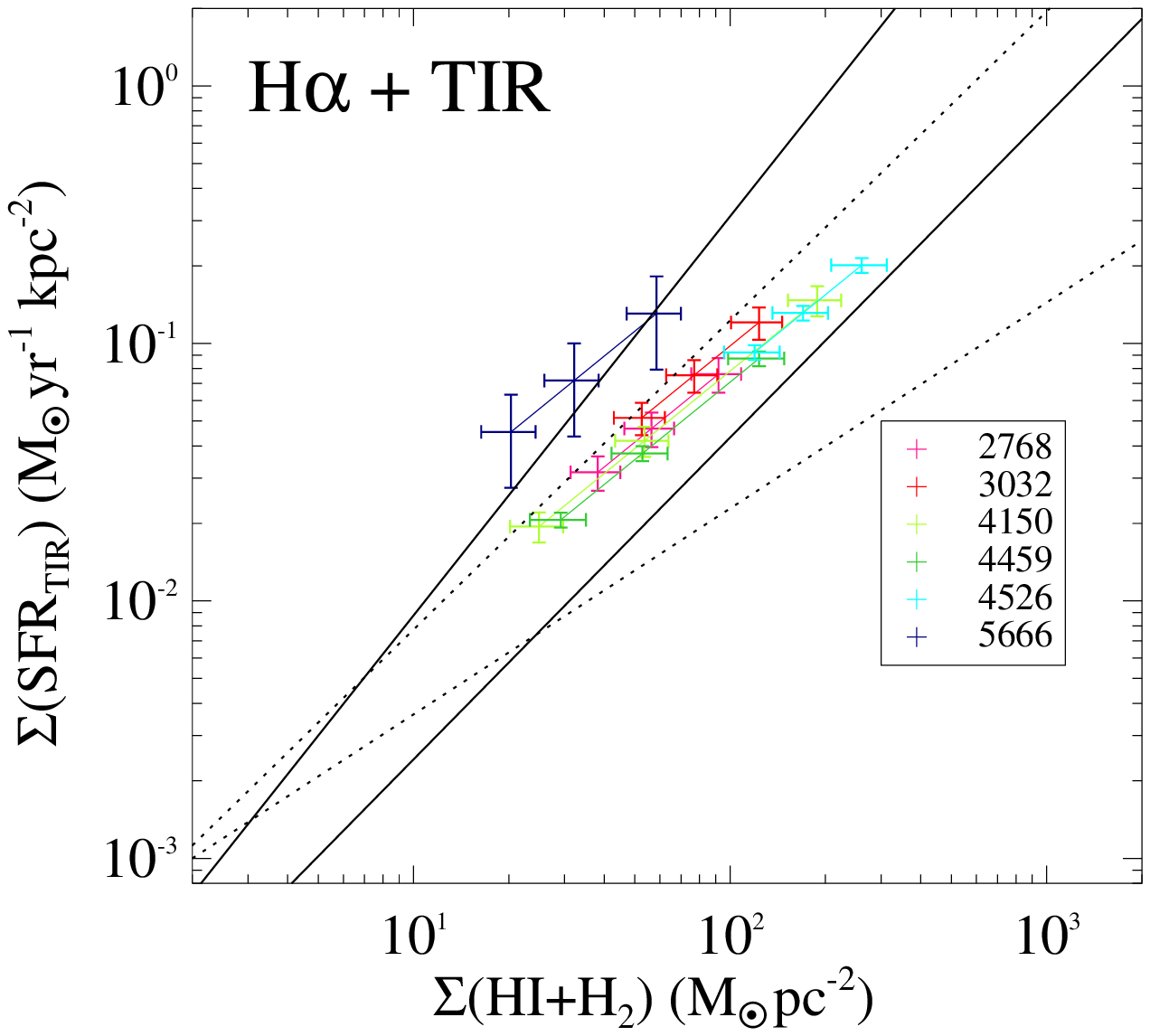}} 
\rotatebox{0}{\includegraphics[clip=true, width=7.5cm]
{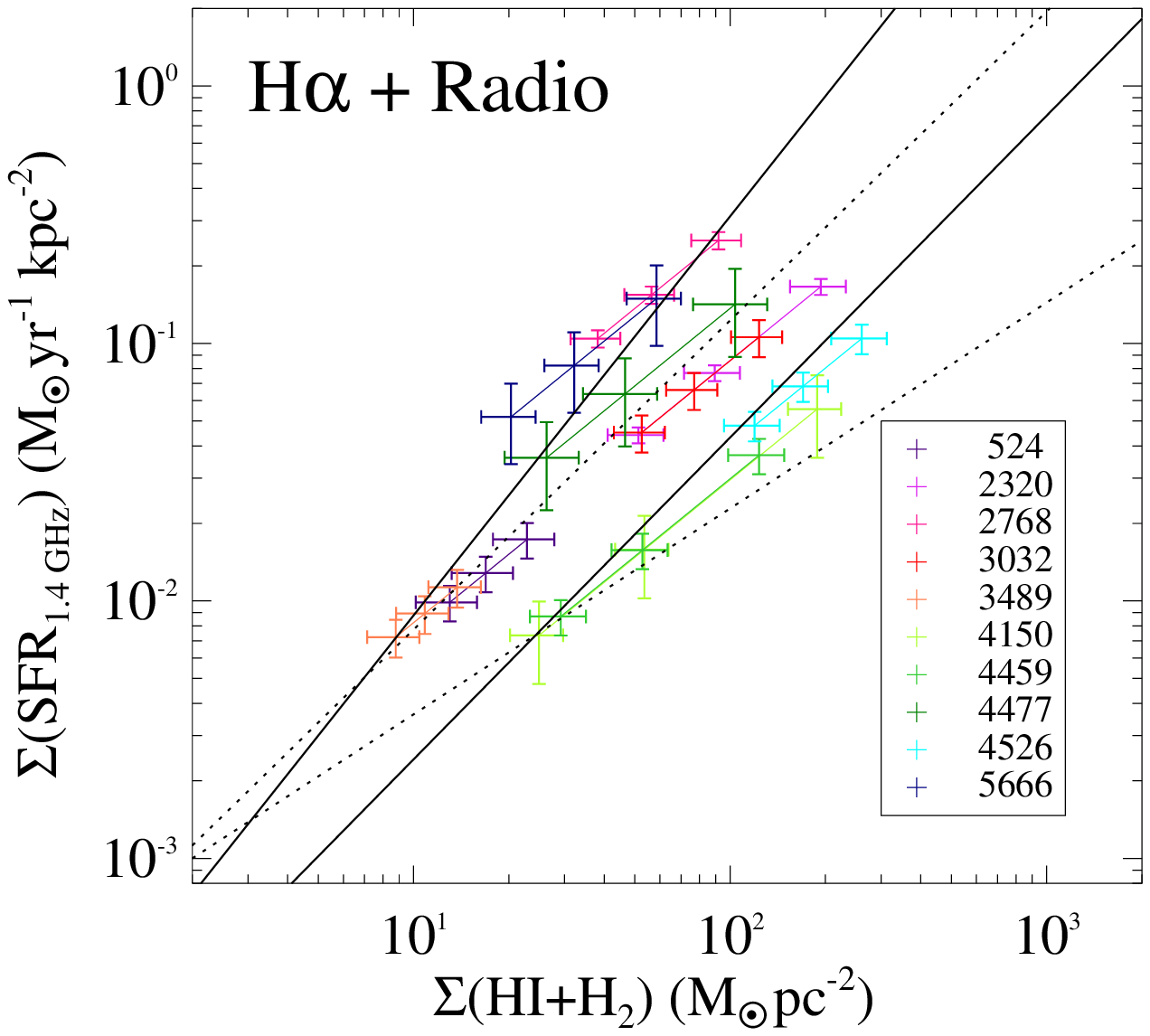}}
\rotatebox{0}{\includegraphics[clip=true,width=7.5cm]
{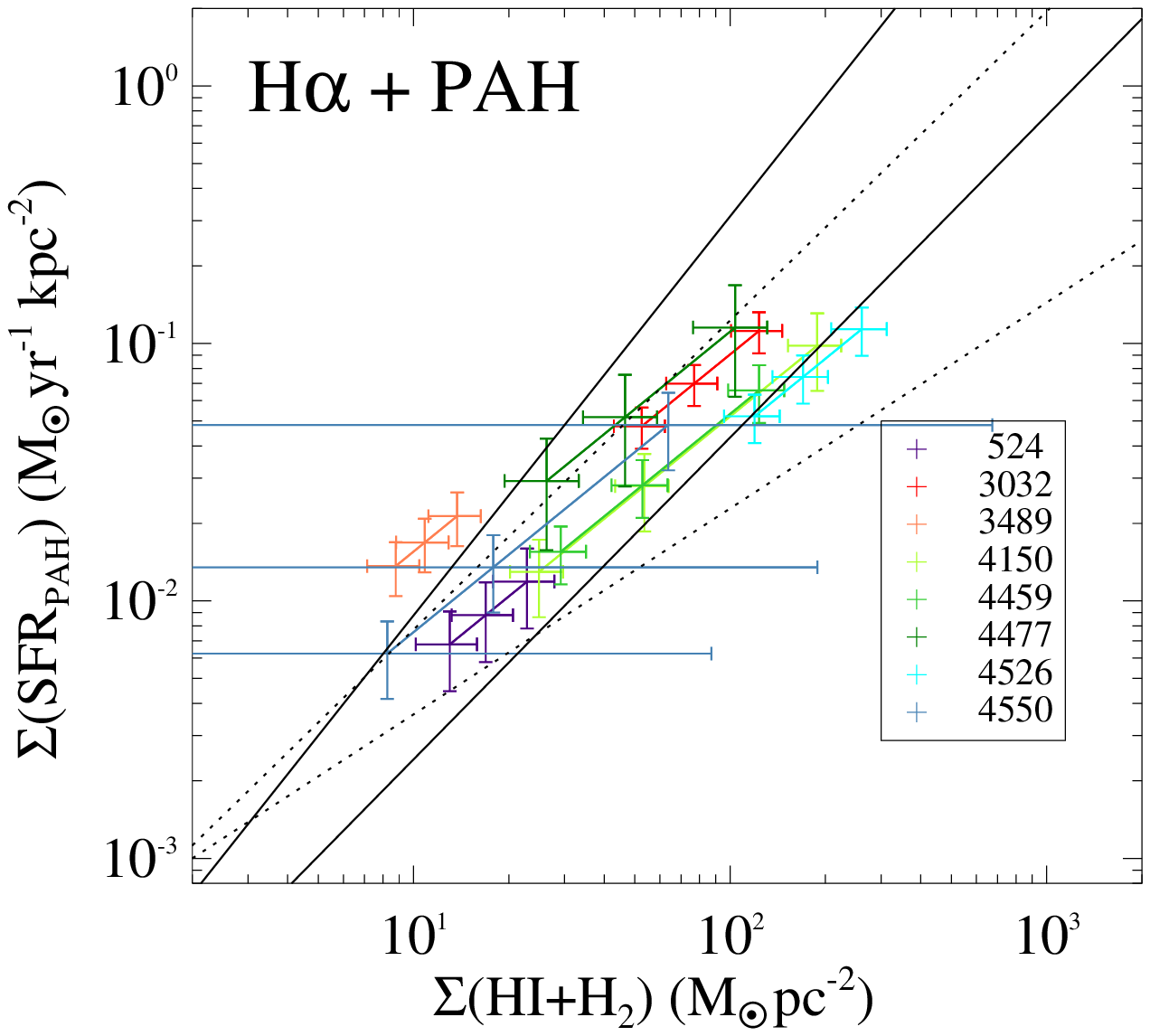}} 
\rotatebox{0}{\includegraphics[clip=true, width=7.5cm]
{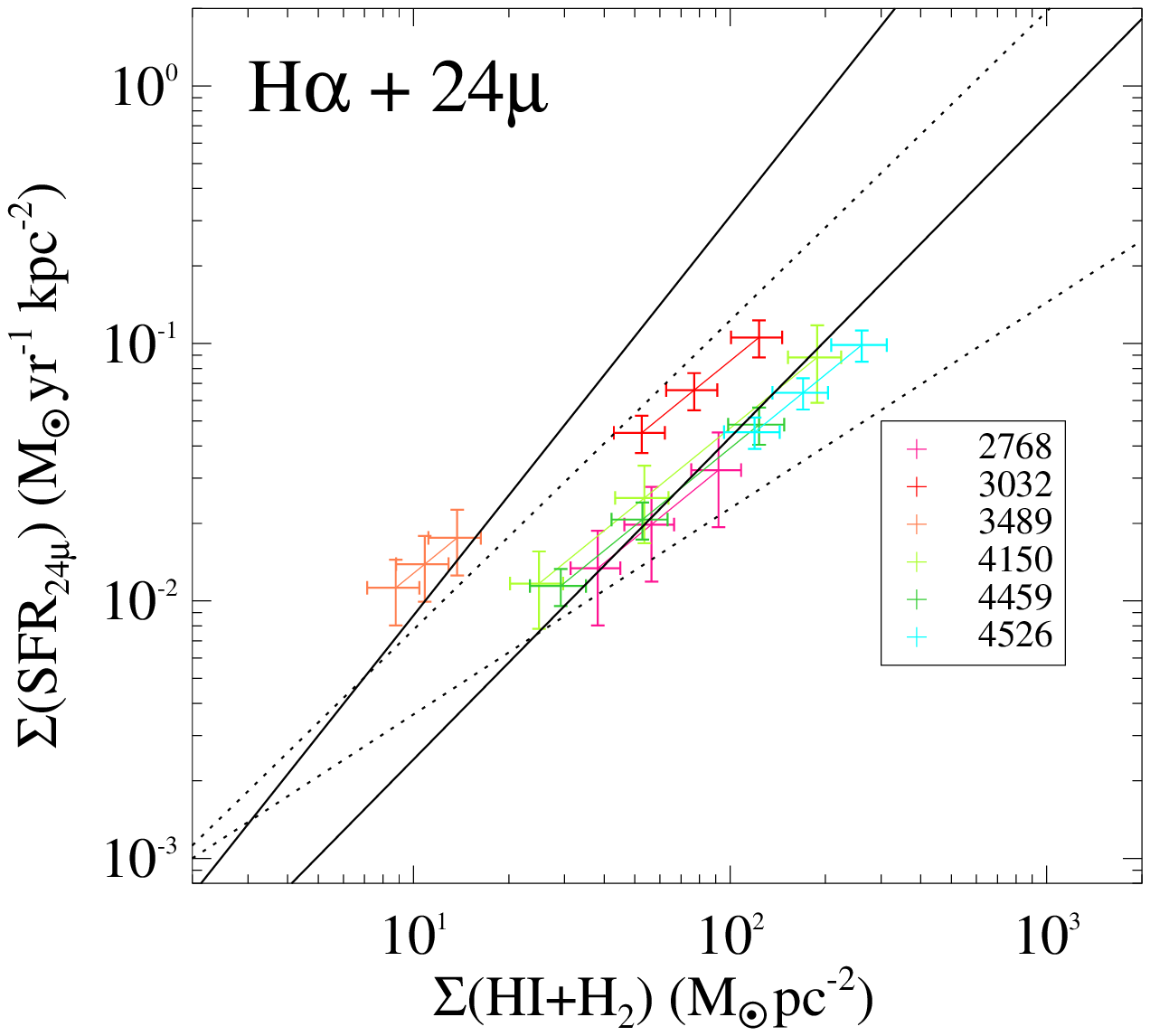}}
\caption{A total (\hi + H$_{2}$) gas-based  comparison of our galaxies with the
  Schmidt-Kennicutt law of \citet{kennicutt98a} (solid lines)
  and the constant SFE 
  prescription of \citet{bigiel08} (dotted lines), both updated for a
  Kroupa IMF and using our  X$_{\mathrm{CO}}$ factor. 
 {\em Top-left:} SFR from H$\alpha$ and TIR emission. {\em
  Top-right:} SFR from H$\alpha$ and 1.4~GHz radio continuum
  emission. {\em Bottom-left:} SFR from H$\alpha$ and 8\mum PAH
  emission. {\em Bottom-right:} SFR from H$\alpha$ and 24\mum emission.}
\label{fig:sflawsHI}
\end{center}
\end{figure*}

 Star formation
laws are usually given with respect to star formation and gas surface
densities and follow the form of $\Sigma(\mathrm{SFR}) \propto 
\Sigma($\hi$\mathrm{+ H_{2}})^{n}$. Using surface densities both removes
the distance dependence of the relation and is physically motivated by
theoretical predictions that star formation depends on gas volume
density \citep[e.g.][]{schmidt59}. For a sample of local star-forming and
starburst galaxies, \citet{kennicutt98a} finds $n=1.4\pm0.25$. The
star formation law with this exponent is commonly referred to as the 
Schmidt-Kennicutt law. But other work suggests a constant star
formation efficiency (SFE), in which $n$ is nearly 1
\citep{young91,young96} and a constant fraction of gas turns into
stars per unit time. For the constant SFE law, we use the
prescription found in \citet{bigiel08}. We
compare our E/S0 galaxies to both the Schmidt-Kennicutt law and this
constant SFE law below.

Average molecular gas surface densities are determined using the
molecular gas radii $R_{\mathrm H_{2}}$ and H$_{2}$ 
masses. Total cold gas masses are first calculated, using
the ratio of central \hi (found in \citealp{morganti06} and Oosterloo et al. in preparation) to H$_{2}$
mass to provide a correction factor for the \hi expected
within the CO radius.  The CO radius is then used to calculate the
average gas surface density using this total cold gas mass. We note that, as for the extinction correction in
Section~7.2.2, the inclusion of \hi is not a large correction. Star
formation surface densities are calculated using total SFRs [pSFR(H$\alpha$) +
  pSFR(obscured)] and the CO radius. These total SFRs can be found in
Table~\ref{tab:sfrs}. 

For each different obscured SFR tracer, we plot $\Sigma(\mathrm{SFR})$ 
against  $\Sigma($\hi$+\mathrm{H_{2}})$  in
Fig.~\ref{fig:sflawsHI}. The Schmidt-Kennicutt law given in
\citet{kennicutt98a} is shown via its upper and lower bounds, after
having been modified for a Kroupa IMF as used in K09 (a multiplicative
factor of 0.69). We also plot the constant SFE law with its bounds as
described in  \citet{bigiel08}. Our 
sample galaxies are plotted  with error bars that represent the errors
in mass (for the x-axis) and SFR (for the y-axis). The errors in the
radius determinations make the star formation and gas surface
densities correlated such that the galaxies move along a line of slope unity
for different radii.  We thus represent the radius errors by
plotting three different data points per galaxy: one using the
measured radius, then one each using the radius plus and minus the
radius error. 

As seen in Fig.~\ref{fig:sflawsHI}, early-type 
galaxies lie almost entirely within the 1$\sigma$ bounds for both the
Schmidt-Kennicutt and constant SFE laws (see also fig.~5 of \citealp{shapiro09}). This good agreement
is surprising - the galaxies without observed young stars or dominant ongoing
star formation lie on the relations just as galaxies with significant
AGN. However, neither law has particularly stringent 
bounds. For example, a galaxy with an average gas density of 70 \Msun
pc$^{-2}$  (typical for our sample) could have a SFR surface density
anywhere between 0.024 and 0.153 \Msun kpc$^{-2}$ by the
Schmidt-Kennicutt law, a range of nearly an order of 
magnitude. AGN, old stars and other sources of ionisation and dust
heating do not contribute enough to move the galaxies outside of the
wide range allowed by the relations. In fact, these same sources may be present
in the galaxies used to define the star-formation laws in the first
place, explaining their large scatter. Taking an example from our
sample, the known radio AGN NGC~2768 does not lie off
the Schmidt-Kennicutt law when its SFR is determined with radio+H$\alpha$
(although it is a bit above the constant SFE range). Similarly, the overestimation of 
SFR by the TIR does not move galaxies completely off either relation. If we
assume that the 24 and 8\mum PAH emission are the most accurate tracers of the
obscured SFR, then galaxies still occupy a large range. 

A lower star formation efficiency is expected for early-type
galaxies based  on the morphological quenching model of
\citet{martig09}, due to the 
lack of a stellar disc to aid the development of gas instabilities. If the
Schmidt-Kennicutt law is the correct description of how stars form
from gas, then we only see some tentative signs that early-type galaxies
may be less efficient at converting gas into stars when using the H$\alpha$ + 24\mum tracer.  No bias towards inefficiency is found with respect to the constant SFE
relation of \citet{bigiel08}.

\section{Conclusions}

We present new molecular gas maps for the three early-type
galaxies NGC~524, NGC~3489 and NGC~4477. The molecular gas
distribution of NGC~3489 shows spiral structure within an S0 galaxy
for the first time. Strong upper limits on the 
molecular gas content of NGC~4278 and NGC~7457 are also reported
(previously considered detections in the literature).  With these new  maps, we now have a
sample of 13 E/S0 galaxies with optical IFU coverage and CO detections, 12 of
which also have CO maps. 

The distribution of molecular gas in our sample
galaxies is generally central, in settled rings or
discs, although there are also a few cases of more externally distributed
molecular gas. Optically-obscuring dust is always found coincident with the
molecular gas distributions, and reveals tightly-wound spiral discs
in the galaxies that contain more than around 
10$^{8}$ M$_{{\tiny \sun}}$ of molecular gas. Average molecular gas
surface densities range  between 10 and
250 M$_{{\tiny \sun}}$ pc$^{-2}$ across the sample. 

Comparing IFU and CO maps, we find that the ionised gas and molecular gas
are connected in early-type galaxies. Ionised gas is present wherever
we detect molecular gas and the two phases always have aligned
rotation. In about half the 
galaxies, the ionised gas is even more extended than the
molecular gas. These are the same galaxies that have high
[O$\:${\small III}]/H$\beta$ ratios, indicating that a process other
than star formation 
dominates their gas ionisation. Young stellar populations are detected
through a mix of optical colours, UV-optical colours and absorption
linestrength maps. Young or intermediate populations are more extended
than the molecular gas in 6/11 galaxies, coincident with the molecular gas in
2/11 galaxies and not present in 3/11 galaxies. 

Combining information on the gas ionisation and stellar populations of our
sample of galaxies, a loose classification is suggested. First are
obviously star-forming galaxies. These galaxies have low [O$\:${\small
    III}]/H$\beta$ emission line ratios and clearly-detected young
stellar populations. They also host larger quantities of molecular gas
($>10^{8}$ M$_{{\tiny \sun}}$). Second are post-starburst galaxies
with high [O$\:${\small III}]/H$\beta$ ratios, but still widespread young
stellar populations. They likely had more widespread and higher
rates of star formation in the recent past, and are now exhausting their
relatively small molecular gas reservoirs ($<10^{8}$ M$_{{\tiny 
    \sun}}$). 
The final category contains galaxies where
any star formation from the molecular gas is not significant enough to
lower the \otohb ratio or produce a detectable young stellar
population. These galaxies have similar properties to E/S0s without any molecular gas. The 13 early-types we study here are almost equally distributed between these three categories. 

E/S0s with molecular gas do not follow the radio-FIR correlation. A few of our sample galaxies host
radio AGN and have excess radio continuum flux compared to FIR flux,
but most E/S0s fall on the FIR-excess side of the relation. The 8
and 24\mum star formation tracers agree well for E/S0 galaxies,
but the TIR overestimates the SFR by an average factor of around 2.8,
presumably because of additional heating by older low-mass stars.  
The pSFRs based on our calculated H$\alpha$ values are very
similar to the values of the IR pSFRs, indicating that our galaxies do
not vary widely in their degree of obscuration and have gas-phase
metallicities more similar to those of spirals than metal-poor
dwarfs. 

We find that our E/S0 galaxies mostly lie within the large ranges of  the
Schmidt-Kennicutt law of \citet{kennicutt98a} and the constant
star-formation efficiency law as described in \citet{bigiel08}. This agreement
suggests that star formation in early-type galaxies with molecular gas is
not extremely different from that observed in spiral or starburst
galaxies.

\section*{Acknowledgements}
Based on observations carried out with the IRAM Plateau de Bure
Interferometer. IRAM is supported by INSU/CNRS (France), MPG (Germany)
and IGN (Spain). We would like to thank Philippe Salome for help with
the reduction of the Plateau de Bure data. We are grateful to
the {\tt SAURON} Team for providing {\tt SAURON} data, especially Marc
Sarzi, Eric Emsellem and Harald Kuntschner who enabled the inclusion
of the {\tt SAURON} maps of NGC~2320 and NGC~5666. We would also like
to thank Michele Cappellari for help with NGC~524's
circular velocities, Marc Sarzi for useful discussions on the gas
ionisation, Kristen Shapiro for thoughts on
the MIR-indicated star formation, and Timothy Davis for helpful
exchanges with regards to CO distributions and kinematics. 

Much of this work was completed for the DPhil thesis of AFC, who would like to thank her examiners, Rob Kennicutt and Steve Rawlings, for their insightful questions which led to the improvement of this work. AFC gratefully acknowledges the receipt of a Rhodes Scholarship that
enabled her DPhil studies at the University of Oxford. LMY
acknowledges partial support from NSF AST-0507432. 

The dust maps used observations made with the NASA/ESA Hubble
Space Telescope, obtained from the data archive at the Space Telescope
Institute. STScI is operated by the association of Universities for
Research in Astronomy, Inc. under the NASA contract  NAS 5-26555. 
The NASA/IPAC Extragalactic Database (NED) is operated by the Jet
Propulsion Laboratory, California Institute of Technology, under
contract with the National Aeronautics and Space Administration.

\bibliography{t.bbl}

\begin{thebibliography}{99}
  
\bibitem[\protect\citeauthoryear{Bacon et al.}{2001}]{bacon01}
  Bacon R. et al., 2001, MNRAS, 326, 23
\bibitem[\protect\citeauthoryear{Baum}{1959}]{baum59}
  Baum C. W., 1959, PASP, 71, 106
\bibitem[\protect\citeauthoryear{Bigiel et al.}{2008}]{bigiel08}
  Bigiel F., Leroy A., Walter F., Brinks E., de Blok W. J. G., Madore
  B., Thornley M. D., 2008, AJ, 136, 2846
\bibitem[\protect\citeauthoryear{Bohlin, Savage \&
    Drake}{1978}]{bohlin78}
  Bohlin R. C., Savage B. D., Drake J. F., ApJ, 224, 132
\bibitem[\protect\citeauthoryear{Bower, Lucey \& Ellis}{1992}]{bower92}
  Bower R. G., Lucey, J. R., Ellis R. S., 1992, MNRAS, 254, 601
\bibitem[\protect\citeauthoryear{Calzetti et al.}{2005}]{calzetti05}
  Calzetti D. et al., 2005, ApJ, 633, 871
\bibitem[\protect\citeauthoryear{Calzetti et al.}{2007}]{calzetti07}
  Calzetti D. et al., 2007, ApJ, 666, 870
\bibitem[\protect\citeauthoryear{Cappellari \&
    Copin}{2003}]{cappellari03}
Cappellari M., Copin Y., 2003, MNRAS, 342, 345
\bibitem[\protect\citeauthoryear{Cappellari \&
    Emsellem}{2004}]{capellari04}
  Cappellari  M., Emsellem E., 2004, PASP, 116, 138
\bibitem[\protect\citeauthoryear{Cappellari et al.}{2006}]{cappellari06}
  Cappellari M. et al., 2006, MNRAS, 366, 1126
\bibitem[\protect\citeauthoryear{Cappellari et al.}{2007}]{cappellari07}
  Cappellari M. et al., 2007, MNRAS, 379, 418
\bibitem[\protect\citeauthoryear{Cardelli, Clayton \& Mathis}{Cardelli
    et al.}{1989}]{cardelli89}
  Cardelli J. A., Clayton G. C., Mathis J. S., ApJ, 345, 245
\bibitem[\protect\citeauthoryear{Colbert \& Mushotzky}{1999}]{colbert99}
  Colbert E. J. M., Mushotzky R. F., 1999, ApJ, 519, 89
\bibitem[\protect\citeauthoryear{Combes, Young \&
    Bureau}{Combes et al.}{2007}]{combes07}
  Combes F., Young L.M., Bureau M., 2007, MNRAS, 377, 1795
\bibitem[\protect\citeauthoryear{Condon et al.}{1991}]{condon91}
  Condon J. J., Anderson M. L., Helou G., 1991, ApJ, 376, 95
\bibitem[\protect\citeauthoryear{Condon et al.}{1998}]{condon98}
  Condon J. J., Cotton W. D., Greisen E. W., Yin Q. F., Perley R. A.,
  Taylor G. B., Broderick J. J., 1998, AJ, 115, 1693
\bibitem[\protect\citeauthoryear{Condon, Cotton \&
    Broderick}{Condon et al.}{2002}]{condon02}
  Condon J. J., Cotton W. D., Broderick J. J., 2002, AJ, 124, 675
\bibitem[\protect\citeauthoryear{Crocker et al.}{2008}]{crocker08}
  Crocker A. F., Bureau M., Young L. M., Combes F., 2008, MNRAS, 386, 1811
\bibitem[\protect\citeauthoryear{Crocker et al.}{2009}]{crocker09}
  Crocker A. F., Jeong H., Komugi S., Combes F., Bureau M., Young
  L. M., Yi S., 2009, MNRAS, 393, 1255
\bibitem[\protect\citeauthoryear{Dale \& Helou}{2002}]{dale02}
  Dale D. A., Helou G., 2002, ApJ, 576, 159
\bibitem[\protect\citeauthoryear{Dale et al.}{2009}]{dale09}
  Dale D. A. et al., 2009, ApJ, 703, 517
\bibitem[\protect\citeauthoryear{de Vaucouleurs et
    al.}{1991}]{devaucouleurs91}
  de Vaucouleurs G., de Vaucouleurs A., Corwin H. G., Buta R. J.,
    Paturel G., Fouque P., 1991, Third Reference Catalog of Bright
    Galaxies, Vols. 1-3, XII. Springer-Verlag, Berlin (RC3)
\bibitem[\protect\citeauthoryear{de Zeeuw et al.}{2002}]{dezeeuw02}
  de Zeeuw T., et al., 2002, MNRAS, 329, 513
\bibitem[\protect\citeauthoryear{di Serego Alighieri et
    al.}{2007}]{diseregoalighieri07}
  di Serego Alighieri S., 2007, A\&A, 474, 851
\bibitem[\protect\citeauthoryear{Djorgovski \&
    Davis}{1987}]{djorgovski87}
  Djorgovski S., Davis M., 1987, ApJ, 313 59
\bibitem[\protect\citeauthoryear{Doi et al.}{2005}]{doi05}
  Doi A., Kameno S., Kohno K., Nakanishi K., Inoue M., 2005, MNRAS,
  363, 692
\bibitem[\protect\citeauthoryear{Donovan et al.}{2009}]{donovan09}
  Donovan J. et al., 2009, ApJ, 137, 5037
\bibitem[\protect\citeauthoryear{Donzelli \&
    Davoust}{2003}]{donzelli03}
  Donzelli C. J., Davoust E., 2003, A\&A, 409, 91
\bibitem[\protect\citeauthoryear{Dressler et al.}{1987}]{dressler87}
  Dressler A., Lynden-Bell D., Burstein D., Davies R. L., Faber S. M.,
  Terlevich R., Wegner G., 1987, ApJ, 313, 42
\bibitem[\protect\citeauthoryear{Emsellem et al.}{2004}]{emsellem04}
  Emsellem E., et al., 2004, MNRAS, 352, 721
\bibitem[\protect\citeauthoryear{Emsellem et al.}{2007}]{emsellem07}
  Emsellem E., et al., 2007, MNRAS, 379, 401
\bibitem[\protect\citeauthoryear{Faber \& Gallagher}{1976}]{faber76}
  Faber S. M., Gallagher J. S., 1976, ApJ, 204, 365
\bibitem[\protect\citeauthoryear{Faber et al.}{1987}]{faber87}
  Faber S. M., Dressler A., Daview R. L., Burstein D., Lynden-Bell D.,
  1987, in Nearly Normal Galaxies: From the Planck Time to the
  Present, ed. S. M., Faber (New York: Springer), 175
\bibitem[\protect\citeauthoryear{Filho, Barthel \& Ho}{2002}]{filho02}
  Filho M. E., Barthel P. D., Ho L. C., 2002, ApJS, 142, 223
\bibitem[\protect\citeauthoryear{Filho et al.}{2004}]{filho04}
  Filho M., Fraternali F., Markoff S., Nagar N., Barthel P., Ho L.,
  Yuan F., 2004, A\&A, 418, 429
\bibitem[\protect\citeauthoryear{Giovannini et al.}{2001}]{giovannini01}
  Giovannini G., Cotton W. D., Feretti L., Lara L., Venturi T., 2001,
  ApJ, 552, 508
\bibitem[\protect\citeauthoryear{Guilloteau \& Lucas}{2000}]{GL} 
  Guilloteau S., Lucas R., 2000, in Mangum J.G., Radford S.J.E., eds,
  ASP Conf. Ser. Vol. 217, Imaging at Radio through Submillimeter
  Wavelengths. Astron. Soc. Pac., San Francisco, p. 299
\bibitem[\protect\citeauthoryear{H\"ogbom}{1974}]{hogbom74}
  H\"ogbom J.A., 1974, A\&AS, 15, 417
\bibitem[\protect\citeauthoryear{Jansen et al.}{2000}]{jansen00}
  Jansen R. A., Fabricant D., Franx M., Caldwell N., 2000, ApJS, 126,
  331
\bibitem[\protect\citeauthoryear{Jeong et al.}{2007}]{jeong07}
  Jeong H., Bureau M., Yi S. K., Kranjović D., Davies R., 2007, MNRAS,
  376, 1021
\bibitem[\protect\citeauthoryear{Jeong et al.}{2009}]{jeong09}
  Jeong H. et al. 2009, MNRAS, 398, 2028
\bibitem[\protect\citeauthoryear{Johnson \& Gottesman}{1979}]{johnson79}
  Johnson D. W., Gottesman S. T., 1979, in Photometry, Kinematics, and
    Dynamics, ed. D. S. Evans (Austin: University of Texas Press),
    p. 57
\bibitem[\protect\citeauthoryear{Jura}{1982}]{jura82}
  Jura M., 1982, ApJ, 254, 70
\bibitem[\protect\citeauthoryear{Kawata, Cen \& Ho}{Kawata et
    al.}{2007}]{kawata07}
  Kawata D., Cen R., Ho L. C., 2007, ApJ, 669, 232
  \bibitem[\protect\citeauthoryear{Kennicutt}{1998a}]{kennicutt98b}
  Kennicutt R. C., 1998a, ARA\&A, 36, 189
\bibitem[\protect\citeauthoryear{Kennicutt}{1998b}]{kennicutt98a}
  Kennicutt R. C., 1998b, ApJ, 498, 541
\bibitem[\protect\citeauthoryear{Kennicutt et al.}{2008}]{kennicutt08}
  Kennicutt R. C., Lee, J. C., Funes S., J., Jos\'e G., Sakai, S.,
  Akiyama S., 2008, ApJS, 178, 247
\bibitem[\protect\citeauthoryear{Kennicutt et al.}{2009}]{kennicutt09}
  Kennicutt R. C. et al., 2009, ApJ, 703, 1672
\bibitem[\protect\citeauthoryear{Kewley et al.}{2002}]{kewley02}
  Kewley L. J., Geller M. J., Jansen R. A., Dopita M. A., 2002, AJ,
  124, 3135
\bibitem[\protect\citeauthoryear{Knapp at al.}{1989}]{knapp89}
  Knapp G. R., Guhathakurta P., Kim D., Jura M.A., 1989, ApJS, 70, 329
\bibitem[\protect\citeauthoryear{Knapp \& Rupen}{1996}]{knapp96}
  Knapp G. R., Rupen M. P., 1996, ApJ, 460, 271
\bibitem[\protect\citeauthoryear{Kormendy et al.}{2009}]{kormendy09}
  Kormendy J., Fisher D. B. Cornell M. E., Bender R., 2009, ApJS, 182,
  216 
\bibitem[\protect\citeauthoryear{Krips et al.}{2007}]{krips07}
  Krips M. et al., 2007, A\&A, 464, 553
\bibitem[\protect\citeauthoryear{Kuntschner et al.}{2006}]{kuntschner06}
  Kuntschner H., et al., 2006, MNRAS, 369, 497 
\bibitem[\protect\citeauthoryear{Kuntschner et al.}{2010}]{kuntschner09}
  Kuntschner H., et al., 2010, MNRAS, accepted
\bibitem[\protect\citeauthoryear{Lee et al.}{2009}]{lee09}
  Lee J. C. et al., 2009, ApJ, 706, 599
\bibitem[\protect\citeauthoryear{Lonsdale Persson \&
    Helou}{1987}]{lonsdale87}
  Lonsdale Persson C. J., Helou G., 1987, ApJ, 314, 513
\bibitem[\protect\citeauthoryear{Lucero \& Young}{2007}]{lucero07}
  Lucero D. M., Young L. M., 2007, AJ, 134, 2148
\bibitem[\protect\citeauthoryear{Lucero \& Young}{2008}]{lucero08}
  Lucero D. M., Young L. M., 2008, AIPC, 1035, 135
\bibitem[\protect\citeauthoryear{Martig et al.}{2009}]{martig09}
  Martig M., Bournaud F., Teyssier R., Dekel A., 2009, ApJ, 707, 250
\bibitem[\protect\citeauthoryear{McDermid et al.}{2006}]{mcdermid06a}
  McDermid R.M., et al., 2006, MNRAS, 373, 906
\bibitem[\protect\citeauthoryear{Morganti et al.}{2006}]{morganti06}
  Morganti R. et al., 2006, MNRAS, 371, 157
\bibitem[\protect\citeauthoryear{Nilson}{1973}]{nilson73}
  Nilson P., 1973, Uppsala General Catalogue of Galaxies. Nova Acta
  Regiae Soc. Sci. Upsaliensis Astronomiska Observatorium, Uppsala
\bibitem[\protect\citeauthoryear{O'Connell}{1999}]{oconnell99}
  O'Connell R. W., 1999, AR\&A, 37, 603
\bibitem[\protect\citeauthoryear{Okuda et al.}{2005}]{okuda05}
  Okuda T., Kohno K., Iguchi S., Nakanishi K., 2005, ApJ, 620, 673
\bibitem[\protect\citeauthoryear{
}{1989}]{osterbrock89}
  Osterbrock D. E., 1989, Astrophysics of Gaseous Nebulae and Active
Galactic Nuclei (Mill Valley: University Science Books)
\bibitem[\protect\citeauthoryear{Price \& Duric}{1992}]{price92}
  Price R., Duric N., 1992, ApJ, 401, 81
\bibitem[\protect\citeauthoryear{Roussel et al.}{2001}]{roussel01}
  Roussel H., Sauvage M., Vigroux L., Bosma A., 2001, A\&A, 372
\bibitem[\protect\citeauthoryear{S\'anchez-Bl\'azquez et
    al.}{2006}]{sanchezblazquez06}
  S\'anchez-Bl\'azquez P. et al., 2006, MNRAS, 371, 703
\bibitem[\protect\citeauthoryear{Sarzi et al.}{2006}]{sarzi06}
  Sarzi M. et al. 2006, MNRAS, 366, 1151
\bibitem[\protect\citeauthoryear{Sarzi et al.}{2010}]{sarzi09}
  Sarzi M. et al. 2010, MNRAS, 402, 2187
\bibitem[\protect\citeauthoryear{Schawinski et
    al.}{2007a}]{schawinski07}
  Schawinski K., Thomas D., Sarzi M., Maraston C., Kaviraj S., Joo
    S.-J., Yi S. K., Silk J., 2007, MNRAS, 382, 1415
 \bibitem[\protect\citeauthoryear{Schawinski et
    al.}{2007b}]{schawinski07b}
   Schawinski K. et al., 2007, ApJS, 173, 512 
\bibitem[\protect\citeauthoryear{Schilizzi et al.}{1983}]{schilizzi83}
  Schilizzi R. T., Fanti C., Fanti R., Parma P., 1983, A\&A, 126, 412
\bibitem[\protect\citeauthoryear{Schinnerer \&
    Scoville}{2002}]{schinnerer02}
  Schinnerer E., Scoville N., 2002, ApJ, 577, L103
\bibitem[\protect\citeauthoryear{Schmidt}{1959}]{schmidt59}
  Schmidt M., 1959, ApJ, 129, 243
\bibitem[\protect\citeauthoryear{Scott et al.}{2009}]{scott09}
Scott N. et al., 2009, MNRAS, 398, 1835
\bibitem[\protect\citeauthoryear{Shapiro et al.}{2010}]{shapiro09}
  Shapiro K., et al., 2010, MNRAS, 402, 2140
\bibitem[\protect\citeauthoryear{Sil'chenko}{2000}]{silchenko00}
  Sil'chenko O. K., 2000, AJ, 120, 741
\bibitem[\protect\citeauthoryear{Taniguchi et al.}{1994}]{taniguchi94}
  Taniguchi Y., Murayama T., Nakai N., Suzuki M., Kameya O., 1994,
  ApJ, 108, 468
\bibitem[\protect\citeauthoryear{Temi, Brighenti \& Matthews}{Temi et
    al.}{2007}]{temi07}
  Temi P., Brighenti F., Matthews W.G., 2007, ApJ, 660, 1215
\bibitem[\protect\citeauthoryear{Temi, Brighenti \& Matthews}{Temi et
    al.}{2009a}]{temi09}
  Temi P., Brighenti F., Matthews W.G., 2009a, ApJ, 695, 1
\bibitem[\protect\citeauthoryear{Temi, Brighenti \& Matthews}{Temi et
    al.}{2009b}]{temi09b}
  Temi P., Brighenti F., Matthews W.G., 2009b, ApJ, 707, 890
\bibitem[\protect\citeauthoryear{Thomas et al.}{2005}]{thomas05}
  Thomas D., Maraston C., Bender R., Mendes de Oliveira C., 2005, ApJ,
  621, 673
\bibitem[\protect\citeauthoryear{Trager et al.}{2000}]{trager00}
  Trager S. C., Faber S. M., Worthey G., Gonz\'alez J. J., 2000, AJ,
  120, 165
\bibitem[\protect\citeauthoryear{Visvanathan \&
    Sandage}{1977}]{visvanathan77}
  Visvanathan N., Sandage A., 1977, ApJ, 216, 214
\bibitem[\protect\citeauthoryear{Walsh et al.}{1989}]{walsh89}
  Walsh D. E. P., Knapp G. R., Wrobel J. M., Kim D.-W., 1989, ApJ,
  337, 209
\bibitem[\protect\citeauthoryear{Welch \& Sage}{2003}]{welch03}
  Welch G. A., Sage L. J., 2003, ApJ, 584, 260
\bibitem[\protect\citeauthoryear{Weijmans et al.}{2008}]{weijmans08}
  Weijmans A.-M., Kranjović D., van de Ven G., Oosterloo T., Morganti
  R., de Zeeuw P. T., 2008, MNRAS, 383, 1343
\bibitem[\protect\citeauthoryear{Wiklind \& Henkel}{1989}]{wiklind89}
  Wiklind T., Henkel C., 1989, A\&A, 225, 1
\bibitem[\protect\citeauthoryear{Wiklind \& Henkel}{1995}]{wiklind95}
  Wiklind T., Henkel C., 1995, A\&A, 297, L71
\bibitem[\protect\citeauthoryear{Wiklind et al.}{1997}]{wiklind97}
  Wiklind T., Combes F., Henkel C., Wyrowski, F., 1997, A\&A, 323, 727
\bibitem[\protect\citeauthoryear{Wu et al.}{2005}]{wu05}
  Wu H., Cao C., Hao C.-N., Liu F.-S., Wang J.-L., Xia X.-Y., Deng
  Z.-G., Young C. K.-S., 2005, ApJ, 632, L79
\bibitem[\protect\citeauthoryear{Yi et al.}{2005}]{yi05}
  Yi, S. K., et al., 2005, ApJ, 610, L111
\bibitem[\protect\citeauthoryear{Young \& Scoville}{1991}]{young91}
  Young J. S., Scoville N. Z., 1991, ARA\&A, 29, 581
\bibitem[\protect\citeauthoryear{Young et al.}{1996}]{young96}
  Young J. S., Allen L., Kenney J. D. P., Lesser A., Rownd, B., 1996,
  AJ, 112, 1903
\bibitem[\protect\citeauthoryear{Young}{2002}]{young02}
  Young L. M., 2002, ApJ, 124, 788
\bibitem[\protect\citeauthoryear{Young}{2005}]{young05}
  Young L. M., 2005, ApJ, 634, 258 
\bibitem[\protect\citeauthoryear{Young, Bureau \& Cappellari}{Young et
  al.}{2008}]{young08}
  Young L. M., Bureau M., Cappellari M., 2008, ApJ, 676, 317
\bibitem[\protect\citeauthoryear{Young, Bendo \&
    Lucero}{Young et al.}{2009}]{young09}
  Young L. M., Bendo G. J., Lucero D. M., 2009, AJ, 137, 3053
\bibitem[\protect\citeauthoryear{Yun, Reddy \& Condon}{Yun et
    al.}{2001}]{yun01}
  Yun M. S., Reddy N. A., Condon J. J., 2001, ApJ, 554, 803

\end{thebibliography}

\appendix

\section{{\tt SAURON} maps of NGC~2320 and NGC~5666}

\begin{figure*}
\begin{center}
\begin{tabular}[c]{cccc}
& ~~Flux &  ~~log(F$_{[\mathrm{O\:{\scriptsize III}}]}$)  & ~~log(F$_{\mathrm{H\beta}}$)  \\
\begin{sideways}
\phantom{0000000000}arcsec
\end {sideways} &
\rotatebox{0}{\includegraphics[width=3.4cm]
  {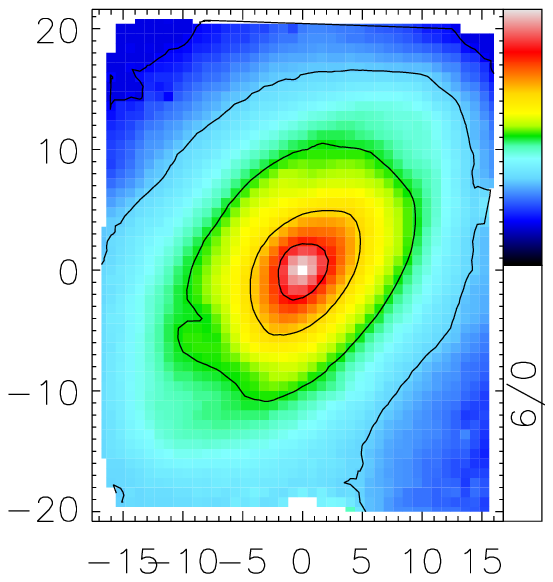}} &
\rotatebox{0}{\includegraphics[width=3.4cm]
  {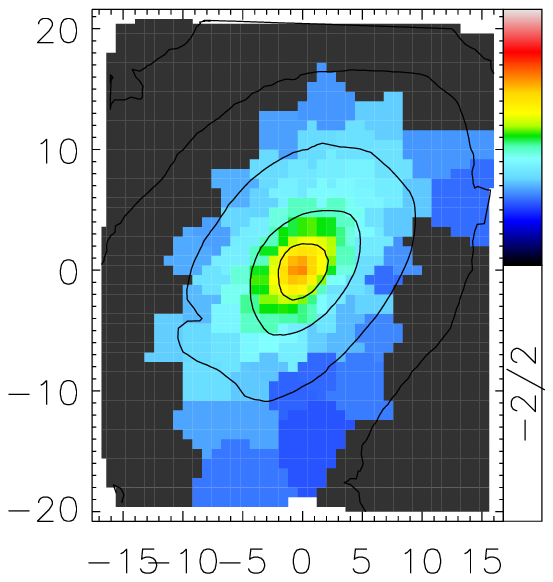}} &
\rotatebox{0}{\includegraphics[width=3.4cm]
  {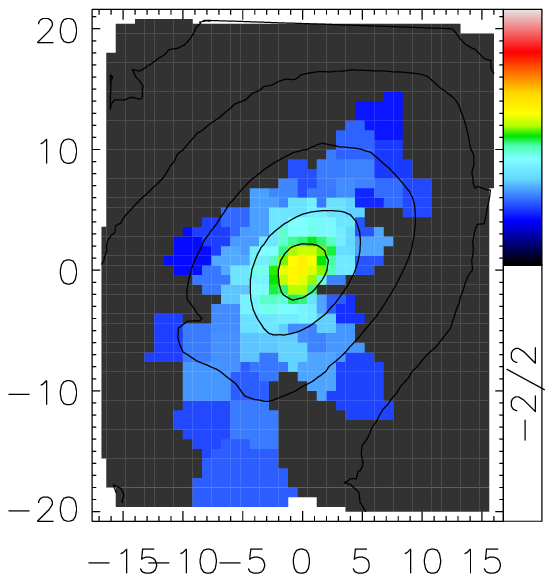}}\\
& ~~Stellar velocity & ~~Gas velocity & ~~log([O$\:${\scriptsize III}]/H$\beta$)\\\begin{sideways}
\phantom{0000000000}arcsec
\end {sideways} &
\rotatebox{0}{\includegraphics[width=3.4cm]
  {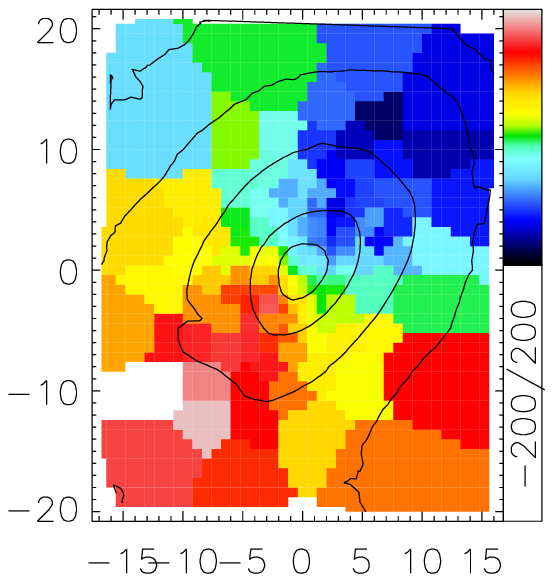}} &
\rotatebox{0}{\includegraphics[width=3.4cm]
  {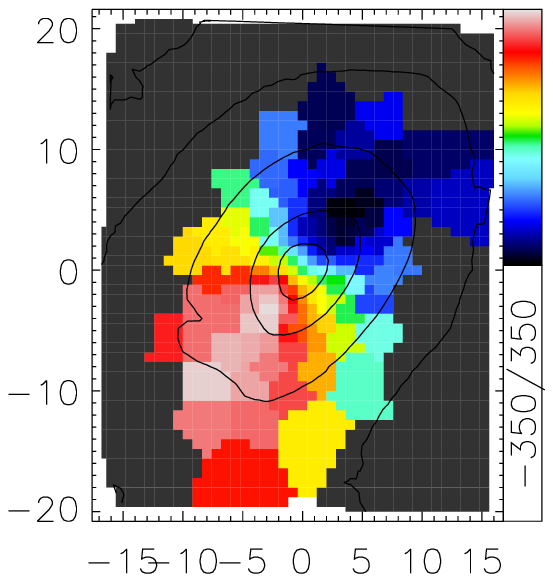}} &
\rotatebox{0}{\includegraphics[width=3.4cm]
  {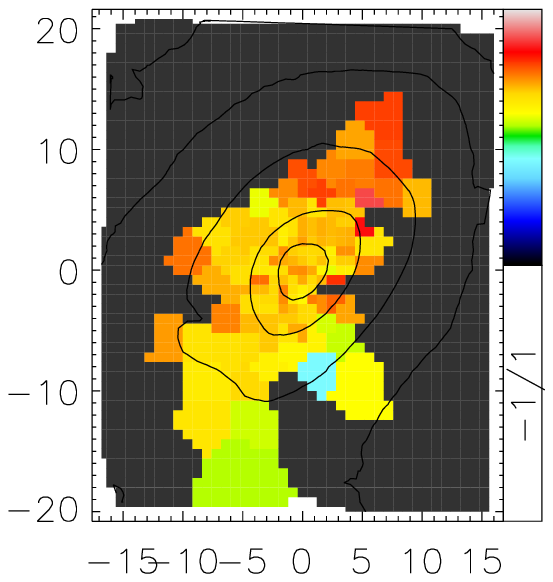}} \\
& ~~Stellar dispersion & ~~Gas dispersion & ~~H$\beta$ linestrength \\
\begin{sideways}
\phantom{0000000000}arcsec
\end {sideways} &
\rotatebox{0}{\includegraphics[width=3.4cm]
  {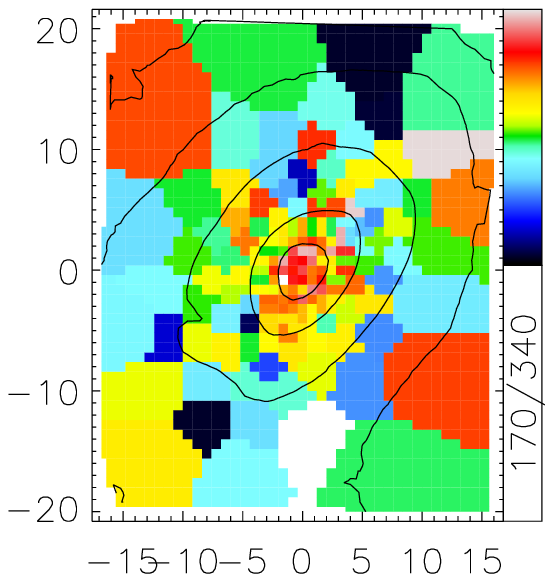}} &
\rotatebox{0}{\includegraphics[width=3.4cm]
  {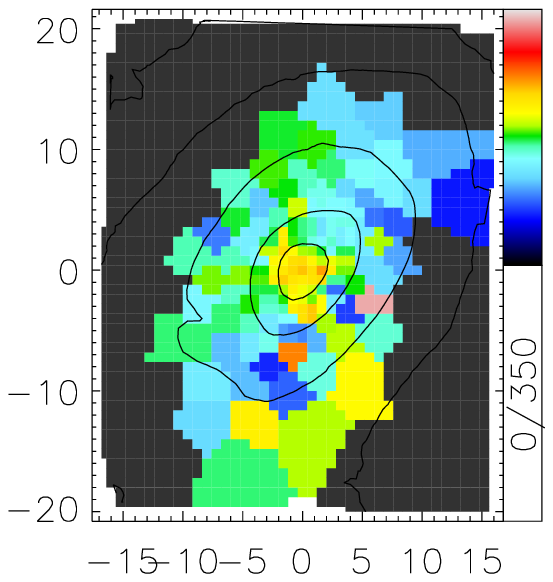}} &
\rotatebox{0}{\includegraphics[width=3.4cm]
  {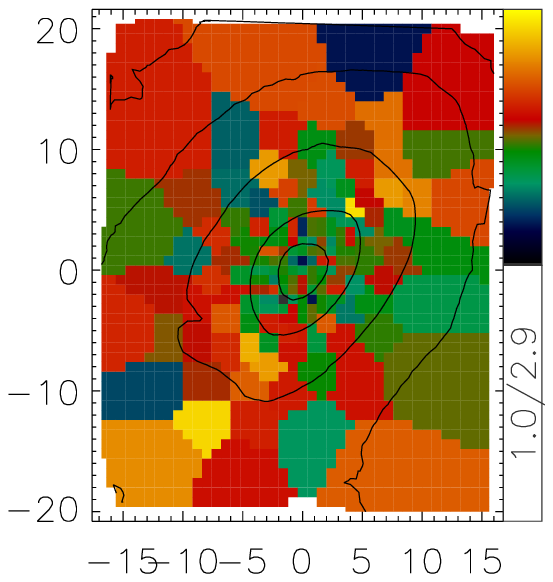}} \\
& ~~Stellar h$_{3}$ & ~~log(EW$_{[\mathrm{O III]}}$) \\
\begin{sideways}
\phantom{0000000000}arcsec
\end {sideways} &
\rotatebox{0}{\includegraphics[width=3.4cm]
  {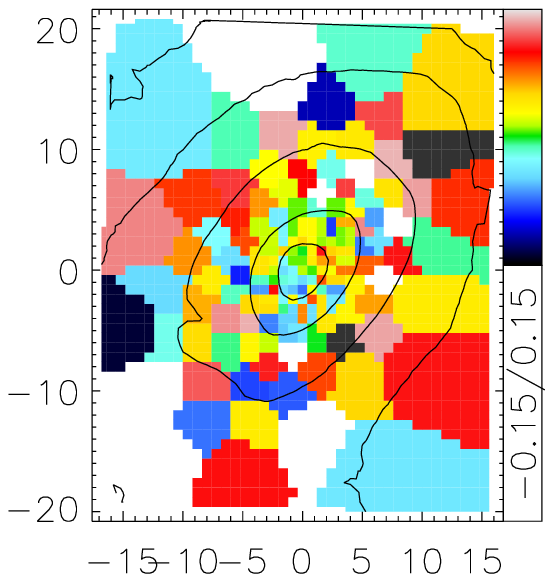}} &
\rotatebox{0}{\includegraphics[width=3.4cm]
  {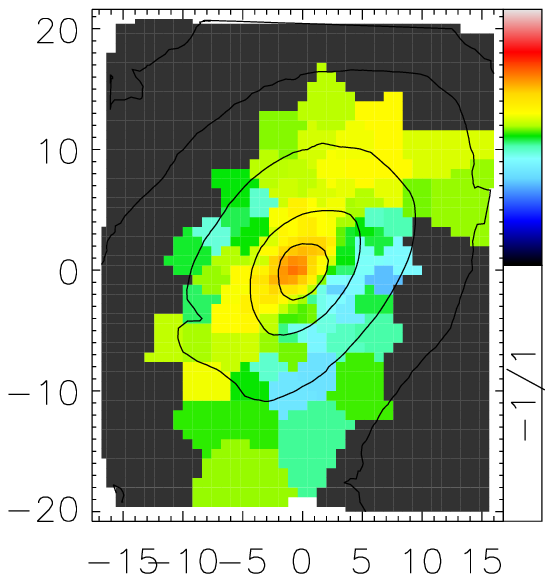}} &
\\
& ~~Stellar h$_{4}$ & ~~log(EW$_{\mathrm H\beta}$) & ~~Fe5015
linestrength \\
\begin{sideways}
\phantom{0000000000}arcsec
\end {sideways} &
\rotatebox{0}{\includegraphics[width=3.4cm]
  {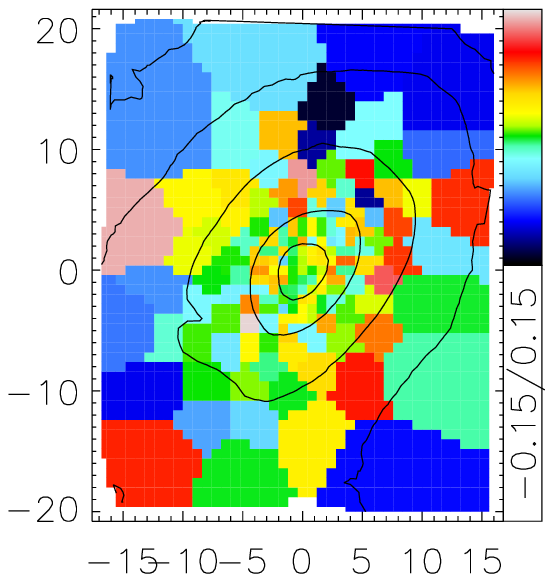}} &
\rotatebox{0}{\includegraphics[width=3.4cm]
  {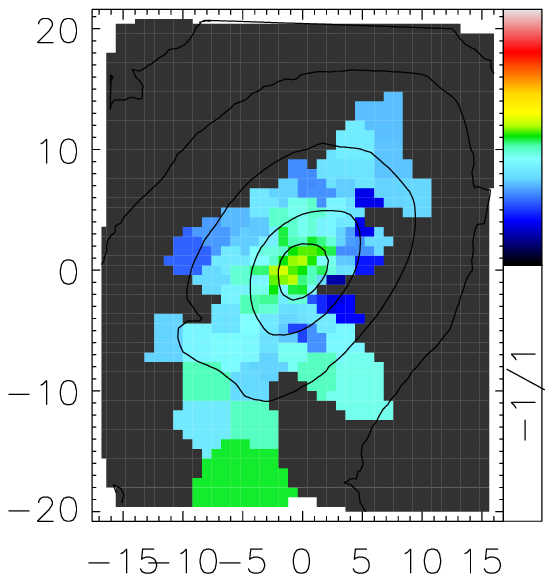}} &
\rotatebox{0}{\includegraphics[width=3.4cm]
  {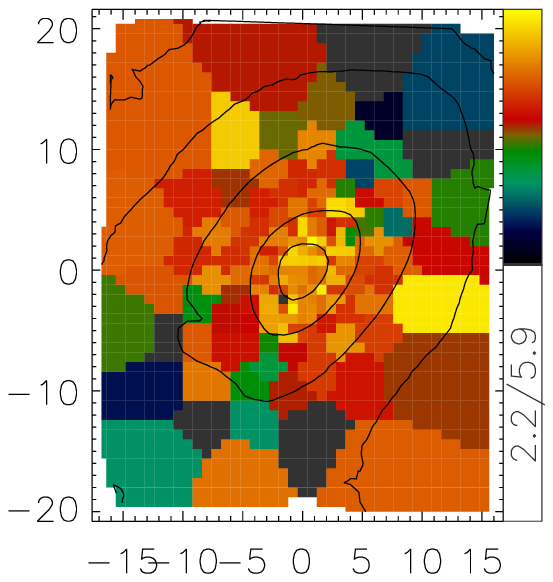}} \\
& arcsec & arcsec & arcsec\\
\\

\end{tabular}
\end{center}
\caption{{\tt SAURON} maps of NGC~2320. From top to bottom, left to right:
  {\it first column} -- reconstructed intensity (arbitrary units),
  mean stellar velocity (km s$^{-1})$, stellar velocity dispersion (km
  s$^{-1}$), stellar Gauss-Hermite coefficients $h_{3}$ and 
  $h_{4}$ (dimensionless); {\it second column} -- [O$\:${\scriptsize
  III}] emission line flux (10$^{-16}$ erg s$^{-1}$ cm$^{-2}$
  arcsec$^{-2}$), [O$\:${\scriptsize III}] velocity (km s$^{-1}$), 
  [O$\:${\scriptsize III}] velocity dispersion (km s$^{-1}$), [O$\:${\scriptsize III}] equivalent
  width (\AA), H$\beta$ equivalent width (\AA); {\it third column}
  -- H$\beta$ emission line flux (10$^{-16}$ erg s$^{-1}$ cm$^{-2}$
  arcsec$^{-2}$), [O$\:${\scriptsize III}]/H$\beta$ line ratio
  (dimensionless), H$\beta$ Lick 
  absorption line index (\AA),
  Fe5015 Lick absorption line index (\AA). The
  isophotes  over-plotted are in half-magnitude steps. Grey bins in the
  emission-line maps indicate that emission was not detected.}
\label{fig:ifu2320}
\end{figure*}

\begin{figure*}
\begin{center}
\begin{tabular}[c]{cccc}
& ~~Flux &  ~~log(F$_{[\mathrm{O\:{\scriptsize III}}]}$)  & ~~log(F$_{\mathrm{H\beta}}$)  \\
\begin{sideways}
\phantom{0000000000}arcsec
\end {sideways} &
\rotatebox{0}{\includegraphics[width=3.8cm]
  {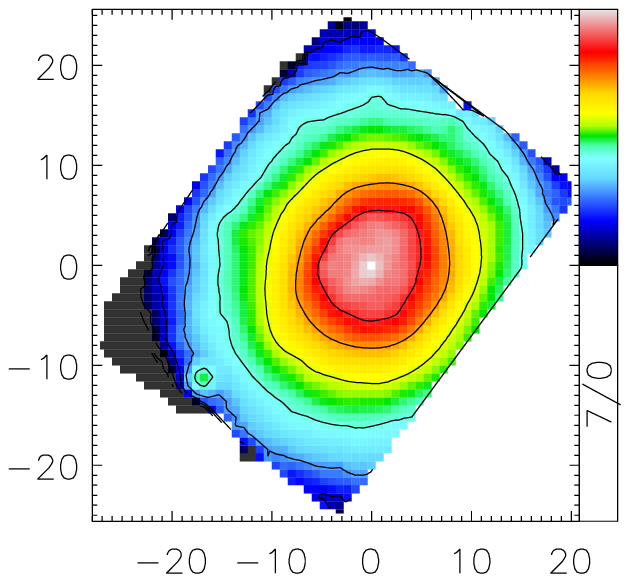}} &
\rotatebox{0}{\includegraphics[width=3.8cm]
  {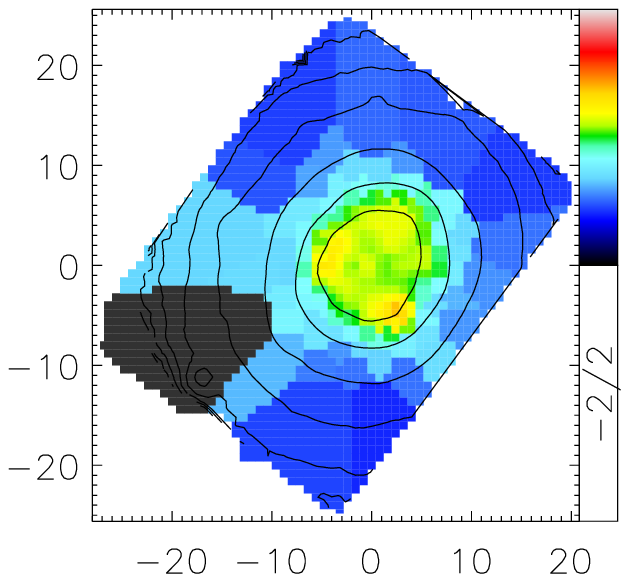}} &
\rotatebox{0}{\includegraphics[width=3.8cm]
  {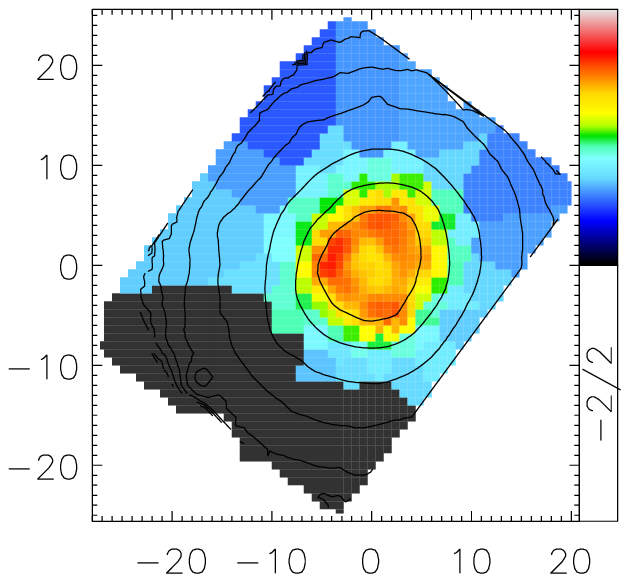}}\\
& ~~Stellar velocity & ~~Gas velocity & ~~log([O$\:${\scriptsize
  III}]/H$\beta$)\\
\begin{sideways}
\phantom{0000000000}arcsec
\end {sideways} &
\rotatebox{0}{\includegraphics[width=3.8cm]
  {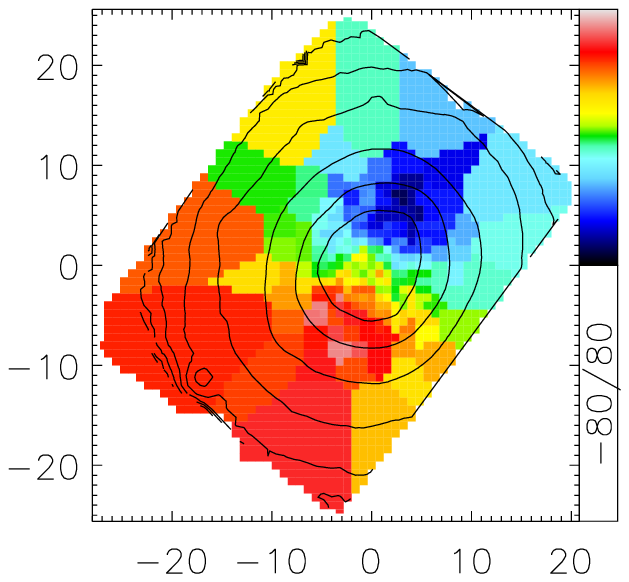}} &
\rotatebox{0}{\includegraphics[width=3.8cm]
  {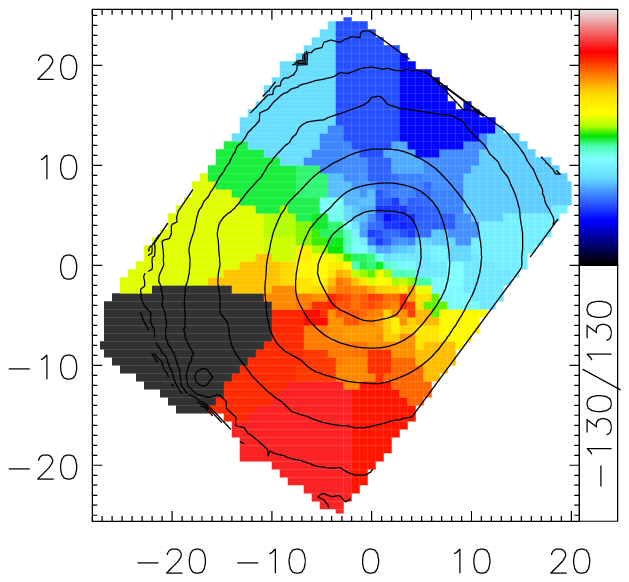}} &
\rotatebox{0}{\includegraphics[width=3.8cm]
  {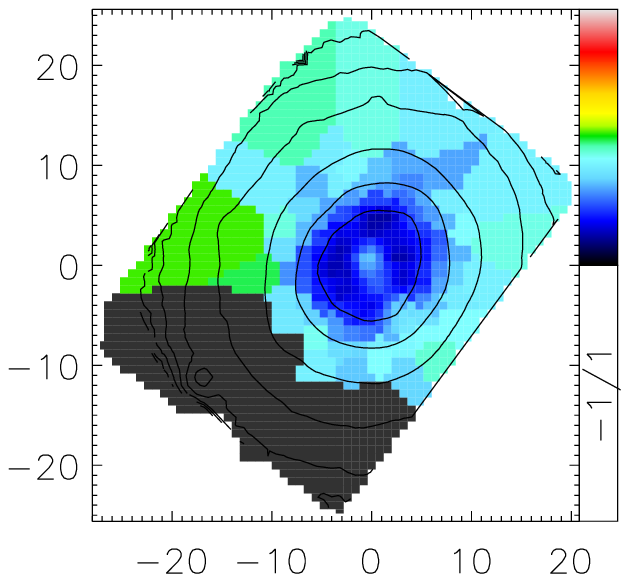}} \\
& ~~Stellar dispersion & ~~Gas dispersion & ~~H$\beta$ linestrength\\
\begin{sideways}
\phantom{0000000000}arcsec
\end {sideways} &
\rotatebox{0}{\includegraphics[width=3.8cm]
  {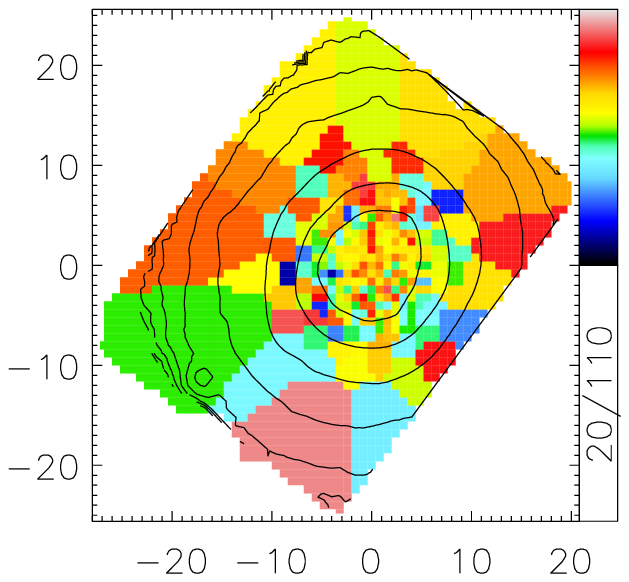}} &
\rotatebox{0}{\includegraphics[width=3.8cm]
  {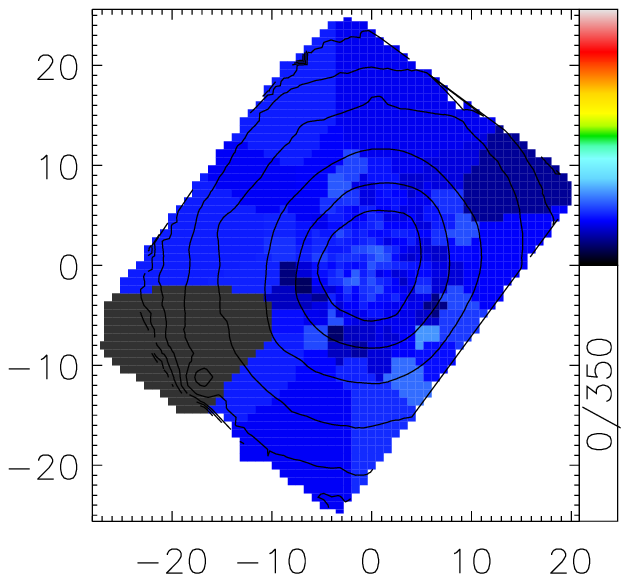}} &
\rotatebox{0}{\includegraphics[width=3.8cm]
  {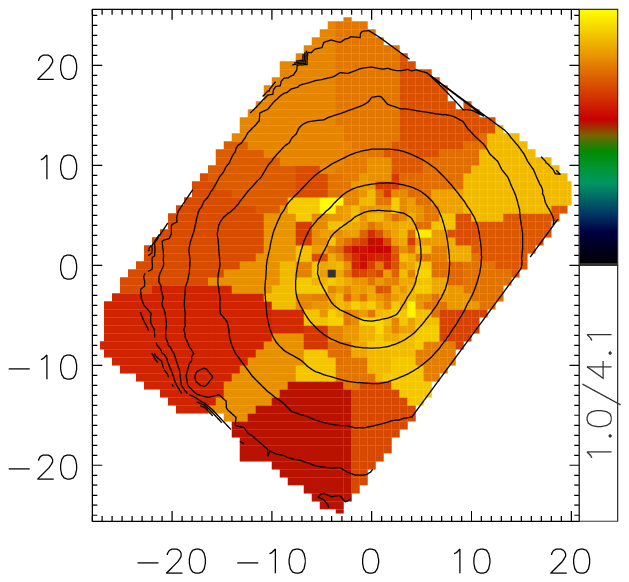}} \\
& ~~Stellar h$_{3}$ & ~~log(EW$_{[\mathrm{O \small III]}}$) & ~~Mgb
  linestrength \\
\begin{sideways}
\phantom{0000000000}arcsec
\end {sideways} &
\rotatebox{0}{\includegraphics[width=3.8cm]
  {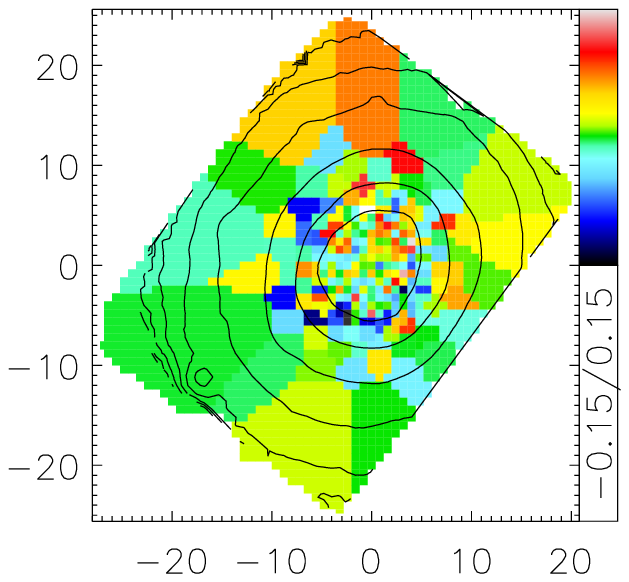}} &
\rotatebox{0}{\includegraphics[width=3.8cm]
  {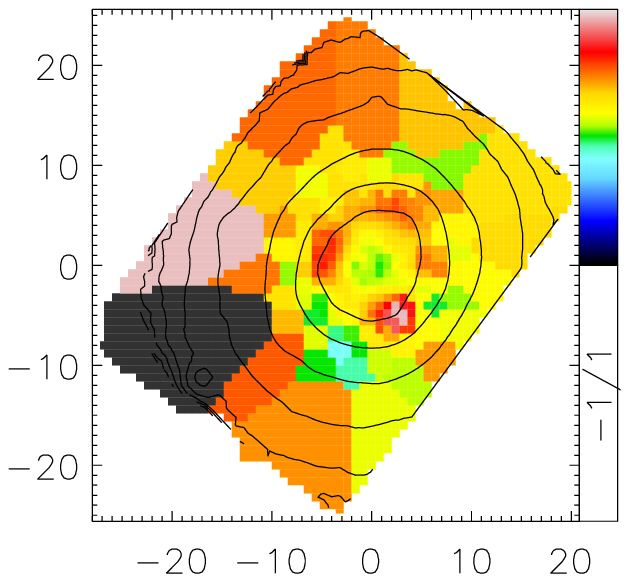}} &
\rotatebox{0}{\includegraphics[width=3.8cm]
  {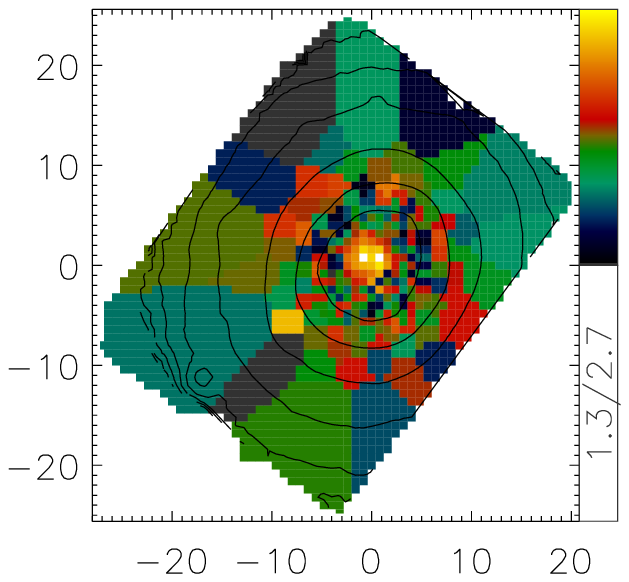}}
\\
& ~~Stellar h$_{4}$ & ~~log(EW$_{\mathrm H\beta}$) & ~~Fe5015
  linestrength\\
\begin{sideways}
\phantom{0000000000}arcsec
\end {sideways} &
\rotatebox{0}{\includegraphics[width=3.8cm]
  {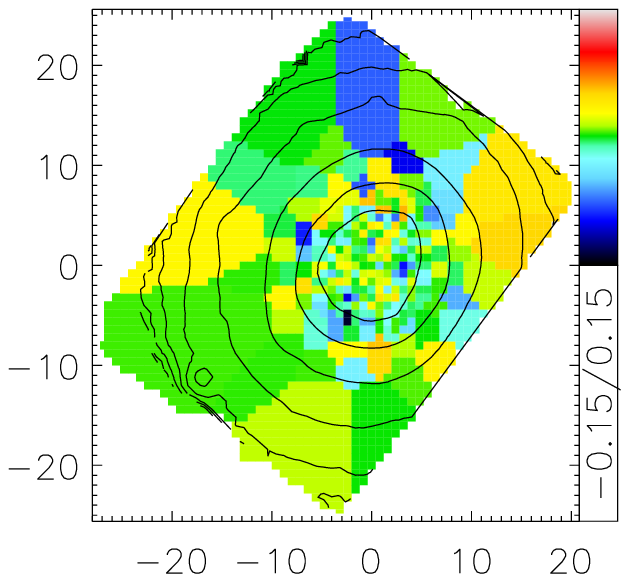}} &
\rotatebox{0}{\includegraphics[width=3.8cm]
  {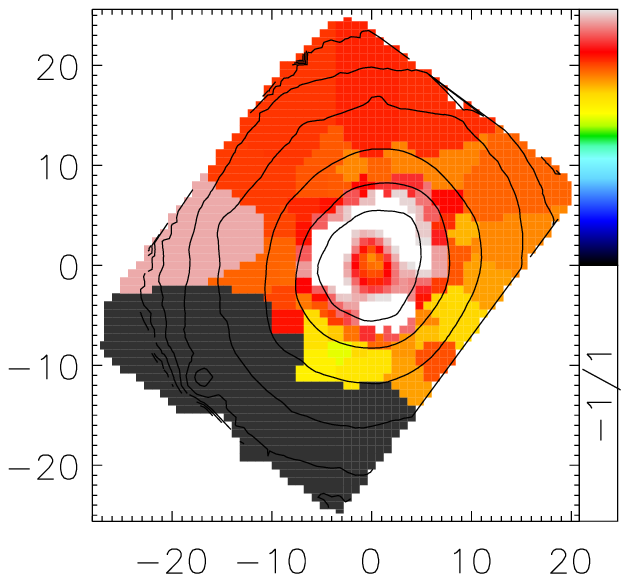}} &
\rotatebox{0}{\includegraphics[width=3.8cm]
  {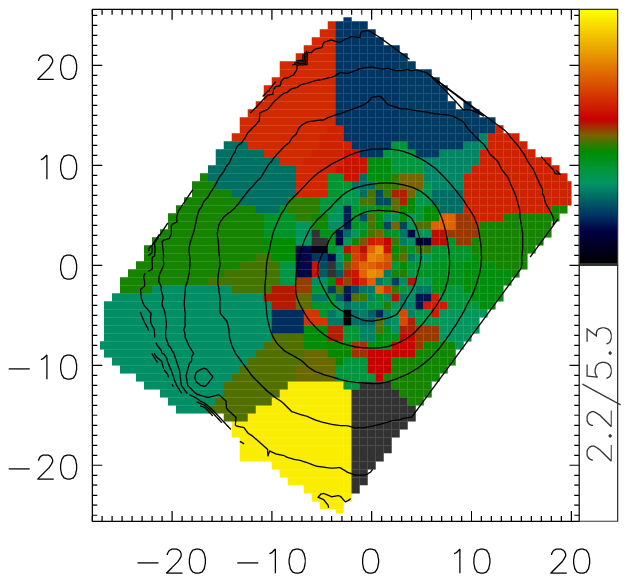}} \\
& arcsec & arcsec & arcsec\\

\end{tabular}
\end{center}
\caption{{\tt SAURON} maps of NGC~5666. From top to bottom, left to right:
  {\it first column} -- reconstructed intensity (arbitrary units),
  mean stellar velocity (km s$^{-1})$, stellar velocity dispersion (km
  s$^{-1}$), stellar Gauss-Hermite coefficients $h_{3}$ and 
  $h_{4}$ (dimensionless); {\it second column} -- [O$\:${\scriptsize
  III}] emission line flux (10$^{-16}$ erg s$^{-1}$ cm$^{-2}$
  arcsec$^{-2}$), [O$\:${\scriptsize III}] velocity (km s$^{-1}$), 
  [O$\:${\scriptsize III}] velocity dispersion (km s$^{-1}$), [O$\:${\scriptsize III}] equivalent
  width (\AA), H$\beta$ equivalent width (\AA); {\it third column}
  -- H$\beta$ emission line flux (10$^{-16}$ erg s$^{-1}$ cm$^{-2}$
  arcsec$^{-2}$), [O$\:${\scriptsize III}]/H$\beta$ line ratio
  (dimensionless), H$\beta$ Lick 
  absorption line index (\AA), Mgb Lick absorption line index (\AA),
  Fe5015 Lick absorption line index (\AA). The
   isophotes over-plotted are in half-magnitude steps. Grey bins in the
  emission-line maps indicate that emission was not detected.}
\label{fig:ifu5666}
\end{figure*}

Being outside of the main {\tt SAURON} sample, the IFU maps of
NGC~2320 and NGC~5666 have not previously been published. We
present these maps in Figs.~\ref{fig:ifu2320} and \ref{fig:ifu5666},
then briefly summarise the properties revealed for each
galaxy.  

Observations of NGC~2320 and NGC~5666
were taken during {\tt SAURON} run 1 and run 8, with 4 $\times$ 1800s
and 3 $\times$ 1800s exposures, respectively. \citet{bacon01}
describes the steps taken in the {\sc XSAURON} software to create the
datacube: bias and dark subtraction, extraction of the spectra,
wavelength calibration, flat-fielding, cosmic ray removal, sky
subtraction and flux calibration. The stellar kinematics are fit using
the penalised pixel fitting method \citep[{\sc pPXF},][]{capellari04}
and the gas emission and kinematics are fit using {\sc GANDALF}
\citep[Gas AND Absorption Line Fitting algorithm;][]{sarzi06}. The
linestrengths are determined as described in \citet{kuntschner06}. Due
to the higher recessional velocity of NGC~2320, the Mgb line was
redshifted out of the {\tt SAURON} wavelength range and is thus not
available. The maps are made using data Voronoi binned spatially
\citep{cappellari03} to a minimum signal to noise of $\approx60$ per spectral pixel.

\subsection{NGC~2320}

The stellar velocity map of NGC~2320 shows clear rotation and its
stellar velocity dispersion map has a central peak (see Fig.~\ref{fig:ifu2320}). The third and
fourth Gauss-Hermite moments are noisy. Based on these data,
\citet{cappellari07} calculate a $\lambda_{R_\mathrm{e}}$ parameter
for NGC~2320 of 0.34, a measure of the
projected stellar angular momentum per unit mass. NGC~2320's
relatively high value puts it into the class of fast-rotating
galaxies, which is surprising given its large mass.  On a diagram of
$\lambda_{{R_\mathrm{e}}}$ versus virial mass, NGC~2320 clearly sticks out;
all other such massive galaxies are slow rotators \citep[see fig. 11
  of][]{emsellem07}. 

The ionised gas in NGC~2320 corotates with the stars. The high central
gas velocity dispersion points to AGN activity, which is supported by
the peaks in the equivalent widths of both the [O$\:${\small III}] and H$\beta$ emission
lines and the compact radio emission observed by \citet{lucero07}. The
[O$\:${\small III}] equivalent width map shows a structure elongated along the
major axis of the galaxy. The [O$\:${\small III}]/H$\beta$ ratio is
high throughout 
the region of detected emission, indicating that star formation is
unlikely to be dominating the ionisation. The luminosity-weighted average H$\beta$ and Fe5015 
absorption linestrengths within {\it R}$_{\mathrm e}$/8 (4.6 arcsec) are
1.80 and 5.11 \AA, respectively. Outside of this radius, the bins become
much bigger and the signal-to-noise is marginal for measuring
linestrengths, as is evidenced by the wide spread in adjacent
bins. Unfortunately, the
Mgb feature is redshifted outside of the {\tt SAURON} spectral range
for this galaxy.

\subsection{NGC~5666}

NGC~5666 has an obvious face-on star-forming ring in its centre (see Fig.~\ref{fig:ifu5666}). The stellar
kinematics show that the rotation of the ring is aligned with the rest
of the galaxy and there are signs of a slightly lower stellar
velocity dispersion within the ring. 

The ring can be seen clearly in the flux and equivalent width maps for
both the H$\beta$ and [O$\:${\small III}] emission lines. Moreover, hot spots can be seen within the ring,
presumably the sites of the most active star formation. The entire
region has a low [O$\:${\small III}]/H$\beta$
ratio that can only be caused by star formation. The gas rotates like
the stars and has no central dispersion peak to suggest an AGN.  

The linestrengths also fit the picture of a star-forming
ring. The H$\beta$ linestrength is high everywhere, but highest in the
ring. Breaking the galaxy into three projected regions of radius under
3 arcsec, radius between 3 and 9 arcsec and radius over 9 arcsec, the
luminosity-averaged H$\beta$ linestrengths are 3.17, 3.56 and
3.33 \AA. Young stars must contribute to this high linestrength everywhere,
but even more so within the ring. Low Mgb and Fe5015
values are found in the ring and outside of it but both clearly peak
in the centre. Mgb is partially age-sensitive and decreases with a
young stellar population, as is clearly seen in the youngest {\tt
  SAURON} galaxies (NGC~3032, NGC~3156, NGC~4150). The Fe5015 index is
less age sensitive and the low values in NGC~5666 may suggest a
genuine low-metallicity population (as in NGC~3032;
\citealp{kuntschner09}).

\label{lastpage}

\end{document}